%% file: DPG_2column.tex
\documentclass[journal]{IEEEtran}
\usepackage{ifpdf}
\usepackage{cite}
\ifCLASSINFOpdf
 \usepackage[pdftex]{graphicx}
 \DeclareGraphicsExtensions{.pdf,.jpeg,.png}
\else
 \usepackage[dvips]{graphicx}
 \DeclareGraphicsExtensions{.eps}
\fi
\usepackage{amssymb}
\usepackage[cmex10]{amsmath}
\usepackage{array}
\usepackage{mdwmath}
\usepackage{mdwtab}
\usepackage{eqparbox}
\usepackage[tight,footnotesize]{subfigure}
\usepackage{stfloats}

\usepackage{bm}
\usepackage{overpic}
\usepackage{color}
\usepackage{dsfont}
\usepackage{amsthm}
\usepackage{setspace}
\usepackage{mathbbol}
\usepackage[ruled,norelsize]{algorithm2e}
\usepackage{bbm}
\usepackage{boxedminipage}

\usepackage{pgfplots} 
\pgfplotsset{compat=newest} 
\pgfplotsset{plot coordinates/math parser=false} 
\newlength\figureheight 
\newlength\figurewidth

\newcommand{\T}{\top}
\newcommand{\mc}{\mathcal}
\newcommand{\defeq}{\triangleq}

\newcommand{\Na}{\mathds{N}}

\newcommand{\Q}{\mathcal{Q}}
\newcommand{\X}{\mathcal{X}}
\newcommand{\U}{\mathcal{U}}
\newcommand{\C}{\mathcal{C}}
\newcommand{\mix}{\mathbf{x}}
\newcommand{\miu}{\mathbf{u}}
\newcommand{\miF}{\mathbf{F}}
\newcommand{\milambda}{\boldsymbol{\lambda}}
\newcommand{\mimu}{\boldsymbol{\mu}}
\newcommand{\migamma}{\boldsymbol{\gamma}}
\newcommand{\midelta}{\boldsymbol{\delta}}
\newcommand{\mieta}{\boldsymbol{\eta}}
\newcommand{\mixi}{\boldsymbol{\xi}}
\newcommand{\miR}{\mathbf{R}}
\newcommand{\miD}{\mathbf{D}}
\newcommand{\miQ}{\mathbf{Q}}
\newcommand{\miB}{\mathbf{B}}
\newcommand{\miC}{\mathbf{C}}
\newcommand{\miA}{\mathbf{A}}
\newcommand{\miP}{\mathbf{P}}
\newcommand{\mic}{\mathbf{c}}
\newcommand{\miM}{\mathbf{M}}
\newcommand{\mia}{\mathbf{a}}

\newtheorem{definition}{Definition}
\newtheorem{assumption}{Assumption}
\newtheorem{lemma}{Lemma}
\newtheorem{theorem}{Theorem}

\newlength{\eqboxstorage}

\newcommand{\textblue}{\textcolor{black}}

\allowdisplaybreaks 

\title{Dynamic Potential Games with Constraints: Fundamentals and Applications in Communications}

\begin{document}

\author{
Santiago~Zazo,~\IEEEmembership{Member,~IEEE,}
~Sergio~Valcarcel~Macua,~\IEEEmembership{Student Member,~IEEE,}
\\
~Matilde~S\'{a}nchez-Fern\'{a}ndez, \IEEEmembership{Senior Member,~IEEE,} 
~Javier Zazo\thanks{This work has been partly funded by the Spanish Ministry of Economy and Competitiveness under the grant TEC2013-46011-C3-1-R, 
by the Spanish Ministry of Science and Innovation with the project ELISA (TEC2014-59255-C3-R3)
and by an FPU doctoral grant to the fourth author.}
\thanks{S. Zazo, S. Valcarcel Macua and J. Zazo are with the Signals, Systems \& Radiocommunications Dept., Universidad Polit\'ecnica de Madrid.  E-mail: {\tt \{santiago,sergio\}@gaps.ssr.upm.es, javier.zazo.ruiz@upm.es}.}
\thanks{M. S\'anchez-Fern\'andez is with the Signal Theory \& Communications Department, Universidad Carlos III de Madrid.  E-mail: {\tt mati@tsc.uc3m.es}.}
}


\maketitle

\begin{abstract}
In a noncooperative dynamic game, 
multiple agents operating in a changing environment aim to optimize their utilities over an infinite time horizon.
Time-varying environments allow to model more realistic scenarios
(e.g., mobile devices equipped with batteries, 
wireless communications over a fading channel, etc.).
However, solving a dynamic game is a difficult task that requires dealing with multiple coupled optimal control problems.
We focus our analysis on a class of problems, named \textit{dynamic potential games}, 
whose solution can be found through a single multivariate optimal control problem.
Our analysis generalizes previous studies 
by considering that the set of environment's states 
and the set of players' actions are constrained, 
as it is required by most of the applications.
We also show that the theoretical results are the natural extension of the analysis for static potential games.
We apply the analysis 
and provide numerical methods to solve four key example problems, with different features each: 
\textit{i)} energy demand control in a smart-grid network, 
\textit{ii)} network flow optimization in which the relays have bounded link capacity and limited battery life,
\textit{iii)} uplink multiple access communication with users that have to optimize the use of their batteries, 
and \textit{iv)} two optimal scheduling games with nonstationary channels.
\end{abstract}


%
\IEEEpeerreviewmaketitle

\section{Introduction}
\label{sec:intro}

\IEEEPARstart{G}{ame} 
theory is a field of mathematics that studies conflict and cooperation between intelligent decision makers \cite{Han2012GameTheory}.
It has become a useful tool for modeling communication and networking problems, 
such as power control and resource sharing (see, e.g.,\cite{Scutari2010}),
wherein the strategies followed by the users (i.e., players) influence each other,
and the actions have to be taken in a decentralized manner.
%
%
However, one main assumption of classic game theory is that the users operate in a \textit{static} environment,
which is not influenced by the players' actions.
This assumption is unrealistic in many communication and networking problems.
For instance, wireless devices have to maximize throughput while facing time-varying fading channels,
and 
mobile devices may have to control their transmitter power while saving their battery level.
These time-varying scenarios can be better modeled by \textit{dynamic games}.

In a \textit{noncooperative dynamic game},
the players compete in a time-varying environment,
which we assume can be characterized by a deterministic discrete-time dynamical system
equipped with a set of states and a Markovian state-transition equation.
Each player has its utility function,
which depends on the current state of the system and the players' current actions.
Both the state and action sets are subject to constraints.
Since the state-transitions induce a notion of time-evolution in the game, 
we consider the general case wherein utilities, state-transition function and constraints can be nonstationary.
A dynamic game starts at an initial state.
Then, the players take some action,
based on the current state of the game,
and receive some utility values. 
Then, the game moves to another state.
This sequence of state-transitions is repeated at every time step
over a (possibly) infinite time horizon.
We consider the case in which the aim of each player is to find the sequence of actions that maximizes its long term cumulative utility,
given other players' sequence of actions.
Thus, a game can be represented as a set of coupled optimal-control-problems (OCP),
which are difficult to solve in general.
Fortunately, there is a class of dynamic games,
named \textit{dynamic potential games} (DPG),
that can be solved through a single multivariate-optimal-control-problem (MOCP).
The benefit of DPG is that solving a single MOCP is generally simpler than solving a set of coupled OCP (see \cite{HernandezLerma2013} for a recent survey on DPG). 

The pioneering work in the field of DPG is that of \cite{Dechert1978},
later extended by \cite{Dechert1997} and \cite{Dechert2000}.
There have been two main approaches to study DPG:
the Euler-Lagrange equations 
and the Pontryagin's maximum (or minimum) principle. 
%
Recent analysis by \cite{HernandezLerma2013} and \cite{Gonzalez-Sanchez2014}
used the Euler-Lagrange with DPG in its \textit{reduced form},
that is when it is possible to isolate the action from the state-transition equation, 
so that the action is expressed as a function of the current and future (i.e., after transition) states.
\textblue{Consider, for example, that the future state is linear in the current action;
then, it is easy to invert the state-transition function and 
rewrite the problem in reduced form, 
with the action expressed as a function of the current and future states.
However, in many cases, 
it is not possible to find such reduced form of the game (i.e., we cannot isolate the action)
because the state-transition function is not invertible
(e.g., when the state transition function is quadratic in the action variable).}
The more general case of DPG in \textit{nonreduced form} was studied 
with the Pontryagin's maximum principle approach
by \cite{Dechert1997} and \cite{Dragone2009} for discrete and continuous time models, respectively. 
However, in all these studies \cite{HernandezLerma2013,Dechert1978,Dechert1997,Dechert2000,Gonzalez-Sanchez2014,Dragone2009}, 
the games have been analyzed without explicitly considering constraints for the state and action sets.

\textblue{Other works that consider potential games with state-dynamics include
\cite{MardenStateBased2012,LiDesigningGames2013,LiDecoupling2014}.
However, these references study the \textit{myopic} problem  
in which the agents aim to maximize their immediate reward.
This is different from DPG, where the agents aim to maximize their long term utility by solving a control problem.}

\textblue{Dynamic games offer two kinds of possible analysis based on the type of control that players use. 
These cases are normally referred to as \emph{open loop} (OL) and \emph{closed loop} (CL) game analysis. 
	In the open loop approach, in order to find the optimal action sequence, 
	the players have to take into account other players' action sequences.
	On the other hand, in a closed loop approach, players find a strategy that is a function of the \emph{state}, i.e., it is a mapping from states to actions.
	Thus, in order to find their optimal policies, they need to know the form of other players' policy functions.
	The OL analysis has, in general, more tractable analysis than the CL analysis.
	Indeed, there are only few CL known solutions for simple games, 	
	such as the fish war example presented in~\cite{Amir2006}, oligopolistic Cournot games~\cite{Dockner1988}, 
	or quadratic games~\cite{Kydland1975}.}

The main theoretical contribution of this work is to analyze DPG with constrained action and state sets,
as it is required by most of applications
(e.g., in a network flow problem, the aggregated throughput of multiple users is bounded by the maximum link capacity;
or in cognitive radio, the aggregated power of all secondary users is bounded by the maximum interference allowed by the primary users).
To do so, we apply the Euler-Lagrange equation to the Lagrangian (as it is customary in the MOCP literature \cite{Sage77}), 
rather than to the utility function (as done by earlier works \cite{HernandezLerma2013} and \cite{Gonzalez-Sanchez2014}).
Using the Lagrangian, we can formulate the optimality condition in the general \textit{nonreduced form} 
(i.e., it is not necessary to isolate the action in the transition equation).
In addition, we establish the existence of a suitable conservative vector field as an easily verifiable condition for a dynamic game to be of the potential type.
To the best of our knowledge, 
this is a novel extension of the conditions established for static games by \cite{Slade1994a} and \cite{Monderer1996}.

The second main contribution of this work is to show that the proposed framework can be applied to several communication and networking problems in a unified manner.
We present four examples with increasing complexity level.  
First, we model the \textit{energy demand control} in a \textit{smart grid network} as a linear-quadratic-dynamic-game (LQDG).
This scenario is illustrative because the analytical solution of an LQDG is known.
The second example is an \textit{optimal network flow problem},
in which there are two levels of relay nodes equipped with finite batteries.
The users aim to maximize their flow while optimizing the use of the nodes' batteries.
This problem illustrates that, 
when the utilities have some separable form, it is straightforward to establish that the problem is a DPG.
However, the analytical solution for this problem is unknown and we have to solve it numerically.
It turns out that,
since all batteries will deplete eventually, the game will get stuck in this depletion-state.
Hence, we can approximate the infinite-horizon MOCP by an effective finite-horizon problem,
which simplifies the numerical computation.
The third example is an \textit{uplink multiple access channel} wherein the users' devices are also equipped with batteries (this example was introduced in the preliminary paper \cite{Zazo2014}).
Again, the simple---but more realistic---extension of battery-usage optimization makes the game dynamic. 
In this example, instead of rewriting the utilities in a separable form,
we perform a very general analysis to establish that the problem is a DPG.
The fourth example studies two decentralized scheduling problems: 
\textit{proportional fair} and \textit{equal rate scheduling}, where multiple users share a time-varying channel
(see the preliminary paper \cite{Zazo2015}). 
This example shows how to use the proposed framework in its most general form.
The problems are nonconcave and the utilities have \textblue{a} nonobvious separable form. 
The problem is nonstationary, with state-transition equation changing with time.
And there is no reason that justifies a finite horizon approximation of the problem, 
so we have to use optimal control methods (e.g., dynamic programming) to solve it numerically.

%
\textbf{Outline:} Sec. \ref{sec:problem-setting} introduces the problem setting,
its solution and the assumptions \textblue{on} which we base our analysis.
In Sec. \ref{sec:static-potential-games},
we review \textit{static} potential games
together with the instrumental notion of conservative vector field.
In Sec. \ref{sec:dynamic-potential-games},
we provide sufficient conditions for a dynamic game with constrained state and action sets to be a DPG, 
and show that a DPG can be solved through and equivalent MOCP.
Sections \ref{sec:smart-grid-model}--\ref{sec:scheduling} deal with application examples,
the methods for solving them, 
and some illustrative simulations.
We provide some conclusions in Sec. \ref{sec:conclusions}.


\section{Problem Setting} 
\label{sec:problem-setting}

Let $\Q \defeq \{1, \ldots, Q \}$ denote the set of players
and
let $\X \subseteq \Re^S$ denote the set of states of the game.
Note that the dimensionality of the state set can be different from the number of players (i.e., $S \neq Q$).
At every time step $t$, 
the state-vector of the game is represented by
$
\mix_t 
\defeq  
	\left(
		x_t^k
	\right)_{k=1}^S
	\in \X
$.
Every player $i\in\Q$ can be influenced only by a subset of states $\X^i \subseteq \X$.
The partition of the state space $\X$
among players is done in the component domain. 
We define $\X(i) \subseteq \{1, \ldots, S \}$
as the subset of indexes of state-vector components that influence player $i$,
then 
$
\mix_t^i 
	\defeq 
		\left( 
			x_t^m 
		\right)_{m \in \X(i)}
$
indicates the value of the state-vector for player $i$
at time $t$.
This generality allows for games in which multiple players are affected by common components of the state vector (e.g., when they share a common resource),
and includes the particular case wherein they share no components. 
We also define
$
\mix_t^{-i} 
	\defeq 
		\left( 
			x_t^l 
		\right)_{l \notin \X(i)}
	\in 
		\X^{-i}
$
for the vector of components that do not influence player $i$,
for some subset 	$\X^{-i} \subseteq \X$.

Let $\U \subseteq \Re^Q$ 
denote the set of actions of all players,
and let $\U^i \subseteq \Re$ stand for the subset of actions of player $i$,
such that $\U \defeq \prod_{i=1}^Q \U^i$.
The extension to higher dimensional action sets is straightforward
(i.e., when $\U^i \subseteq \Re^{A^i}$),
but we restrict to scalar actions in order to simplify notation
(the general case will be introduced when necessary for some of the application examples).
We write $u_t^i \in \U^i$ the action variable of player $i$ at time $t$,
such that the vector
$\miu_t 
	\defeq 
		\left(
			u_t^1, \ldots, u_t^Q
		\right)
\in \U
$
contains the actions of all players.
We also define 
$\miu_t^{-i} 
	\defeq 
		\left(
			u_t^1, \ldots, u_t^{i-1}, u_t^{i+1}, \ldots, u_t^Q
		\right)
\in 
	\U^{-i} 
\defeq
	\prod_{j \neq i} \U^j$ as
the vector of all players' actions except that of player $i$.
Hence, by slightly abusing notation, 
we can rewrite
$\miu_t 
	=
		\left(
			u_t^i, \miu_t^{-i}
		\right)
$.

The state transitions are determined by
$f: \X \times \U \times \Na \rightarrow \X$,
such that the nonstationary Markovian dynamic equation of the game is
$\mix_{t+1} = f(\mix_t, \miu_t, t)$,
which can be split among components:
$x_{t+1}^k = f^k(\mix_t, \miu_t, t)$
for $k=1,\ldots,S$,
such that 
$
f
\defeq
	\left(
		f^k
	\right)_{k=1}^S
$.
The dynamic is Markovian because the state transition to $\mix_{t+1}$ depends on the current state-action pair $(\mix_t, \miu_t)$, 
rather than on the whole history of state-action pairs 
$\{ (\mix_0, \miu_0), \ldots (\mix_t, \miu_t) \} $.
\textblue{We remark that $f$ corresponds to a nonreduced form, 
such that there is no function $\varphi$ such that $\miu_t = \varphi (\mix_t, \mix_{t+1}, t)$.}

We include a vector of $C$ nonstationary constraints
$g \defeq \left( g^{c} \right)_{c=1}^C$,
as it is required by most applications,
and define the sets 
$
\C_t
\defeq
	\{
		\X \times \U
	\}
	\cap
	\{ 
		( \mix_t, \miu_t):
			g ( \mix_t, \miu_t, t) \le 0
	\} 
	\cap
	\{ 
		( \mix_t, \miu_t):
			\mix_{t+1} = f ( \mix_t, \miu_t, t)
	\} 
$.

Each player has its nonstationary utility function
$\pi^i: \X^i \times \U \times \Na \rightarrow \Re$,
such that, at every time $t$, each player receives
a utility value
equal to
$
\pi^i(\mix_t^i, u_t^i, \miu_t^{-i}, t) 
$.
The aim of player $i$
is to find the sequence of actions $\{ u_0^i, \ldots, u_t^i,\ldots \}$ that maximizes its long term cumulative utility,
given other players' sequence of actions $\{ \miu_0^{-i}, \ldots, \miu_t^{-i},\ldots\}$.
Thus, a discrete-time infinite-horizon noncooperative nonstationary Markovian dynamic game can be represented as a set of $Q$ coupled optimal control problems:
\begin{IEEEeqnarray}{rCl}
\begin{aligned}
\mc{G}_1
:\\
\:
\forall i\in\Q
\end{aligned}
\;\;
	\begin{aligned}
		\underset{ \{ u_t^i \} \in \prod_{t=0}^{\infty}\U^i }{\rm maximize} 	
						&\quad 	\sum_{t=0}^{\infty} 
									\beta^t 
									\pi^i(\mix_t^i, u_t^i, \miu_t^{-i}, t)
\\
		{\rm s.t.} 		&\quad 	\mix_{t+1} 	= f ( \mix_t , \miu_t, t)
								,\;\; \mix_0 \text{ given}
\quad
\\
						&\quad	g (\mix_t, \miu_t, t) \le 0
	\end{aligned}
\label{eq:dynamic-game-problem}
\end{IEEEeqnarray}
where $0 < \beta < 1$ is the discount factor that bounds the cumulative utility
\textblue{(for simplicity, we define the same $\beta$ for every player).}
Note that, 
since the players can share state-vector components,
the constraints may affect every player's feasible region.
Problem \eqref{eq:dynamic-game-problem} 
is infinite-horizon because the reward is accumulated over infinite time steps.

%
%
The solution concept of problem \eqref{eq:dynamic-game-problem} in which we are interested is the Nash Equilibrium (NE) of the game, which is defined as follows.
\begin{definition}
A solution of problem \eqref{eq:dynamic-game-problem},
known as a Nash Equilibrium (NE), 
is a feasible sequence of actions
$
\{ \miu_t^{\star} \}_{t=0}^{\infty}
$
that satisfies the following condition 
for every player $i \in \Q$:
\begin{IEEEeqnarray}{rCl}
	\sum_{t=0}^{\infty} \beta^t \pi^i(\mix_t^i, u_t^{\star i}, \miu_t^{\star -i}, t)
&
	\ge
&
	\sum_{t=0}^{\infty} \beta^t \pi^i(\mix_t^i, u_t^i, \miu_t^{\star -i}, t)
\notag	\\
	\forall (\mix_t, \miu_t) 
&
	\in 
&
	\C_t
\label{eq:nash-equilibrium}
\end{IEEEeqnarray}
\label{def:nash-equilibrium}
\end{definition}
%
%

%
We consider the following assumptions:
\begin{assumption}
The utilities $\pi^i$ are twice continuously differentiable in $\X \times \U $. 
\label{as:differentiability}
\end{assumption}
\begin{assumption}
The state and action spaces, 
$\X$ and $\U$, 
are open and convex subsets of a real vector space.
\label{as:convex-state-and-action-sets}
\end{assumption}
\begin{assumption}
The state-transition function $f$ and the 
constraints $g$ are continuously differentiable 
in $\X \times \U $
and satisfy some regularity conditions.
\label{as:qualified-constraints}
\end{assumption}
%
%
%
%
In general,
finding an NE of problem \eqref{eq:dynamic-game-problem} is a difficult task because the utilities, dynamic equation and constraints of the individual optimal control problems (OCP) are coupled among players.
However, when problem \eqref{eq:dynamic-game-problem} is a DPG, 
we can solve it through an equivalent MOCP---as opposed to a set of coupled univariate OCP.
\textblue{We use Assumptions \ref{as:differentiability} and \ref {as:convex-state-and-action-sets} 
to obtain a verifiable condition for problem \eqref{eq:dynamic-game-problem} to be a DPG.
Assumption \ref{as:qualified-constraints}  is required
to introduce the conditions that guarantee equivalence between the solution of 
the MOCP and an NE of the original DPG.
In particular, 
since we derive the KKT optimality conditions for both problems 
(namely the DPG and the MOCP),
some regularity conditions
(such as 
Slater's,
the linear independence of gradients
or the Mangasarian-Fromovitz 
constraint qualifications)
are required to ensure 
that the KKT conditions hold at the optimal points
and that feasible dual variables exist
(see, e.g., \cite[Sec. 3.3.5]{bertsekas1999nonlinear},
\cite{wang2013constraint}).
Finally,
we introduce one further assumption in Sec. \ref{sec:dynamic-potential-games}
to ensure existence of a solution to the MOCP
and, hence, existence of an NE of the DPG.
}

This equivalence between DPG and MOCP
generalizes the well studied but simpler case of static potential games \cite{Slade1994a,Monderer1996},
which is reviewed in the following section.

\section{Overview of Static Potential Games}
\label{sec:static-potential-games}
Static games are a simplified version of dynamic games in the sense that there are neither states, nor system dynamics. 
The aim of each player $i$,
given other players' actions $\miu^{-i}$,
is to choose an action $u^i \in \U^i$
that maximizes its utility function:
\begin{IEEEeqnarray}{rCl}
\mc{G}_2
:
\:
\forall i\in\Q
\quad
	\begin{aligned}
		\underset{ u^i  \in \U^i }{\rm maximize} 	
						&\quad 	\pi^i(u^i, \miu^{-i})
\\
		{\rm s.t.} 		&\quad 	g(\miu) \le 0
	\end{aligned}
\label{eq:static-game-problem}
\end{IEEEeqnarray}
where (similar to dynamic games but removing the time-dependence subscript) $u^i \in \U^i$ refers to the action of player $i$;
and $\miu^{-i} = (u^j)_{j \in \Q : j\neq i}$
is the set of actions of the rest of agents,
such that $\miu = (u^i, \miu^{-i}) \in \U$ denotes the set of all players' actions.
We assume $\U \subseteq \Re^Q $ to be open and convex.

In general, finding or even characterizing the set of equilibrium points (e.g., in terms of existence or uniqueness) of problem \eqref{eq:static-game-problem} is difficult.
Fortunately, there are particular cases of this problem 
for which the analysis is greatly simplified.
Potential games is one of these cases.
\begin{definition}
Let Assumptions \ref{as:differentiability}--\ref{as:convex-state-and-action-sets} hold.
Then, problem \eqref{eq:static-game-problem} is called a static potential game if
there is a function $\Pi : \U \rightarrow \Re$,
named the \textit{potential},
that satisfies the following condition for every player \cite{Monderer1996}:
\begin{IEEEeqnarray}{rCl}
\begin{aligned}
\pi^i(u^i, \miu^{-i}) - \pi^i(v^i, \miu^{-i}) 
&= 
\Pi(u^i, \miu^{-i}) - \Pi(v^i, \miu^{-i}) 
\;\;
\\
\forall u^i, v^i \in \U^i
&,\;\;
\forall i \in \Q
\end{aligned}
\label{eq:condition-static-potential-game}
\end{IEEEeqnarray}
\end{definition}

\textblue{Under Assumptions \ref{as:differentiability}--\ref{as:convex-state-and-action-sets},}
it can be shown 
(see, e.g., \cite[Lemma 4.4]{Monderer1996})
that a necessary and sufficient condition for a static game to be potential is the following:
\begin{IEEEeqnarray}{rCl}
\frac{\partial \pi^i (\miu)}{ \partial u^i} 
=
\frac{\partial \Pi (\miu)}{ \partial u^i}
,\quad \forall i \in \Q
\label{eq:first-order-condition-static-potential-game}
\end{IEEEeqnarray}

\textblue{We can gain insight on potential games by relating
\eqref{eq:first-order-condition-static-potential-game}
to the concept of \textit{conservative vector field}.
The following lemma will be useful to this end.}

\begin{lemma}
\label{lemma:conservative-vector-field}
Let $\miF (\miu) = (F_1(\miu), \ldots, F_Q(\miu))$ be a vector field with continuous derivatives
defined over an open convex set $\U \in \Re^Q $. 
The following conditions on $\miF$ are equivalent:
\begin{enumerate}
\item There exists a scalar potential function $\Pi(\miu)$ such that
	$ \miF (\miu) = \nabla \Pi(\miu)$, where $\nabla$ is the gradient.
\item
\textblue{The partial derivatives satisfy
\begin{IEEEeqnarray}{rCl}
		\frac{\partial F_j (\miu) }{\partial u^i }  
			= 
		\frac{\partial F_i (\miu) }{\partial u^j }
,\;\;
\forall \miu \in \U, 
\; i,j = 1, \ldots, Q
\end{IEEEeqnarray}
}
\vspace{-1em}
\item 
\textblue{Let $\mia$ be a fixed point of $\U$. 
For any piecewise smooth path $\mixi$ joining $\mia$ with $\miu$,
we have 
$\Pi(\miu) = \int_{\mia}^{\miu} \miF(\mixi) \cdot d \mixi $.}
\end{enumerate}
A vector field satisfying these conditions is called conservative.
\end{lemma}
\begin{IEEEproof}
See, e.g., 
\textblue{
\cite[Theorems 10.4, 10.5 and 10.9]{apostol1969calculus}.
}
\end{IEEEproof}
Let us define a vector field with components the partial derivatives of the players' utilities:
\begin{IEEEeqnarray}{rCl}
	\miF (\miu)
	&\defeq&
		\left(
			\frac{ \partial \pi^1 (\miu) }
				 { \partial u^1 } 
			, \ldots ,
			\frac{ \partial \pi^Q (\miu) }
				 { \partial u^Q } 
		\right)
\label{eq:static-vector-field}
\end{IEEEeqnarray}
\textblue{
Let us rewrite \eqref{eq:static-vector-field} more compactly
as
$
\miF (\miu) = \nabla \Pi (\miu)
$
so that Lemma \ref{lemma:conservative-vector-field}.1 holds.
Then,
we have that
$
\frac{\partial \pi^i (\miu)}{ \partial u^i} 
=
\frac{\partial \Pi (\miu)}{ \partial u^i}
$,
$
\forall i \in \Q
$.
Note that this is exactly condition given by \eqref{eq:first-order-condition-static-potential-game}.
It follows from Lemma \ref{lemma:conservative-vector-field}.2
that a necessary, sufficient and also easily verifiable condition 
for problem \eqref{eq:static-game-problem} to be a static potential game is given by:
\begin{IEEEeqnarray}{rCl}
\frac{\partial^2 \pi^i (\miu)}{ \partial u^i \partial u^j} 
=
\frac{\partial^2 \pi^j (\miu)}{ \partial u^i \partial u^j}
,\quad \forall i,j \in \Q
\label{eq:second-order-condition-static-potential-game}
\end{IEEEeqnarray}
Finally, 
Lemma \ref{lemma:conservative-vector-field}.3 is useful since
we can find the potential function $\Pi$ by solving the line integral of the field:
\begin{IEEEeqnarray}{rCl}
\Pi(\miu) 
	= 
	\int_0^1 
		\sum_{i=1}^Q 
				\frac{
					\partial \pi^i (\xi^i (\lambda), \miu^{-i}) 
				}{ \partial u^i }
				\frac{d \xi^i (\lambda) }{ d \lambda} d\lambda
\label{eq:potential-function-static-game}
\end{IEEEeqnarray}
where 
$
\mixi 
\defeq 
	\left(
		\xi^i
	\right)_{i\in\Q}
$
is a piecewise smooth path in $\U$
that connects the initial and final conditions: 
$\mixi(0) = \mia$, $\mixi(1)  = \miu$.}

%
Once we have found $\Pi$,
it can be seen \cite{Slade1994a} that 
necessary conditions for $\miu^{\star}$ to be an equilibrium of the game \eqref{eq:static-game-problem} are also necessary conditions for the following optimization problem:
\begin{IEEEeqnarray}{rCl}
\mc{P}_1:
	\begin{aligned}
		\underset{ \miu  \in \U }{\rm maximize} 	
						&\quad 	\Pi(\miu)
\\
		{\rm s.t.} 		&\quad 	g(\miu) \le 0
	\end{aligned}
\label{eq:static-optimization-problem}
\end{IEEEeqnarray}
Indeed, optimization theorems concerning existence and convergence can now be applied to game \eqref{eq:static-game-problem}. 
In particular, 
reference \cite{Slade1994a} showed that the local maxima of the potential function 
are a subset of the NE of the game. 
Furthermore, in the case that all players' utilities are quasi-concave, 
the maximum is unique and coincides with the stable equilibrium of the game. 

This same approach can be extended to dynamic games.
Nevertheless, instead of obtaining an analogous optimization problem,
DPG will yield an analogous MOCP.

\section{Dynamic Potential Games with Constraints}
\label{sec:dynamic-potential-games}

This section introduces the main theoretical contribution of the paper:
we establish conditions under which
we can find an NE of problem \eqref{eq:dynamic-game-problem} 
by solving an alternative MOCP,
instead of having to solve the set of coupled infinite horizon OCP with coupled constraints. 
First, we introduce the definition of a DPG
and show conditions 
for problem \eqref{eq:dynamic-game-problem} to belong to this class. 
Then, we introduce the alternative MOCP and prove that its solution is an NE of the game.

\begin{definition}
Problem \eqref{eq:dynamic-game-problem} is called a DPG if
there is a function $\Pi : \X \times \U \times \Na \rightarrow \Re$,
named the \textit{potential},
that satisfies the following condition for every player $i \in \Q$:
\begin{IEEEeqnarray}{rCl}
\sum_{t=0}^{\infty}
&&
	\beta^t
	\left(
		\pi^i(\mix_t^i, u^i_t, \miu^{-i}_t, t) - \pi^i(\mix_t^i, v^i_t, \miu^{-i}_t, t) 
	\right)
\notag \\
&&=
\:
\sum_{t=0}^{\infty}
	\beta^t
	\left(
		\Pi(\mix_t, u^i_t, \miu^{-i}_t, t) - \Pi(\mix_t, v^i_t, \miu^{-i}_t, t) 
	\right)
\notag \\
&&\:
\forall \mix_t \in \X
,\;\;
\forall u^i_t, v^i_t \in \U^i
\label{eq:condition-definition-DPG}
\end{IEEEeqnarray}
\label{def:dynamic-potential-game}
\end{definition}
\textblue{Note that, although the potential function $\Pi$ is defined for the larger set
	$\X \times \U \times \Na$, 
	the local objective $\pi^i$ is only defined over its local subset $\X^i \times \U \times \Na$.
	Therefore,
	we only have to check whether condition \eqref{eq:condition-definition-DPG} is 
	satisfied in each players' subset.}

The following three lemmas give conditions under which problem \eqref{eq:dynamic-game-problem} is a DPG
(i.e., it satisfies Definition \ref{def:dynamic-potential-game}).
%
%
\begin{lemma}
\label{lemma:dynamic-potential-games-1}
Problem \eqref{eq:dynamic-game-problem} is a DPG
if there exists some function $\Pi\left( \mix_t, \miu_t, t \right)$
that satisfies
\begin{IEEEeqnarray}{rCl}
	\frac 	{\partial \pi^i \left( \mix_t^i, \miu_t, t \right) }
			{\partial x_t^m}
&=&
	\frac	{\partial \Pi \left( \mix_t, \miu_t, t \right)}
			{\partial x_t^m}
\notag\\
	\frac 	{\partial \pi^i \left( \mix_t^i, \miu_t, t \right) }
			{\partial u_t^i}
&=&
	\frac	{\partial \Pi \left( \mix_t, \miu_t, t \right)}
			{\partial u_t^i}
\notag\\
	\forall m \in \X(i)
,\;\;
	\forall i \in \Q
,
&&
	t = 0, \ldots, \infty
\label{eq:condition-dynamic-potential-1}
\end{IEEEeqnarray}
\end{lemma}
%
%
\begin{IEEEproof}
We simply extend to dynamic games
the argument for static games 
due to \cite[Prop. 1]{Slade1994a}.
From 
\eqref{eq:condition-dynamic-potential-1} 
and Assumption \ref{as:differentiability}
we have:
\begin{IEEEeqnarray}{rCl}
	\frac 	{\partial }
			{\partial x_t^m}
			\left( 
				\Pi \left( \mix_t, \miu_t, t \right)
				-
				\pi^i \left( \mix_t^i, \miu_t, t \right)
			\right)
&=&
	0
		,\;\;
		\forall m \in \X(i)
\quad
\\
	\frac 	{\partial }
			{\partial u_t^i}
			\left( 
				\Pi \left( \mix_t, \miu_t, t \right)
				-
				\pi^i \left( \mix_t^i, \miu_t, t \right)
			\right)
&=&
	0
\end{IEEEeqnarray}
This means that the difference between the potential and each player's utility 
depends neither on $x_t^m$ nor $u_t^i$. 
Thus, we can express this difference as
\begin{IEEEeqnarray}{rCl}
\Pi 
\left( \mix_t, u^i, \miu_t^{-i}, t \right)
&
-
&
\pi^i \left( \mix^{i}_t, u^i_t, \miu^{-i}, t \right)
\notag\\
&
=
&
	\Theta(\mix_t^{-i}, \miu_t^{-i}, t)
	\quad
	\forall u^i \in \U^i
\label{eq:slade's-argument}
\end{IEEEeqnarray}
for some function 
$\Theta: \X ^{-i} \times \U^{-i} \times \Na \rightarrow \Re$.
\textblue{Since \eqref{eq:slade's-argument} is satisfied for every $u^{i} \in \U^i$,
we can subtract two versions of \eqref{eq:slade's-argument} with
actions $u^{i}$ and $v^{i}$ in $\U^i$.
Then, by arranging terms and summing over all $t$,
we obtain \eqref{eq:condition-definition-DPG}.}
%
\end{IEEEproof}
%
%

Condition \eqref{eq:condition-dynamic-potential-1} is usually difficult to check in practice because we do not know $\Pi$ beforehand.
Fortunately, 
there are cases in which the player's utilities have some separable structure 
that allows us to easily deduce that the game is of the potential type,
as it is explained in the following lemma.

%
%
\begin{lemma}
\label{lemma:dynamic-potential-games-2}
Problem \eqref{eq:dynamic-game-problem} is a DPG
if 
the utility function of every player 
$i \in \Q$
can be expressed as the sum of 
a term that is common to all players
plus another term that depends neither on its own action, 
nor on its own state-components:
\begin{IEEEeqnarray}{rCl}
\pi^i \left( \mix^{i}_t, u^i_t, \miu_t^{-i}, t \right)
&=&
	\Pi \left( \mix_t, u^i_t, \miu_t^{-i}, t \right)
	+
	\Theta(\mix_t^{-i}, \miu_t^{-i}, t)
\quad\;\;\:
\label{eq:condition-dynamic-potential-2}
\end{IEEEeqnarray}
\end{lemma}
%
%
\begin{IEEEproof}
By taking the partial derivative of 
\eqref{eq:condition-dynamic-potential-2}
we obtain \eqref{eq:condition-dynamic-potential-1}.
Therefore, 
we can apply Lemma \ref{lemma:dynamic-potential-games-1} 
(see also \cite[Prop. 1]{Slade1994a}).
\end{IEEEproof}
%
%
However, 
posing the utility in the separable structure \eqref{eq:condition-dynamic-potential-2}
may be difficult.
We need a more general framework that allows us to check whether problem \eqref{def:dynamic-potential-game} is a DPG when the player's utilities 
have \textblue{a} nonobvious separable structure.
This framework is formally introduced in the following lemma.

%
%
\begin{lemma}
\label{lemma:dynamic-potential-games-3}
Problem \eqref{eq:dynamic-game-problem} is a DPG
if 
all players' utilities satisfy the following conditions, $\forall i,j \in \Q$,
\textblue{
$\forall m \in \X(i)$, 
$\forall n \in \X(j)$}:
\begin{IEEEeqnarray}{rCl}
	\frac{ \partial^2 \pi^i (\mix_t^i, \miu_t, t) }
		 { \partial 
		 	\textblue{
		 		x^m_t 
		 	}
		 	\partial u^j _t} 
	&=&
	\frac{ \partial^2 \pi^j (\mix_t^j, \miu_t, t) }
		 { \partial 
		 	\textblue{
		 		x^n_t 
		 	}
		   \partial u^i_t } 
\label{eq:conservative-field-condition-1}
\\
	\frac{ \partial^2 \pi^i (\mix_t^i, \miu_t, t) }
		 { \partial 
		 	\textblue{
		 		x^m_t
			 	\partial x^n_t } 
		 	}
	&=&
	\frac{ \partial^2 \pi^j (\mix_t^j, \miu_t, t) }
		 { \partial 
		 	\textblue{
		 		x^n_t
			 	\partial x^m_t } 
		 	}
\label{eq:conservative-field-condition-2}
\\
	\frac{ \partial^2 \pi^i (\mix_t^i, \miu_t, t) }
		 { \partial u^i_t \partial u^j_t } 
	&=&
	\frac{ \partial^2 \pi^j (\mix_t^j, \miu_t, t) }
		 { \partial u^j_t \partial u^i_t } 
\label{eq:conservative-field-condition-3}
\end{IEEEeqnarray}
\end{lemma}

\begin{IEEEproof}
Under Assumption \ref{as:differentiability},
we can introduce the following vector field:
\begin{IEEEeqnarray}{rCl}
\miF
	&
	\defeq
	&
		\Bigg(
			\nabla_{\mix_t^1} \pi^1 (\mix_t^1, \miu_t, t)^\T
			, \ldots ,
			\nabla_{\mix_t^Q} \pi^Q (\mix_t^Q, \miu_t, t)^\T
\notag\\
	&&
			,
			\frac{ \partial \pi^1 (\mix_t^1, \miu_t, t) }
				 { \partial u^1_t } 
			, \ldots ,
			\frac{ \partial \pi^Q (\mix_t^Q, \miu_t, t) }
				 { \partial u^Q_t } 
		\Bigg)
\label{eq:vector-field-dynamic-game-definition}
\end{IEEEeqnarray}
where 
$
\nabla_{\mix_t^i} \pi^i (\mix_t^i, \miu_t, t)
=
\left(
	\frac{ \partial \pi^i (\mix_t^i, \miu_t, t) }
				 { \partial x^m_t } 
\right)_{m \in \X(i)}
$.
From Lemma \ref{lemma:dynamic-potential-games-1},
we can express \eqref{eq:vector-field-dynamic-game-definition} as
\begin{IEEEeqnarray}{rCl}
\miF
	&=&
		\nabla \Pi (\mix_t, \miu_t, t)
\label{eq:vector-field-dynamic-game-from-potential}
\end{IEEEeqnarray}
From 
Assumption \ref{as:convex-state-and-action-sets}
and
Lemma \ref{lemma:conservative-vector-field}.1,
we know that $\miF$ is conservative.
Hence, Lemma \ref{lemma:conservative-vector-field}.2 establishes that 
the second partial derivatives must satisfy 
\eqref{eq:conservative-field-condition-1}--\eqref{eq:conservative-field-condition-3}.
\end{IEEEproof}

Introduce the following MOCP:
\begin{IEEEeqnarray}{rCl}
\mc{P}_2:
	\begin{aligned}
	\underset{ \{ \miu_t \} \in \prod_{t=0}^{\infty}\U }{\rm maximize} 	
						&\quad 	\sum_{t=0}^{\infty} 
									\beta^t 
									\Pi(\mix_t, \miu_t, t)
\\
		{\rm s.t.} 		&\quad 	\mix_{t+1} 	= f ( \mix_t , \miu_t, t)
								,\;\; \mix_0 \;\; {\rm given}
\\						&\quad
								g(\mix_t, \miu_t, t) \le 0
	\end{aligned}
\label{eq:equivalent-mocp-problem}
\end{IEEEeqnarray}
\textblue{Let us consider the following assumption,
which is needed for establishing equivalence between a DPG and the MOCP \eqref{eq:equivalent-mocp-problem}.
\begin{assumption}
The MOCP \eqref{eq:equivalent-mocp-problem} has a nonempty solution set.
\label{as:existence-MOCP-solution}
\end{assumption}
}
\textblue{Sufficient---and easily verifiable---conditions 
to satisfy Assumption \ref{as:existence-MOCP-solution}
are given by the following lemma,
which is a standard result in optimal control theory.}
\begin{lemma}
\label{lemma:existence}
\textblue{
Let $\Pi : \X \times \U \times \Na \rightarrow [-\infty, \infty)$
be a proper continuous function.
And let any one of the following conditions hold
for $t = 1,\ldots,\infty$:
}
\begin{enumerate}
\item 
	\textblue{
		The constraint sets 	$\C_t$ are bounded.
	}
\item
\textblue{
$
\Pi (\mix_t, \miu_t, t) \rightarrow -\infty
\:
$
as
$
\:
\|(\mix_t, \miu_t)\| \rightarrow \infty
$
(coercive).
}
\item 
\textblue{
There exists a scalar $M$ such that the level sets, 
defined by
$
\{ 
(\mix_t, \miu_t, t)  
	|  
	\Pi(\mix_t, \miu_t, t) \ge M
\}_{t=1}^{\infty}
$,
are nonempty and bounded.
}
\end{enumerate}
\textblue{
Then, 
$\forall \mix_0 \in \X$,
there exists an optimal sequence of actions 
$
\{ 
	\miu_t^{\star}
\}_{t=0}^{\infty}$
that is solution to the MOCP \eqref{eq:equivalent-mocp-problem}.
Moreover, 
there exists an optimal policy 
$\phi^{\star} : \X \times \Na \rightarrow \U$,
which is a mapping from states to optimal actions, 
such that
when applied over the state-trajectory
$
\{ 
	\mix_t
\}_{t=0}^{\infty}$, 
it provides an 
optimal sequence of actions
$
\{ 
	\miu_t^{\star}
	\defeq
		\phi^{\star}( \mix_t, t)
\}_{t=0}^{\infty}$.
}
\end{lemma}
\begin{IEEEproof}
\textblue{
Since $\Pi$ is proper, 
it has some nonempty level set.
Since $\Pi$ is continuous, its bounded level sets are compact.
Hence, we can use \cite[Prop. 3.1.7]{Bertsekas2007}
(see, also \cite[Sections 1.2 and 3.6]{Bertsekas2007})
to establish existence of an optimal policy.
%
}
\end{IEEEproof}

The main theoretical result of this work is that 
we can find an NE of a DPG 
by solving the MOCP \eqref{eq:equivalent-mocp-problem}.
This is proved in the following theorem.
%
%
\begin{theorem}
If problem \eqref{eq:dynamic-game-problem} is a DPG,
under Assumptions \ref{as:differentiability}--\ref{as:existence-MOCP-solution}, 
the solution of the MOCP \eqref{eq:equivalent-mocp-problem}
is an NE of \eqref{eq:dynamic-game-problem}
when the objective function of the MOCP is given by
\begin{IEEEeqnarray}{rCl}
\Pi(\mix_t, \miu_t, t) 
	=
	\int_0^1 
		\sum_{i=1}^Q 
	&&
		\Bigg (
				\sum_{m \in \X(i)}
				\frac{
					\partial \pi^i (\mieta (\lambda), \miu_t, t ) 
				}{ \partial x^m_t }
				\frac{d \eta^m (\lambda) }{ d \lambda} 
\notag\\
&&		
			+		
\:
				\frac{
					\partial \pi^i (\mix_t, \mixi (\lambda), t ) 
				}{ \partial u^i_t }
				\frac{d \xi^i (\lambda) }{ d \lambda} 
		\Bigg )
		d\lambda
\label{eq:potential-function-dynamic-game}
\end{IEEEeqnarray}
where 
$
\mieta (\lambda) 
\defeq 
	\left(
		\eta^k (\lambda) 
	\right)_{k=1}^{S}
$,
$
\mixi (\lambda) 
\defeq 
	\left(
		\xi^i (\lambda)
	\right)_{i=1}^Q
$,
and
$\mieta(0)$-$\mixi(0)$ and $\mieta(1)$-$\mixi(1)$
correspond to the initial and final state-action conditions,
respectively.
\label{theorem:game-and-mocp-equivalence}
\end{theorem}
%
\textblue{The usefulness of Theorem \ref{theorem:game-and-mocp-equivalence} is that,
in order to find an NE of \eqref{eq:dynamic-game-problem},
instead of solving several coupled control problems, 
we can check whether \eqref{eq:dynamic-game-problem} is a DPG
(i.e., anyone of Lemmas \ref{lemma:dynamic-potential-games-1}--\ref{lemma:dynamic-potential-games-3} holds).
If so, we can find an NE by computing the potential function \eqref{eq:potential-function-dynamic-game}
and, then, by solving the equivalent MOCP \eqref{eq:equivalent-mocp-problem}.}
%
\begin{IEEEproof}
The proof is structured in five steps.
First, we compute the Euler equation of the Lagrangian of the dynamic game 
and derive the KKT optimality conditions.
\textblue{Assumption \ref{as:qualified-constraints} is required to ensure that 
the KKT conditions hold at the optimal point
and that there exist feasible dual variables \cite[Prop. 3.3.8]{bertsekas1999nonlinear}.
}
Second, we study when the necessary optimality conditions of the game become equal to those of the MOCP.
Third, we show that having the same necessary optimality conditions is sufficient condition for the dynamic game to be potential.
\textblue{Fourth, having established that the dynamic game is a DPG
we show that the solution to the MOCP
(whose existence is guaranteed by Assumption \ref{as:existence-MOCP-solution})
is also an NE of the DPG.}
Finally, we derive the per stage utility of the MOCP 
as the potential function of a suitable vector field.
We proceed to explain the details.

First, 
for problem \eqref{eq:dynamic-game-problem}, 
introduce each player's Lagrangian 
$\forall i \in \Q$:
\begin{IEEEeqnarray}{rCl}
\mc{L}^{i}
\big(
	\mix_{t},
	&&
	\miu_{t}, \milambda_{t}^{i}, \mimu_{t}^{i}
\big)
=
	\sum_{t=0}^\infty
		\beta^{t}
		\Big(
			\pi^{i} \left( \mix_{t}^i,\miu_{t},t \right)
\notag\\
&&
\quad
			+
\:
			\milambda_t^{i^{\T}}
			\left(
				f \left( \mix_{t},\miu_{t},t \right)
				-
				\mix_{t+1}
			\right)
			+
			\mimu_{t}^{i^{\T}}
			g \left( \mix_{t},\miu_{t},t \right)
		\Big)
\notag\\
&&
	=
\:
	\sum_{t=0}^\infty
		\beta^{t}
		\Phi^i
		\left(
			\mix_{t},\miu_{t},t,\milambda_{t}^{i}, \mimu_{t}^{i}
		\right)
\label{eq:lagrangian-game}
\end{IEEEeqnarray}
where $\milambda_{t}^{i} \defeq \left( \lambda_{t}^{ik} \right)_{k=1}^{S}$ 
and 
$\mimu_{t}^{i} \defeq \left( \mu_{t}^{ic} \right)_{c=1}^{C}$ 
are the corresponding vectors of multipliers,
and we introduced the shorthand:
%
\begin{IEEEeqnarray}{rCl}
\Phi^i
	\big(
		\mix_{t},
&&
		\miu_{t},t, \milambda_{t}^{i}, \mimu_{t}^{i}
	\big)
\defeq 
			\pi^{i} \left( \mix_{t}^i,\miu_{t},t \right)
\notag\\
&&
			+
\:
			\milambda_t^{i^{\T}}
			\left(
				f \left( \mix_{t},\miu_{t},t \right)
				-
				\mix_{t+1}
			\right)
			+
			\mimu_{t}^{i^{\T}}
			g \left( \mix_{t},\miu_{t},t \right)
\qquad
\label{eq:instantaneous-lagrange}
\end{IEEEeqnarray}
The discrete time Euler-Lagrange equations \cite[Sec. 6.1]{Sage77}
applied to each player's Lagrangian are given by:
\begin{IEEEeqnarray}{rCl}
&&
\frac	{\partial \Phi^{i}
			\left(
				\mix_{t-1},\miu_{t-1},t-1,\milambda_{t-1}^{i},\mimu_{t-1}^{i}
			\right)
		}
		{\partial x_{t}^{m}}
\qquad\qquad\qquad\qquad\quad
\notag\\
&&\qquad\qquad
+
\:
\frac	{ \partial \Phi^{i}
			\left(
				\mix_{t},\miu_{t},t,\milambda_{t}^{i},\mimu_{t}^{i}
			\right)}
		{ \partial x_{t}^{m}}
=
		0 
,\quad \forall m \in \X(i)
\label{eq:game-euler-equations-state}
\\
&&
\frac	{\partial \Phi^{i}
			\left(
				\mix_{t-1},\miu_{t-1},t-1,\milambda_{t-1}^{i},\mimu_{t-1}^{i}
			\right)
		}
		{\partial u_{t}^{i}}
\notag\\
&&\qquad\qquad
+
\:
\frac	{ \partial \Phi^{i}
			\left(
				\mix_{t},\miu_{t},t,\milambda_{t}^{i},\mimu_{t}^{i}
			\right)}
		{ \partial u_{t}^{i}}
=
		0 
\label{eq:game-euler-equations-action}
\end{IEEEeqnarray}
Actually, 
note that \eqref{eq:game-euler-equations-state}--\eqref{eq:game-euler-equations-action} are 
the Euler-Lagrange equations in a more general form than the standard \textit{reduced form}.
As mentioned in Sec. \ref{sec:problem-setting}
(see also, e.g., \cite[Sec. 6.1]{Sage77}, \cite{HernandezLerma2013}),
in the standard reduced form, 
the current action can be posed as a function of the current and future states:
$\miu_t = \varphi(\mix_t, \mix_{t+1}, t)$, 
for some function 
$
\varphi :\X \times \X \times \Na \rightarrow \U
$.
The reason why we introduced this general form of the Euler-Lagrange equations is that 
such function $\varphi$ may not exist
for an arbitrary state-transition function $f$.
By substituting \eqref{eq:instantaneous-lagrange}
into \eqref{eq:game-euler-equations-state}--\eqref{eq:game-euler-equations-action}, 
and adding the corresponding constraints, 
we obtain the KKT conditions of the game 
for every player $i \in \Q$, 
the state-components $m \in \X(i)$,
and all extra constraints:
\begin{IEEEeqnarray}{rCl}
\frac	{ \partial \pi^{i}
			\left(\mix_{t}^i,\miu_{t},t\right)}
		{ \partial x_{t}^{m}}
&
+
&
\sum_{k =1}^S
	\lambda_{t}^{i k}
	\frac 	{ \partial f^{k}
				\left(
					\mix_{t},\miu_{t},t
				\right)
			}
			{ \partial x_{t}^{m}}
\notag\\
&
+
&
\sum_{c=1}^{C}
	\mu_{t}^{ic}
	\frac	{ \partial g^{c} \left( \mix_{t},\miu_{t},t \right)}
			{ \partial x_{t}^{m} }
- 
\lambda_{t-1}^{im}
=
	0
\qquad\;\;
\label{eq:optimality-conditions-game-1}
\\
\frac	{ \partial \pi^{i} \left( \mix_{t}^i,\miu_{t},t \right)}
		{ \partial u_{t}^{i} }
&
+
&
\sum_{k =1}^S
	\lambda_{t}^{i k}
	\frac 	{ \partial f^{k} \left( \mix_{t},\miu_{t},t \right) }
			{ \partial u_{t}^{i} }
\notag\\
&
	+
&
\sum_{c=1}^{C}
	\mu_{t}^{ic}
	\frac	{ \partial g^{c} \left( \mix_{t},\miu_{t},t \right)}
			{ \partial u_{t}^{i} }
=
	0
\label{eq:optimality-conditions-game-2}
\\
\mix_{t+1}
&
	=
&
	f \left(\mix_{t},\miu_{t},t \right)
,
\quad 
	g \left( \mix_{t},\miu_{t},t \right)
	\leq 0
\label{eq:optimality-conditions-game-3}
\\
\mimu_{t}^{i}
&
	\leq 
&
	0
,
\quad 
	\mimu_{t}^{i^\T}
	g \left(\mix_{t},\miu_{t},t\right) 
	= 
		0 
\label{eq:optimality-conditions-game-4}
\end{IEEEeqnarray}
%

Second, we find the KKT conditions of the MOCP.
To do so, we obtain the Lagrangian of \eqref{eq:equivalent-mocp-problem}:
\begin{IEEEeqnarray}{rCl}
\mc{L}^{\Pi}
	(
 		\mix_{t},
&&
		\miu_{t}, \migamma_{t},\midelta_{t}
	)
=
\sum_{t=0}^{\infty}
		\beta^{t}
			\Big(
				\Pi
				\left( \mix_{t},\miu_{t},t \right)
\notag\\
&&
				+
\:
				\migamma_{t}^{\T}
					\left(
						f \left( \mix_{t},\miu_{t},t \right)
						-
						\mix_{t+1}
					\right)
				+
				\midelta_{t}^{\T} 
				g \left( \mix_{t},\miu_{t},t \right)
			\Big)
\quad
\label{eq:lagrangian-mocp}		
\end{IEEEeqnarray}
where 
$ \migamma_{t} \defeq \left( \gamma_t^{k} \right)_{k=1}^S $ 
and 
$ 
\midelta_{t} \defeq \left( \delta_t^{c} 	\right)_{c=1}^C 
$
are the corresponding multipliers.
Again, from \eqref{eq:lagrangian-mocp} we derive the Euler-Lagrange equations, 
which, together with the corresponding constraints,
yield the KKT system of optimality conditions 
for all state-components, 
$ m=1, \ldots, S $,
and 
all actions,
$ i=1, \ldots, Q $:
\begin{IEEEeqnarray}{rCl}
\frac	{ \partial \Pi
			\left( \mix_{t},\miu_{t},t\right )
		}
		{ \partial x_{t}^{m}}
&
+
&
\sum_{k=1}^S
	\gamma_{t}^{k}
	\frac 	{ \partial f^{k}
				\left(
					\mix_{t},\miu_{t},t
				\right)
			}
			{ \partial x_{t}^{m}}
\notag\\
&&
+
\:
\sum_{c=1}^C
	\delta_{t}^{c} 
	\frac	{ \partial g^{c} \left( \mix_{t},\miu_{t},t \right)}
			{ \partial x_{t}^{m} }
-\gamma_{t-1}^{m}
=
	0
\qquad
\label{eq:optimality-conditions-mocp-1}
\\
\frac	{ \partial \Pi
			\left(\mix_{t},\miu_{t},t\right)}
		{ \partial u_{t}^{i} }
&
+
&
\sum_{k=1}^S
	\gamma_{t}^{k}
	\frac 	{ \partial f^{k}
				\left(
					\mix_{t},\miu_{t},t
				\right)
			}
			{ \partial u_{t}^{i} }
\notag\\			
&&
+
\:
\sum_{c=1}^C
	\delta_{t}^{c} 
	\frac	{ \partial g^{c} \left( \mix_{t},\miu_{t},t \right)}
			{ \partial u_{t}^{i} }
=
	0
\label{eq:optimality-conditions-mocp-2}
\\
\mix_{t+1}
&
	=
&
	f \left(\mix_{t},\miu_{t},t \right)
,\quad 
g \left( \mix_{t},\miu_{t},t \right)
	\leq 0
\label{eq:optimality-conditions-mocp-3}
\\
\midelta_{t}
&
	\leq 
&
	0
,\quad 
\midelta_{t}^{\T}
g \left(\mix_{t},\miu_{t},t\right)
	= 0 
\label{eq:optimality-conditions-mocp-4}
\end{IEEEeqnarray}
In order for the MOCP \eqref{eq:equivalent-mocp-problem}
to have the same optimality conditions as the game \eqref{eq:dynamic-game-problem}, 
by comparing
\eqref{eq:optimality-conditions-game-1}--\eqref{eq:optimality-conditions-game-4}
with 
\eqref{eq:optimality-conditions-mocp-1}--\eqref{eq:optimality-conditions-mocp-4},
we conclude that the following conditions must be satisfied
$\forall i \in \Q$:
\begin{IEEEeqnarray}{rCl}
\frac	{ \partial\pi^{i}\left(\mix_{t}^i,\miu_{t},t\right) }
		{ \partial x_{t}^{m} }
&=&
\frac	{ \partial\Pi \left( \mix_{t},\miu_{t},t \right) }
		{ \partial x_{t}^{m} }
,\quad 
	\forall m \in \X(i)
\label{eq:equivalence-condition-1}
\\
\frac	{ \partial\pi^{i}\left(\mix_{t}^i,\miu_{t},t\right) }
		{ \partial u_{t}^{i} }
&=&
\frac	{ \partial\Pi \left( \mix_{t},\miu_{t},t \right) }
		{ \partial u_{t}^{i} }
\label{eq:equivalence-condition-2}
\\
\milambda_t^{i}  	= \migamma_t
\;,	
&&
\;
	\mimu_t^{i}	=\midelta_t
\label{eq:equivalence-condition-3}
\end{IEEEeqnarray}

Third, 
when conditions 
\eqref{eq:equivalence-condition-1}--\eqref{eq:equivalence-condition-2} are satisfied, 
Lemma \ref{lemma:dynamic-potential-games-1} states that problem \eqref{eq:dynamic-game-problem} is a DPG.

Fourth, 
note that condition \eqref{eq:equivalence-condition-3} represents a feasible point of the game.
The reason is that if there exists an optimal primal variable,
then the 
existence of dual variables in the MOCP is guaranteed by suitable regularity conditions.  
\textblue{Since the existence of optimal primal variables of the MOCP 
is ensured by Assumption \ref{as:existence-MOCP-solution},
the regularity conditions established by Assumption \ref{as:qualified-constraints} 
guarantee that there exist some $\migamma_t$ and $\midelta_t$ that satisfy the KKT conditions of the MOCP.}
Substituting these dual variables of the MOCP in place of the individual 
$\milambda^i_t$
and 
$\mimu^i_t$
in 
\eqref{eq:optimality-conditions-game-1}--\eqref{eq:optimality-conditions-game-4}
for every $i \in \Q$,
results in a system of equations where the only unknowns are the user strategies. 
This system has exactly the same structure as the one already presented for the MOCP in the primal variables. 
Therefore, the MOCP primal solution also satisfies the KKT conditions of the DPG.
Indeed, it is straightforward to see that an optimal solution of the MOCP is also an NE of the game. 
Let $\{ \miu^{\star}_t \}_{t=0}^{\infty}$ denote the MOCP solution,
so that it satisfies the following inequality
$\forall u_t^i \in \U^i$:
\begin{IEEEeqnarray}{rCl}
\sum_{t=0}^{\infty}
	\beta^t
	\Pi(\mix_t, u^{i \star}_t, \miu^{\star-i}_t, t)
\ge 
\sum_{t=0}^{\infty}
	\beta^t
	\Pi(\mix_t, u^i_t, 
	\textblue{ 
		\miu^{\star-i}_t
	}	
	, t) 
\quad
\label{eq:inequality-solution-MOCP}
\end{IEEEeqnarray}	
From Definition \ref{def:dynamic-potential-game},
we conclude that the MOCP optimal solution 
is also an NE of game \eqref{eq:dynamic-game-problem}.
The opposite may not be true in general. 
Indeed, this solution, 
in which dual variables are shared between players, 
is only a subclass of the possible NE of the game. 
Nevertheless, other NE that do not share this property have been referred to as \textit{unstable} by \cite{Slade1994a} for static games.

Fifth, although we have shown that we can find an NE of the DPG by solving a MOCP,
we still need to find the objective of the MOCP.
In order to find $\Pi$, 
we deduce from 
\eqref{eq:equivalence-condition-1},
\eqref{eq:equivalence-condition-2},
\eqref{eq:vector-field-dynamic-game-definition}
and 
\eqref{eq:vector-field-dynamic-game-from-potential}
that the vector field \eqref{eq:vector-field-dynamic-game-definition}
can be expressed as 
\begin{IEEEeqnarray}{rCL}
\miF 
	&\defeq &
		\nabla \Pi \left(\mix_{t},\miu_{t},t\right)
\label{eq:vector-field-as-gradient-of-potential}
\end{IEEEeqnarray}
Lemma \ref{lemma:conservative-vector-field}
establishes that $\miF$ is conservative.
Thus, the objective of the MOCP is the potential of the field,
which can be computed through the line integral \eqref{eq:potential-function-dynamic-game}. 
\end{IEEEproof}
%
%

%
%
In the next sections, 
we show how to apply this methodology---of solving DPG through an equivalent MOCP---to different practical problems.

\section{Energy Demand in the Smart Grid \\ as a Linear Quadratic Dynamic Game} 
\label{sec:smart-grid-model}

Our first example consists in a linear-quadratic-dynamic-game (LQDG) 
that solves a smart grid resource allocation problem.
LQDG are convenient because they are amenable to analytical and closed form solutions \cite[Ch. 6]{Basar1999}. 
Our analysis is novel though.
To the best of our knowledge, 
LQDG have not been studied under the easier DPG framework before.

\subsection{Energy demand control DPG and equivalent MOCP}
\label{ssec:smart-grid-scenario}

Consider a community of $Q$ users (i.e., players) that use the smart grid resources in different activities 
(like communications, heating, lighting, home appliances or production needs). 
Suppose that the electrical grid has $S$ types of energy resources 
(such as rechargeable batteries, coal, fuel, hydroelectric power or biomass). 
The state of the game $ \mix_{t} \in \Re^{S}$ is the total amount of overall resources in the smart grid at time $t$.
All players share all components of the state-vector 
(i.e., $\X^i = \X$ and $\X(i) = \{1, \ldots, S \}$, $\forall i \in \Q$).
The amount of resources consumed or contributed by player $i$ at time $t$ is denoted by the action vector 
$ \miu^{i}_{t} \in 	\Re^{A^i}$,
where $A^i$ is the number of activities.

The expenditure and contribution of each player $i$ is weighted by matrix
$\miB^{i} \in \Re^{S \times A^i}$. 
Also, resources can be autonomously recharged/depleted,
which is modeled by a shared matrix 
$\miC \in \Re^{S\times S}$.
Thus, the state transition of the system 
is $f = \miC \mix_t + \sum_{i \in \Q} \miB^i \miu_t^i$.

We consider two cost terms:
unsatisfied demand 
and unbalanced resources.
Given the available resources $\mix_t$, 
every player $i$ will have a target demand 
$ \miD^i \mix_t$ 
that it wants to satisfy,
for some demand matrix $\miD^i \in \Re^{A^i \times S}$. 
The disutility from an unsatisfied demand is modelled by the quadratic form 
$
\left(
	\miD^{i} \mix_{t} - \miu_{t}^{i}
\right)^{\T}
\miQ^{i}
\left(
	\miD^{i} \mix_{t} - \miu_{t}^{i}
\right)$,
with demand cost matrix $\miQ^{i} \in \Re^{A^i \times A^i}$.
In addition, 
the available resources should be just enough to satisfy the demand.
There is a cost for having too little (e.g., productivity decrease) or too much (e.g., storage costs) resources. 
This cost can be modeled as another quadratic form:
$
\left( 
	\mix_{t} - \mix_{t-1}
\right)^{\T}
\miR
\left( 
	\mix_{t} - \mix_{t-1}
\right)$, 
with unbalanced resources cost matrix $\miR \in \Re^{S \times S}$.
In order to pose the game as a maximization problem,
we assume 
$\lbrace \miQ^i \rbrace_{\forall i \in \Q}$ to be negative definite matrices, 
and $\miR$ a negative semidefinite matrix
(this is represented by $\miQ^i \prec 0$
and $\miR \preceq 0$).

The dynamic energy demand control game is given by the following
coupled optimal control problems:
\begin{IEEEeqnarray}{rCl}
\begin{aligned}
\mc{G}_3
: 
\\
\forall i\in\Q
\end{aligned}
	\begin{aligned}
		\underset{ \{ u_t^i \} \in \prod_{t=0}^{\infty}\U^i }{\rm maximize} 	
			& 	\sum_{t=0}^{\infty} 
						\beta^t 
						\Big(
							\left( \mix_t - \mix_{t-1} 	\right)^\T
								\miR
								\left( \mix_t - \mix_{t-1} 	\right)
\\
			&
								+
								\left( \miD^i \mix_t - \miu_{t}^i	\right)^\T
								\miQ^i
								\left( \miD^i \mix_t - \miu_{t}^i 	\right)
						\Big)
\quad\;\;\:
\\
		{\rm s.t.} 		&\;\; 	\mix_{t+1} 	
									= 
										\miC \mix_t 
										+ 
										\sum_{i=1}^Q
											\miB^i \miu_t^i
								,\;\;
								\mix_0 \text{ given} 
	\end{aligned}
\label{eq:smart-grid-problem}
\end{IEEEeqnarray}
By defining augmented state and action vectors:
\begin{IEEEeqnarray}{rCl}
\widetilde{\mix}_t^\T
&
\defeq 
&
	\left[
		\mix_t^\T,
		\mix_{t-1}^\T,
	\right]^\T
,\quad
\widetilde{\miu}_t^i
\defeq 
	\miD^i \mix_t
	-
	\miu_{t}^i
\end{IEEEeqnarray}
we can rewrite 
\eqref{eq:smart-grid-problem} 
in the standard linear-quadratic form:
\begin{IEEEeqnarray}{rCl}
	\begin{aligned}
		\underset{ \{ u_t^i \} \in \prod_{t=0}^{\infty}\U^i }{\rm maximize} 	
						&\quad 	\sum_{t=0}^{\infty} 
									\beta^t 
									\left(
										\widetilde{\mix}_t^\T
										\widetilde{\miR}
										\widetilde{\mix}_t
										+
										\widetilde{\miu}_t^{i^\T}
										\miQ^i
										\widetilde{\miu}_t^i
									\right)
\\
		{\rm s.t.} 		&\quad 	
							\widetilde{\mix}_{t+1} 	
									= 
										\miA \widetilde{\mix}_t 
										- 
										\sum_{i=1}^Q
											\widetilde{\miB}^i 
											\widetilde{\miu}_t^i
								,\quad 
								\mix_0 \text{ given} 
	\end{aligned}
\label{eq:smart-grid-problem-LQ-form}
\end{IEEEeqnarray}
where
\begin{IEEEeqnarray}{rCl}
\miA 
&
	\defeq 
&
		\begin{bmatrix}
			\miC
			+
			\sum_{i\in\Q}
				\miB^{i}
				\miD^{i} 
			& \mathbf{0}_{S \times S}
			\\
			\mathbf{I}_S
			& \mathbf{0}_{S \times S}
		\end{bmatrix} 
,\quad
\widetilde{\miB}^i
	\defeq 
		\begin{bmatrix}
			\miB^i
			\\
			\mathbf{0}_{S \times A^i}
		\end{bmatrix}
\quad
\\
\widetilde{\miR}
&
	\defeq
&
		\begin{bmatrix}
			\miR 
			& -\miR
			\\
			-\miR 
			& \miR
		\end{bmatrix}
\label{eq:smart-grid-problem-LQ-form-matrices}
\end{IEEEeqnarray}
and where 
$\mathbf{I}_{S}$ 
and
$\mathbf{0}_{S \times S}$ 
denote the
identity and null matrices of size $S \times S$,
respectively.
LQDG games in the form \eqref{eq:smart-grid-problem-LQ-form} 
have been presented in \cite[Ch. 6]{Basar1999},
where an NE is found by 
\textit{i)} solving the system of coupled finite horizon OCP,
\textit{ii)} finding the limit of this solution as the horizon tends to infinity,
and then 
\textit{iii)} verifying that this limiting solution  
provides a
NE solution for the infinite-horizon game.
Here we follow a different and simpler approach. 
First, we show that problem \eqref{eq:smart-grid-problem-LQ-form} can be expressed in the separable form \eqref{eq:condition-dynamic-potential-2}:
\begin{IEEEeqnarray}{rCl}
\pi^i 
	(\widetilde{\mix}_t, 
&&	
	\widetilde{\miu}_t) 
=
	\widetilde{\mix}_t^\T \widetilde{\miR} \widetilde{\mix}_t	
	+
	\widetilde{\miu}_t^{i^\T} \miQ^i \widetilde{\miu}_t^{i}
\notag\\
&&=\:
	\widetilde{\mix}_t^\T \widetilde{\miR} \widetilde{\mix}_t
	+
	\sum_{p\in \Q} 
		\widetilde{\miu}_t^{p^\T} \miQ^p \widetilde{\miu}_t^{p}
	-
	\sum_{j\in \Q : j\neq i}
		\widetilde{\miu}_t^{j ^\T} \miQ^j \widetilde{\miu}_t^{j}
\qquad
\label{eq:smart-grid-augmented-local-utility}
\end{IEEEeqnarray}
We identify the potential and separable functions in \eqref{eq:smart-grid-augmented-local-utility}:
\begin{IEEEeqnarray}{rCl}
\Pi \left(\widetilde{\mix}_{t},\widetilde{\miu}_{t},t \right)
&=&
	\widetilde{\mix}_t^\T \widetilde{\miR} \widetilde{\mix}_t
	+
	\sum_{p\in \Q} 
		\widetilde{\miu}_t^{p^\T} \miQ^p \widetilde{\miu}_t^{p}
\\
\Theta(\widetilde{\miu}_t^{-i},t )
&=&
	-
	\sum_{j\in \Q : j\neq i}
		\widetilde{\miu}_t^{j ^\T} \miQ^j \widetilde{\miu}_t^{j}
\label{eq:smart-grid-potential}
\end{IEEEeqnarray}
From Lemma \ref{lemma:dynamic-potential-games-2},
we conclude that problem \eqref{eq:smart-grid-problem} is a DPG.
\textblue{Note also that 
Assumptions \ref{as:differentiability}--\ref{as:convex-state-and-action-sets} hold.
Moreover, the objective in \eqref{eq:smart-grid-problem-LQ-form} is concave and
the state dynamics---which is the only equality constraint---are linear.
Therefore, Slater's constraint qualification is satisfied
and Assumption \ref{as:qualified-constraints} holds.
In addition, 
the matrices 
$\miQ^i \prec 0$, $\forall i \in \Q$, 
and $\miR \preceq 0$
make the potential \eqref{eq:smart-grid-potential} coercive.
Hence, Lemma \ref{lemma:existence} states that Assumption \ref{as:existence-MOCP-solution} is satisfied.
Since Assumptions \ref{as:differentiability}--\ref{as:existence-MOCP-solution} hold,}
Theorem \ref{theorem:game-and-mocp-equivalence}
establishes that we can find an NE of \eqref{eq:smart-grid-problem} 
by solving an equivalent MOCP:
\begin{IEEEeqnarray}{rCl}
\mc{P}_3:
\begin{aligned}
	\underset{ \{ \miu_t \} \in \prod_{t=0}^{\infty}\U }{\rm maximize} 	
		&\quad 	
			V(\widetilde{\mix}_0)
			\defeq
				\sum_{t=0}^{\infty} 
					\beta^t 
					\Big(
						\widetilde{\mix}_t^\T \widetilde{\miR} \widetilde{\mix}_t
\\
&
\qquad\qquad\qquad\quad
						+
						\sum_{p\in \Q} 
							\widetilde{\miu}_t^{p^\T} \miQ^p \widetilde{\miu}_t^{p}				
					\Big)
\\
		{\rm s.t.} 		&\quad 	
				\widetilde{\mix}_{t+1} 	= \miA \widetilde{\mix}_t 
								-
								\sum_{i=1}^Q
									\widetilde{\miB}^i 
									\widetilde{\miu}_t^i
				,\;\; \widetilde{\mix}_0 \;\; {\rm given}
	\end{aligned}
\label{eq:smart-grid-mocp-problem}
\end{IEEEeqnarray}
%
%
where the cumulative objective 
function $V$ is known as \textit{value function} in the optimal control literature (see, e.g., \cite{Bertsekas2007}).
Let 
$
\widetilde{\miu}_t \defeq \left( \widetilde{\miu}_t^i \right)_{i =1}^Q
$
be the vector of all players' augmented actions. 
Aggregate all players' demand matrices in a block diagonal matrix 
$\miQ \defeq \operatorname{diag}
	\left(
		\miQ^1, \ldots, \miQ^Q
	\right)$ 
of size 
$\sum_{i=1}^Q A^i \times \sum_{i=1}^Q A^i$,
and aggregate all players' expenditure weighting matrices
in a $S \times \sum_{i=1}^Q A^i$ thick matrix
$
\widetilde{\miB}
\defeq
	\left(
		\widetilde{\miB}^1, \ldots,
		\widetilde{\miB}^Q
	\right)$.
Then, we can rewrite the value and transition functions as follows:
\begin{IEEEeqnarray}{rCl}
V(\widetilde{\mix}_0) 
&=& 
	\sum_{t=0}^{\infty}
		\beta^t
		\left(
			\widetilde{\miu}_{t}^{\T}
			\miQ
			\widetilde{\miu}_{t}
			+ 
			\widetilde{\mix}_{t}^{\T}
			\widetilde{\miR} 
			\widetilde{\mix}_{t}
		\right)
\label{eq:smart-grid-value-function}
\\
\widetilde{\mix}_{t+1}
&=&
	\miA \widetilde{\mix}_{t}
	- 
	\miB \widetilde{\miu}_{t}
\label{eq:smart-grid-transition-model}
\end{IEEEeqnarray}
%

%
\subsection{Analytical solution to the MOCP and simulation results}
\label{ssed:smart-grid-analytical-solution}
%

It is well known 
that the value function satisfies 
a recursive relationship, known as Bellman equation
(see, e.g., \cite{Bertsekas2007}):
\begin{IEEEeqnarray}{rCl}
V(\widetilde{\mix_t}) 
&=& 
	\beta^t
	\left(
		\widetilde{\miu}_{t}^{\T}
		\miQ
		\widetilde{\miu}_{t}
		+ 
		\widetilde{\mix}_{t}^{\T}
		\widetilde{\miR} 
		\widetilde{\mix}_{t}
	\right)
	+
	\beta^{t+1}
	V(\widetilde{\mix}_{t+1}) 
\quad
\label{smart-grid-bellman-equation}
\end{IEEEeqnarray}
Moreover,
for an LQ control problem, 
it is known \cite[Ch. 6]{Basar1999} that 
the optimal value function 
can be expressed as a quadratic form of the state:
\begin{IEEEeqnarray}{rCl}
V ( \widetilde{\mix}_t)
= 
	\widetilde{\mix}_t^{\T}
	\miP
	\widetilde{\mix}_t
\label{eq:smart-grid-analytical-value-function}
\end{IEEEeqnarray} 
for some negative semidefinite matrix $\miP$.
We can use \eqref{eq:smart-grid-analytical-value-function} to find a closed form expression for the sequence of optimal actions
as follows.
Expand  
\eqref{eq:smart-grid-transition-model} and 
\eqref{eq:smart-grid-analytical-value-function}
into \eqref{smart-grid-bellman-equation}:
\begin{IEEEeqnarray}{rCl}
V(\widetilde{\mix_t}) 
&=& 
	\beta^t
	\left(
		\widetilde{\miu}_{t}^{\T}
		\miQ
		\widetilde{\miu}_{t}
		+ 
		\widetilde{\mix}_{t}^{\T}
		\widetilde{\miR} 
		\widetilde{\mix}_{t}
	\right)
\notag\\
&&
	+
\:
	\beta^{t+1}
	\left( 
		\miA \widetilde{\mix}_{t} - \miB \widetilde{\miu}_{t}
	\right)^{\T}
	\miP
	\left( 
		\miA \widetilde{\mix}_{t} - \miB \widetilde{\miu}_{t}
	\right)
\label{smart-grid-expanded-value-function}
\end{IEEEeqnarray}
Now, 
we just have to maximize \eqref{smart-grid-expanded-value-function} 
over $\widetilde{\miu}_t$.
Since $\miQ$ and $\miP$ are negative definite and semidefinite matrices,
respectively, 
a necessary and sufficient condition for the maximum is
\begin{IEEEeqnarray}{rCl}
\nabla_{ \widetilde{\miu}_{t}} V( \widetilde{\mix}_{t})
&=&
	\beta^t
		\miQ \widetilde{\miu}_{t}
		-
		\beta^{t+1}
		\miB^{\T}
		\miP
		\left(
			\miA
			\widetilde{\mix}_{t}
			-
			\miB 
			\widetilde{\miu}_{t}
		\right)
=
0
\quad
\label{eq:smart-grid-maximum-value-function}
\end{IEEEeqnarray}
From \eqref{eq:smart-grid-maximum-value-function}, 
we obtain an analytical expression for the optimal action at any time step:
\begin{IEEEeqnarray}{rCl}
\widetilde{\miu}_{t} 
=
\beta
\left(
	\miQ
	+
	\beta
	\miB^{\T}
	\miP
	\miB 
\right)^{-1}
\miB^{\T}
\miP
\miA
\widetilde{\mix}_{t}
\label{eq:smart-grid-optimal-action}
\end{IEEEeqnarray}

If we are also interested in finding the optimal value, 
we can expand \eqref{eq:smart-grid-optimal-action} into 
\eqref{smart-grid-expanded-value-function}
and isolate $\miP$:
\begin{IEEEeqnarray}{rCl}
\miP
&
=
&
	\widetilde{\miR}
	+
	\beta
	\miA^{\T}
	\miP \miA
\notag\\
&&
	-
\:
	\beta^{2}
	\miA^{\T}
	\miP \miB
	\left(
		\miQ
		+
		\beta
		\miB^{\T}
		\miP\miB
	\right)^{-1}
	\miB^{\T}
	\miP \miA
\label{eq:dare-P}
\end{IEEEeqnarray}
Note that \eqref{eq:dare-P} is a discrete algebraic Riccati equation,
which is known to be a contraction mapping
if $\miQ \prec 0$,
$\widetilde{\miR} \preceq 0$
and the spectral radius of $\miA$ is smaller than one 
\cite[Ch. 5]{Ljungqvist2012}
(the analysis can be performed under weaker conditions though \cite{Bertsekas2007,Sargent1988}). 
When \eqref{eq:dare-P} is a contraction,
it has a unique solution $\miP^{\star}$
that can be approximated by iterating the following fixed point equation,
such that 
$
\lim_{n \rightarrow \infty} \miP_n = \miP^{\star}
$:
\begin{IEEEeqnarray}{rCl}
\miP_{n+1}
&
=
&
	\widetilde{\miR}
	+
	\beta
	\miA^{\T}
	\miP_n
	\miA
\notag\\
&&
	-
\:
	\beta^{2}
	\miA^{\T}
	\miP_n
	\miB
	\left(
		\miQ
		+
		\beta
		\miB^{\T}
		\miP_n
		\miB
	\right)^{-1}
	\miB^{\T}
	\miP_n
	\miA
\quad\;\;
\label{eq:fixed-point-iteration-P}
\end{IEEEeqnarray}

We have simulated the smart grid model for 
$Q = 8$ players, 
$S = 4$ resources, 
$A^i=6$ activities for every player, 
random negative definite matrices $\miQ^{i}$, $\forall i \in \Q$, 
and random negative semidefinite matrix $\miR$ 
(to build these negative matrices we build an intermediate matrix, e.g., $\miR_{\rm int}$, by drawing random numbers from a uniform distribution, 
with support $[0,10]$ for $\miQ^{i}$
and $[0,5]$ for $\miR$,
and compute $\miR = -\miR_{\rm int}^\T \miR_{\rm int}$). 
Matrices $\miC$, $\miB^{i}$ and $\miD^i$ are also random with elements drawn from the spherical normal distribution.
Finally, the initial state was set to a vector of ones, 
and discount factor $\beta=0.9$. 

\begin{figure}[!ht]
\centering
\input{lq_utility.tikz}
\\
\hspace{2.4pt}
\input{lq_coefficients.tikz}
\vspace{-0.3cm}\caption{Dynamic smart grid scenario with $Q=8$ players.
(Top) Instant utility values of players.
(Bottom) Players' decision coefficients evolution in time.} 
\label{Fig:smart-grid-model-results}
\end{figure}
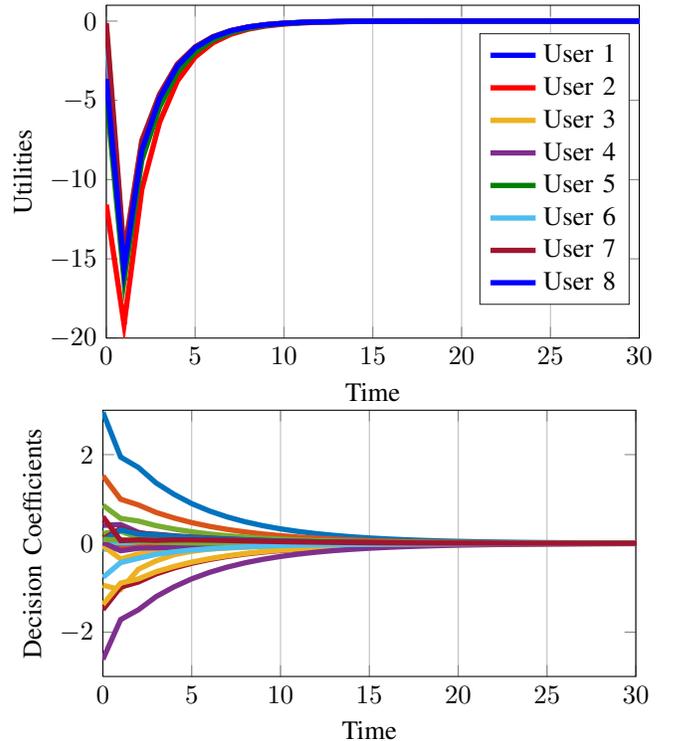

Figure \ref{Fig:smart-grid-model-results}-Top shows the instant utilities per player over time. 
Recall that the utilities have been defined as negative costs.
Therefore, each player's utility starts being a negative value and converges to zero with time.
This behaviour illustrates that all players attain an NE in which they are able to satisfy their demand as well as to hold just enough available resources.
Figure \ref{Fig:smart-grid-model-results}-Bottom shows the evolution of the part of the cost corresponding to the individual coefficients 
$
\widetilde{\miu}_t^i 
= 
\miD^i \mix_t - \miu_t^i
$. 
These coefficients represent the mismatch among target demand, 
$\miD^i \mix_t$, 
and the actual player activities $\miu_t^i$. 
We can see that the agents adjust their actions $\miu_t^i$ to satisfy the target demand.
The equilibrium between target demand and players' activities is an expected consequence of the stability of the LQ game in infinite horizon \cite[Ch. 6]{Basar1999}.

\section{Network Flow Control: Infinite Horizon Approximated by a Finite Horizon Dynamic Game} 
\label{sec:network-flow}

Several works 
(see, e.g., \cite{Low1999,Mo2000,Wang2003,Cjiang2007})
have considered network flow control as an optimization problem 
wherein each source is characterized by a utility function that depends on the transmission rate,
and the goal is to maximize the aggregated utility.
We generalize the standard model 
by considering that the nodes are equipped with batteries that are depleted proportionally to the outgoing flow.
In addition we consider several layers of relay nodes, 
each one with multiple links,
so there are several paths between source and destination. 
When the batteries are completely depleted, 
no more transmissions are allowed and the game is over. 
Hence, although we formulate the problem as an infinite horizon dynamic game,
the effective time horizon---before the batteries deplete---is finite.
This problem has no known analytical solution,
but the utilities are concave.
Therefore,
the finite horizon approximation is convenient 
because we can solve an equivalent concave optimization problem, 
significantly reducing the computational load
with respect to other optimal control algorithms (e.g., dynamic programming).

\subsection{Network flow control dynamic game and equivalent MOCP}
\label{ssec:network-flow-scenario}

Let $u_t^{ia}$ denote the flow along path $a$ for user $i$ at time $t$.
Suppose there are $A^i$ possible paths for each player $i \in \Q$,
so that
$
\miu_t^i 
\defeq 
	\left(
		u_t^{ia}
	\right)_{a=1}^{A^i}
$
denotes the $i$-th player's action vector.
Let $A = \sum_{i=1}^Q A^i$ denote the total number of available paths.

Suppose there are $S$ relay nodes.
Let $x_t^k$ denote the battery level of relay node $k$.
The state of the game is given by 
$
\mix_t 
\defeq 
	\left( 
		x_t^k
	\right)_{k=1}^S
$,
such that all players share all components of the state-vector
(i.e., $\X^i = \X$ and $\X(i) = \{1, \ldots, S \}$, $\forall i \in \Q$).
The battery level evolves with the following state-transition equation
for all components
$
	k=1,\ldots,S
$:
\begin{IEEEeqnarray}{rCl}
\label{eq:network-flow-dynamics}
x_{t+1}^k 
= 
	x_{t}^k 
	- 
	\delta 
	\sum_{i=1}^Q
	\sum_{u^{ia} \in F_k}
		u_t^{ia}
,\quad
	x_0^k = B^k_{\max}
\end{IEEEeqnarray}
where 
$F_k$ denotes the subset of flows through node $k$,
$B^k_{\max}$ is a positive scalar that stands for the maximum battery level of node $k$,
and $\delta$ is a proportional factor.

Similar to the standard static flow control problem,
each player intends to maximize a concave function $\Gamma: \U^i \rightarrow \Re$ of the sum of rates across all available paths.
This function $\Gamma$ can take different forms depending on the scenario under study,
like the square root \cite{Nedic2010}
or a capacity form.
In addition to the transmission rate, 
we include the relay nodes' battery level in each player's utility,
weighted by some positive parameter $\alpha$.
The combination of these two objectives can be understood as the player aiming to maximize its total transmission rate, 
while saving the batteries of the relays.

There is some capacity constraint of the maximum aggregated rate at every relay and destination node.
Let $\mic_{\max} \in \Re^{L}$ denote the vector with maximum capacities, 
where $L$ is the number of relays plus destination nodes.
Let $\miM = \left[ m_{l a} \right]$ denote the ${L \times A}$ matrix that define the aggregated flows for each relay and destination node, 
such that element $m_{l a} = 1$, if flow node $a$ is aggregated in node $l$,
and $m_{l a} = 0$ otherwise.

The dynamic network flow control game is given by the following
set of coupled OCP:
\begin{IEEEeqnarray}{rCl}
\begin{aligned}
\mc{G}_4
:
\\
\forall i\in\Q
\end{aligned}
	\begin{aligned}
		\underset{ \{ u_t^i \} \in \prod_{t=0}^{\infty}\U^i }{\rm maximize} 	
			&\quad 	
				\sum_{t=0}^{\infty} 
					\beta^t 
					\left(
						\Gamma \left( \sum_{a=1}^{A^i} u_t^{ia} \right)
						+
						\alpha
						\sum_{k=1}^S 
							x_t^k
					\right)
\\
		{\rm s.t.} 		
			&\quad
				x_{t+1}^k 
				= 
					x_{t}^k 
					- 
					\delta 
					\sum_{i =1}^{Q}
						\sum_{u^{ia} \in F_k}
							u_t^{ia}
\\
			&\quad
				x_0^k = B^k_{\max}
			,\;\; 
				0 \le x_t^k \le B^k_{\max}
\\
			&\quad
				\miM \miu_t \le \mic_{\max}
			,\;\;
				u_t^{ia} \ge 0
\\
			&\quad
				k=1, \ldots, S
			,\;\; 
				a=1, \ldots, A^i
	\end{aligned}
\label{eq:network-flow-problem}
\end{IEEEeqnarray}

Note that each player's utility can be expressed in separable form:
\begin{IEEEeqnarray}{rCl}
\pi^i 
	( 
&&
		\mix^i_t, 
		\miu, 
		t
	)
	\defeq
		\Gamma \left( \sum_{a=1}^{A^i} u_t^{ia} \right)
		+
		\alpha
		\sum_{k=1}^S 
			x_t^k
\notag\\
&&
	=\:
		\sum_{i \in \Q}
			\Gamma \left( \sum_{a=1}^{A^i} u_t^{i a} \right)
			+
			\alpha
			\sum_{k=1}^S 
				x_t^k
			-
			\sum_{j\in \Q : j\neq i}
				\Gamma \left( \sum_{a=1}^{A^j} u_t^{j a} \right)
\notag\\
\end{IEEEeqnarray}
Therefore, Lemma \ref{lemma:dynamic-potential-games-2} establishes that problem \eqref{eq:network-flow-problem} 
is a DPG, 
with potential function given by:
\begin{IEEEeqnarray}{rCl}
\Pi (\mix_t, \miu_t, t)
	=
			\sum_{i \in \Q}
			\Gamma \left( \sum_{a=1}^{A^i} u_t^{i a} \right)
			+
			\alpha
			\sum_{k=1}^S 
				x_t^k
\end{IEEEeqnarray}

\textblue{
Before applying Theorem \ref{theorem:game-and-mocp-equivalence},
we have to check whether 
Assumptions \ref{as:differentiability}--\ref{as:existence-MOCP-solution}
are satisfied.
We follow \cite{Nedic2010} and choose $\Gamma(\cdot) \defeq \sqrt{(\epsilon + \cdot)}$
(where $\epsilon>0$ is only added to avoid differentiability issues when $u_t^{ia}=0$).
Let $\X$  and $\U^i$ be open convex sets 
containing the Cartesian products of intervals $[0, B^k_{\max}]$ and $[0, \infty)$, respectively.
It follows that 
Assumptions \ref{as:differentiability}--\ref{as:convex-state-and-action-sets} hold.
Moreover, since $\Gamma$ is concave
and problem \eqref{eq:network-flow-problem}
has linear equality constraints and concave inequality constraints,
Slater's condition holds, 
i.e., Assumption \ref{as:qualified-constraints} is satisfied.
Finally, 
since the constraint set in \eqref{eq:network-flow-problem} is compact,
Lemma \ref{lemma:existence}.1 states that Assumption \ref{as:existence-MOCP-solution} holds.
Hence,}
Theorem \ref{theorem:game-and-mocp-equivalence}
establishes that we can find an NE of \eqref{eq:network-flow-problem} 
by solving the following MOCP:
\begin{IEEEeqnarray}{rCl}
\mc{P}_4:
\begin{aligned}
		\underset{ \{ \miu_t \} \in \prod_{t=0}^{\infty}\U }{\rm maximize} 	
				&\quad 	
				\sum_{t=0}^{\infty} 
					\beta^t 
					\left(
						\sum_{i \in \Q}
							\Gamma \left( \sum_{a=1}^{A^i} u_t^{i a} \right)
							+
							\alpha
							\sum_{k=1}^S 
								x_t^k
					\right)
\\
		{\rm s.t.} 		
			&\quad
				x_{t+1}^k 
				= 
					x_{t}^k 
					- 
					\delta 
					\sum_{i \in \Q}
						\sum_{u^{ia} \in F_k}
							u_t^{ia}
\\
			&\quad
					x_0^k = B^k_{\max}
			,\;\; 
				0 \le x_t^k \le B^k_{\max}
\\
			&\quad
				\miM \miu_t \le \mic_{\max}
			,\;\;
				\miu_t \ge 0
\\
			&\quad
				k=1, \ldots, S
	\end{aligned}
\label{eq:network-flow-mocp-problem}
\end{IEEEeqnarray}
%

%
\subsection{Finite horizon approximation and simulation results}
\label{ssec:network-flow-simulation}
%

As opposed to the LQ smart-grid problem,
there is not known closed form solution for problem \eqref{eq:network-flow-mocp-problem}.
Thus, we have to rely on numerical methods to solve the MOCP.
Suppose that we set the weight parameter $\alpha$ in $\Pi$ low enough to incentivize some positive transmission.
Eventually, the nodes' batteries will be depleted,
so the system will get stuck in an equilibrium state, 
with no further state transitions.
Thus, we can approximate the infinite-horizon problem \eqref{eq:network-flow-mocp-problem} as a finite-horizon problem,
with horizon bounded by the time-step at which all batteries have been depleted.
Moreover, in our setting, we have assumed $\Gamma$ to be concave.
Therefore, we can effectively solve \eqref{eq:network-flow-mocp-problem}
with convex optimization solvers
(we use the software described in \cite{GrantCVX2014}).
The benefit of using a convex optimization solver is that standard optimal control algorithms 
are computationally demanding when the state and action spaces are subsets of vector spaces.

\begin{figure}[ht]
\centering
\includegraphics[width=0.9\linewidth]{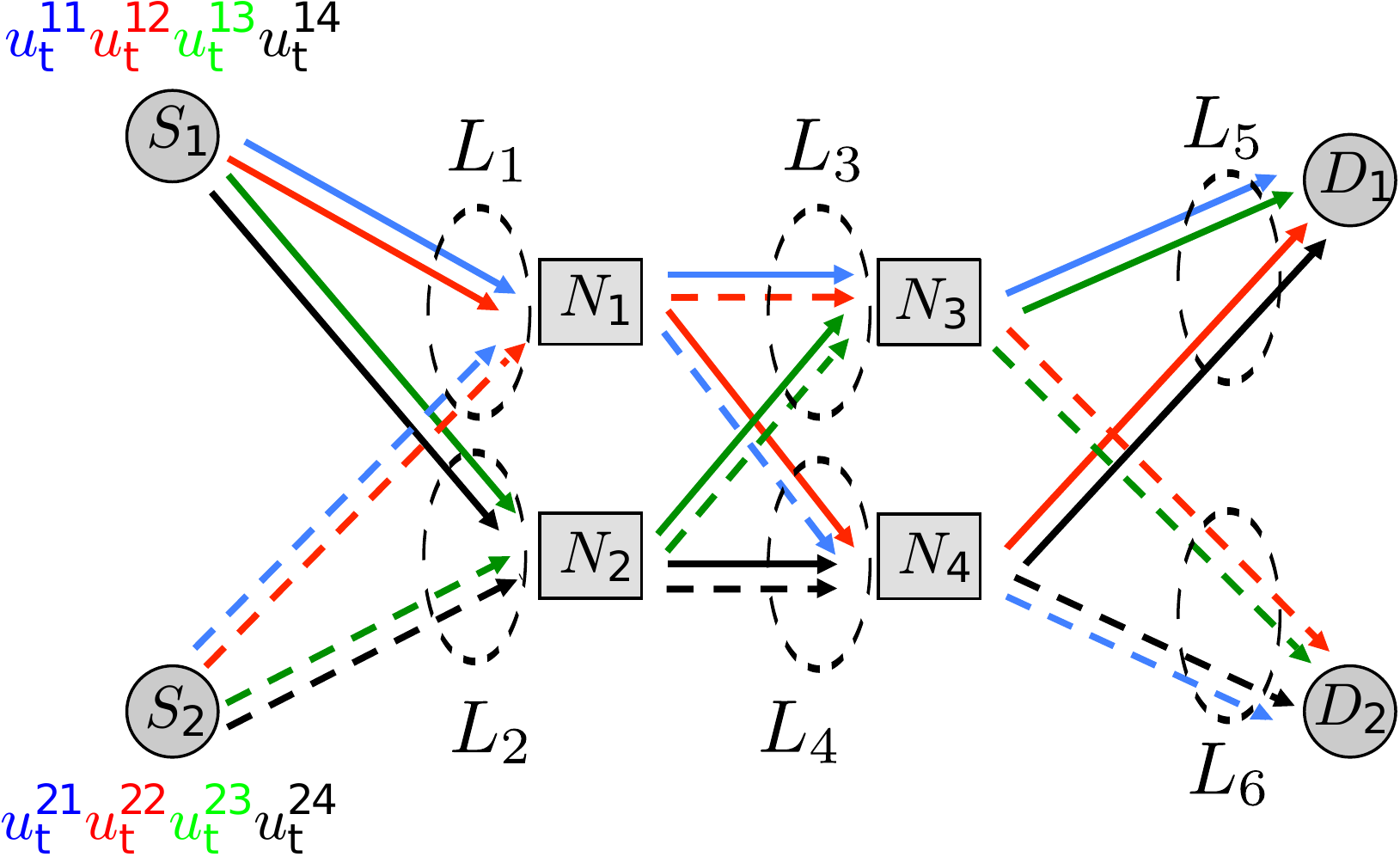}
\caption{
Network scenario for two users and two levels of relying nodes.
Player $S_1$ aims to transmit to destination $D_1$,
while $S_2$ aims to transmit to destination $D_2$.
They have to share relay nodes $N_1, \ldots, N_4$.
We denote the $L=6$ aggregated flows as $L_1, \ldots, L_6$.
}
\label{Fig:network-flow-network}
\end{figure}

For our numerical experiment, 
we consider $Q=2$ players that share a network of $S=4$ relay nodes, 
organized in two layers (see Figure \ref{Fig:network-flow-network}).
In this particular setting, 
each player is allowed to use four paths, 
$A^1 = A^2 = 4$.
The connectivity matrix $\miM$ can be obtained from Figure \ref{Fig:network-flow-network}.
The battery is initialized to $B_{\max}=1$ for the four relay nodes, 
we set the depleting factor $\delta=0.05$,
discount factor $\beta=0.9$,
the weight $\alpha=1$, 
$\epsilon=0.001$
and the vector of maximum capacities 
\textblue{$\mic_{\max} = [0.5, 0.15, 0.5, 0.15, 0.4, 0.4]^\T$}.

\begin{figure}[h!t]
\centering
\hspace{5pt}
\input{flow_aggregated_2.tikz}
\\
\input{flow_flows_2.tikz}
\\
\input{flow_batteries_2.tikz}
\vspace{-0.65cm}\caption{Network flow control with $Q=2$ players, $S=4$ relay nodes
and $A^1=A^2=4$ available paths per node.
(Top) Aggregated flow rates at $L_1,\ldots,L_6$.
(Middle) Flow for each of the $A=8$ available paths.
(Bottom) Battery level in each of the $S=4$ relay nodes.}
\label{Fig:network-flow-network-results}
\end{figure}
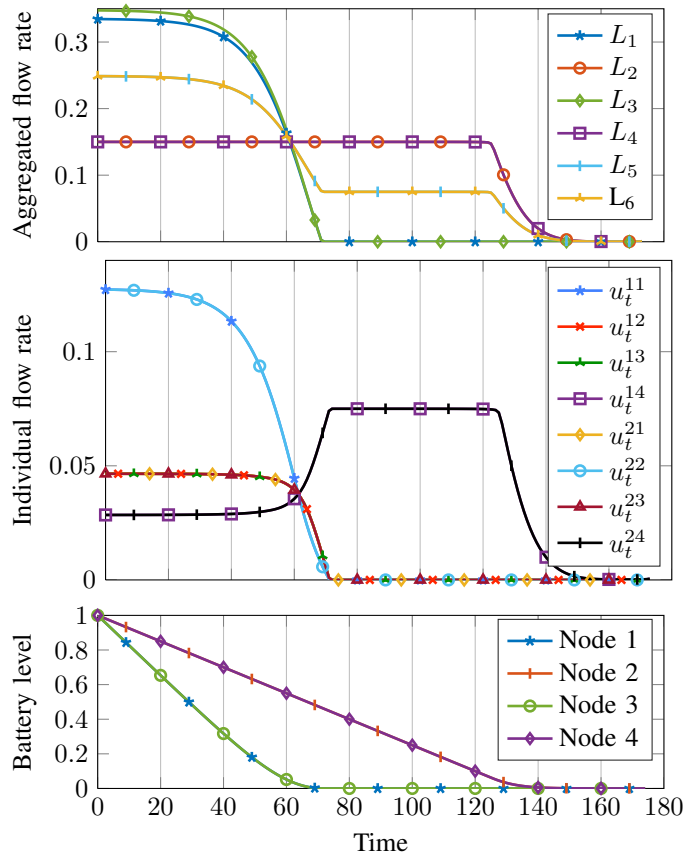

\textblue{Figure \ref{Fig:network-flow-network-results} shows 
the evolution of the $L=6$ aggregated flows,
the $A=8$ flows
and the battery of each of the $N=4$ relay nodes.
%
%
Since we have included the battery level of the relay nodes in the users' utilities
(i.e., $\alpha > 0$), 
the users have an extra incentive to limit their flow rate. 
Thus, there are two effective reasons to limit the flow rate:  
satisfy the problem constraints 
and 
save battery.
We can see that 
the aggregated flows with higher maximum capacity are not saturated
($L_1 < 0.5, L_3 < 0.5, L_4 < 0.4$, and $L_6<0.4$).
The reason is that the users have limited their individual flow rates in order to save relays' batteries. 
On the other hand, the aggregated flows with lower maximum capacity  
are saturated ($L_2=L_4=0.15$)
because the capacity constraint is more restrictive than 
the self-limitation incentive.
When the batteries of the nodes with higher maximum capacity 
($N_1, N_3$) are depleted (around $t=70$),
the flows through these nodes stop.
This allows the other flows ($u_t^{14}, u_t^{24}$) to transmit at a higher rate.
At this time, the capacity constraint in $L_2,L_4$ is more restrictive than 
the self-limitation incentive for saving the batteries, 
so that the users transmit at the maximum rate allowed by the capacity constraints
(note that $L_2=L_4=0.15$ remains constant).
When the battery of every node is depleted,
none of the users is allowed to transmit anymore
and the system enters in an equilibrium state.}

\textblue{We remark that the solution obtained is an NE based on an OL game analysis.}
%
%
	\textblue{
	Finally, 
	the results shown in
	Figure \ref{Fig:network-flow-network-results}
	have been obtained with a \textit{centralized} convex optimization algorithm,
	meaning that it should be run off-line by the system designer,
	before deploying the real system.
	Alternatively,
	we could have used the distributed algorithms proposed by reference \cite{FacchineiDecomposition2011},
	enabling the players to solve the finite horizon approximation of problem 
	\eqref{eq:network-flow-mocp-problem} in a decentralized manner,
	even with the coupled capacity constraints.}

%
\section{Dynamic Multiple Access Channel: Nonseparable utilities}
\label{sec:MAC}
%

In this section,
we consider an uplink scenario in which every user $i \in \Q$ 
independently chooses its transmitter power, $u_{t}^{i}$, aiming to achieve the maximum rate allowed by the channel \cite{Zazo2014}. 
If multiple users transmit at the same time, 
they will interfere each other, 
which will decrease their rate,
so that they have to find an equilibrium.
Let $R_t^i$ denote the rate achieved by user $i$ with normalized noise at time $t$:
\begin{IEEEeqnarray} {rCl}
\label{eq:transmission-rate}
R_{t}^{i} 
\defeq
	\log 
		\left(
			1
			+
			\frac
				{
					\left| h^{i} \right|^{2}
					u_{t}^{i}
				}
				{
					1
					+
					\sum_{j\in \Q : j\neq i}
						\left| h^{j} \right|^{2}
						u_{t}^{j}
				}
		\right)
\end{IEEEeqnarray}
where $h^{i}$ denotes the fading channel coefficient of user $i$.
%

%
\subsection{Multiple access channel DPG and equivalent MOCP}
\label{ssec:mac-dpg-mocp}
%

Let 
$
x_{t}^i 
\in 
	\left[
		0, B^i_{\max}
	\right] 
$ 
denote the battery level for each player $i \in \Q$,
which is discharged proportionally to the transmitted power $u_t^i$.
The state of the system is given by 
the vector with all individual battery levels:
$
\mix_t 
= 
\left(
	x_t^i
\right)_{i\in\Q} 
	\in 
		\X 
$. 
Thus, each player is only affected by its own battery, 
such that $S=Q$, $\X(i) = \{ i \}$
and $\mix_t^i = x_t^i$.
%
Suppose the agents aim to maximize its transmission rate, 
while also saving their battery.
This scenario yields the following dynamic game:
\begin{IEEEeqnarray}{rCl}
\begin{aligned}
	\mc{G}_5
	:
\\
	\forall i\in\Q
\end{aligned}
\;
	\begin{aligned}
		\underset{ \{ u_t^i \} \in \prod_{t=0}^{\infty}\U^i }{\rm maximize} 	
			&\quad 	
				\sum_{t=0}^{\infty} 
					\beta^t 
					\left(
						R_t^i
						+
						\alpha x_t^i
					\right)
\\
		{\rm s.t.} 		
			&\quad
				x_{t+1}^i
				= 
					x_{t}^i
					- 
					\delta u_t^{i}
			,\quad 
					x_0^i = B^{i}_{\max}
\quad\;\;
\\
			&\quad
				0 \le u_t^{i} \le P^i_{\max}
			,\;\; 
				0 \le x_t^i \le B^{i}_{\max}
	\end{aligned}
\label{eq:multiple-access-problem}
\end{IEEEeqnarray}
where 
$\alpha$ is the weight given for saving the battery,
$\delta$ is the discharging factor, 
and $P^i_{\max}$ and $B^i_{\max}$ denote the maximum transmitter power and maximum available battery level for node $i$, respectively.
Problem \eqref{eq:multiple-access-problem} is a dynamic infinite-horizon extension of the static problem proposed in \cite{Scutari2006}.

Instead of looking for a separable structure in the players' utilities,
we show that Lemma \ref{lemma:dynamic-potential-games-3} holds
and, hence, problem \eqref{eq:multiple-access-problem} is a DPG:
\begin{IEEEeqnarray}{rCl}
	\frac{ \partial^2 \pi^i (x_t^i, \miu_t, t) }
		 { \partial x^i_t \partial u^j_t } 
	&=&
	\frac{ \partial^2 \pi^j (x_t^i, \miu_t, t) }
		 { \partial x^j_t \partial u^i_t } 
	=
		0
\label{eq:mac-conservative-condition-1}
\\
	\frac{ \partial^2 \pi^i (x_t^i, \miu_t, t) }
		 { \partial x^i_t \partial x^j_t } 
	&=&
	\frac{ \partial^2 \pi^j (x_t^i, \miu_t, t) }
		 { \partial x^j_t \partial x^i_t } 
	=
		0
\label{eq:mac-conservative-condition-2}
\\
	\frac{ \partial^2 \pi^i (x_t^i, \miu_t, t) }
		 { \partial u^i_t \partial u^j_t } 
	&=&
	\frac{ \partial^2 \pi^j (x_t^i, \miu_t, t) }
		 { \partial u^j_t \partial u^i_t } 
	=
		\frac	{
					-
					\left| h^i \right|^2 
					\left| h^j \right|^2
				}
				{	
					\left(	
						1
						+
						\sum_{p \in \Q}
							\left| h^p \right|^2
							u_t^p
					\right)^2
				}
\notag\\
\label{eq:mac-conservative-condition-3}
\end{IEEEeqnarray}
\textblue{In order to find an equivalent MOCP,
let us define $\X^i$ and $\U^i$ as open convex sets containing the closed intervals $	\left[0, B^i_{\max}	\right]$ and $[0, P_{\max}^i ]$, respectively,
so that 
Assumptions \ref{as:differentiability}--\ref{as:convex-state-and-action-sets} hold.
Derive the potential function from \eqref{eq:potential-function-dynamic-game}:
\begin{IEEEeqnarray}{rCl}
\Pi(\mix_t,\miu_t,t) 
&
=
&
	\log
		\left(
			1
			+
			\sum_{i=1}^Q
				|h^i|^2
				u_t^i
		 \right)
		 +
		 \alpha
		 \sum_{i=1}^Q
		 	x_t^i
\label{eq:multiple-access-potential-function}
\end{IEEEeqnarray}
Since \eqref{eq:multiple-access-potential-function} is concave
and all equality and inequality constraints in \eqref{eq:multiple-access-problem} are linear,
Assumption \ref{as:qualified-constraints} is satisfied through Slater's condition.
Moreover,
since the constraint set is compact
and the potential is continuous,
Lemma \ref{lemma:existence}.1 establishes that
Assumption \ref{as:existence-MOCP-solution} holds.
Therefore,} Theorem \eqref{theorem:game-and-mocp-equivalence} states that
we can find an NE of \eqref{eq:multiple-access-problem} 
by solving the following MOCP:
\begin{IEEEeqnarray}{rCl}
\mc{P}_5:
\begin{aligned}
		\underset{ \{ \miu_t \} \in \prod_{t=0}^{\infty}\U }{\rm maximize} 	
				&\quad 	
				\sum_{t=0}^{\infty} 
					\beta^t 
					\Bigg(
						\log
							\left(
								1
								+
								\sum_{i=1}^Q
									|h^i|^2
									u_t^i
							 \right)
\\
&\qquad\qquad
							 +
							 \alpha
							 \sum_{i=1}^Q
							 	x_t^i
					\Bigg)
\\
		{\rm s.t.} 		
			&\quad
				x_{t+1}^i
				= 
					x_{t}^i
					- 
					\delta u_t^{i}
			,\quad 
					x_0^i = B^{i}_{\max}
\\
			&\quad
				0 \le u_t^{i} \le P^i_{\max}
			,\;\; 
				0 \le x_t^i \le B^{i}_{\max}
\\
			&\quad
				\forall i \in \Q 
	\end{aligned}
\label{eq:multiple-access-mocp-problem}
\end{IEEEeqnarray}

%
\subsection{Simulation results}
\label{ssec:MAC-simulation}
%

%
Similar to Sec. \ref{ssec:network-flow-simulation},
the system reaches an equilibrium state
when the batteries have been depleted.
Thus, the solution can be approximated by solving a finite horizon problem.
Moreover, since the problem is concave,  
we can use convex optimization software, like \cite{GrantCVX2014}.
Alternatively, we could solve the KKT system with an efficient ad-hoc distributed algorithm, 
like in \cite{Zazo2014}.

We simulated an scenario with $Q=4$ users. 
We set the maximum battery level $B^{i}_{\max}=33$ for all users, 
the maximum power allowed per user $P^{i}_{\max}=5$ for all users, 
the weight battery utility factor $\alpha=0.001$,
the transmitter power battery depletion factor $\delta = 1$,
and the discount factor $\beta=0.95$.
The channel gains are
$|h^1| = 2.019$.
$|h^2| = 1.002$
$|h^3| = 0.514$,
and
$|h^4| = 0.308$.

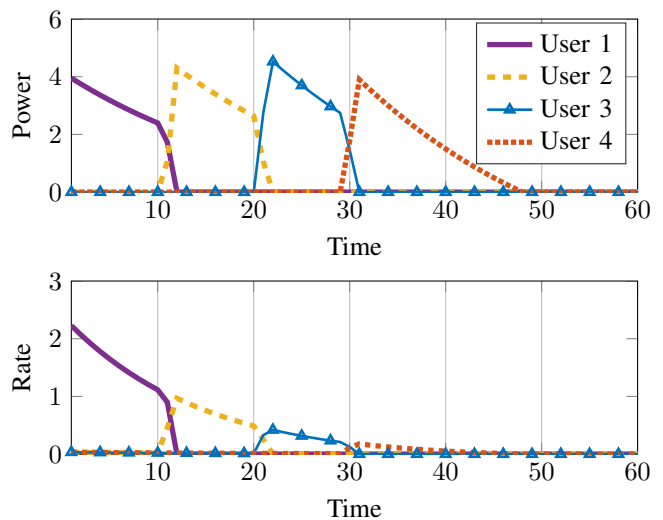
\begin{figure}[!ht]
\input{MAC_powers.tikz}
\\
\input{MAC_utilities.tikz}
\vspace{-0.3cm}
\caption{Dynamic multiple access scenario with $Q=4$ users.
(Top) Sequence of transmitter power chosen by every user. 
(Bottom) Evolution of the transmission rates.
}
\label{Fig:mac-simulations}
\end{figure}

Figure \ref{Fig:mac-simulations} shows appealing results: 
the solution of the MOCP---which is an NE of the game---is actually a schedule.
In other words, instead of creating interference among users, 
they wait until the users with higher channel-gain have depleted their batteries.


%

%
\section{Optimal scheduling: Nonstationary problem with dynamic programming solution}
\label{sec:scheduling}
%

In this section we present the most general form of the proposed framework,
and show its applicability to two scheduling problems.
First, one of the games has nonseparable utilities, 
so we have to verify second order conditions \eqref{eq:conservative-field-condition-1}--\eqref{eq:conservative-field-condition-3}. 
Second, neither the equivalent MOCP can be approximated by a finite horizon problem,
nor the utilities are concave.
Thus, we cannot rely upon convex optimization software and we have to use optimal control methods, 
like dynamic programming \cite{Bertsekas2007}.
Finally, we consider a nonstationary scenario,
in which the channel coefficients evolve with time.
This makes the state-transition equations 
(and the utility for the equal rate problem) 
depend not only on the current state, but also on time.
This problem was introduced in the preliminary paper \cite{Zazo2015}.

%
\subsection{Proportional fair and equal rate scheduling games and their equivalent MOCP}
\label{ssec:scheduling-dynamic-game-and-equivalent-mocp}
%

Let us redefine the rate achieved by user $i$ at time $t$,
so that we consider nonstationary channel coefficients:
\begin{IEEEeqnarray} {rCl}
\label{eq:nonstationary-transmission-rate}
R_{t}^{i} 
\defeq
	\log 
		\left(
			1
			+
			\frac
				{
					\left| h^{i}_t \right|^{2}
					u_{t}^{i}
				}
				{
					1
					+
					\sum_{j\in \Q : j\neq i}
						\left| h^{j}_t \right|^{2}
						u_{t}^{j}
				}
		\right)
\end{IEEEeqnarray}
where $u_t^i$ is the transmitter power of player $i$, 
and
$|h_t^i|$ is its time-varying channel coefficient.

We propose two different scheduling games,
namely,
\textit{proportional fair} 
and 
\textit{equal rate}
scheduling.

\subsubsection{Proportional fair scheduling}
\label{ssec:proportional-fair}

Proportional fair is a compromise-based scheduling algorithm. 
It aims to maintain a balance between two competing interests: 
trying to maximize total throughput while, at the same time, 
guaranteeing a minimal level of service for all users \cite{Kelly1997,Kelly1998,ZhouFair2011}.

In order to achieve this tradeoff, we propose the following game:
\begin{IEEEeqnarray}{rCl}
\begin{aligned}
	\mc{G}_6
	:
\\
	\forall i\in\Q
\end{aligned}
	\begin{aligned}
		\underset{ \{ u_t^i \} \in \prod_{t=0}^{\infty}\U^i }{\rm maximize} 	
			&\quad 	
				\sum_{t=0}^{\infty} 
					\beta^t 
					x_t^i
\\
		{\rm s.t.} 		
			&\quad
				x_{t+1}^i
				= 
				\left(
					1 
					- 
					\frac{1}{t}
				\right)
				x_t^i
				+
				\frac{R_t^i}{t}
\\
			&\quad
				x_0^i = 0
			,\;\: 
				0 \le u_t^{i} \le P^i_{\max}
	\end{aligned}
\quad\;\;
\label{eq:proportional-fair-scheduling-problem}
\end{IEEEeqnarray}
where the state of the system is the vector of all players' average rates
$\mix_t = \left( x_t^i \right)_{i \in \Q}$.
Since each player aims to maximize its own average rate, 
the state-components are unshared among players:
$S=Q$ and $\X(i) = \{ i \}$.

In order to show that problem \eqref{eq:proportional-fair-scheduling-problem}
is a DPG, 
we evaluate Lemma \ref{lemma:dynamic-potential-games-2} with positive result,
and obtain $\Pi$ from \eqref{eq:condition-dynamic-potential-2}:
\begin{IEEEeqnarray}{rCl}
\Pi(\mix_t,\miu_t,t) 
&
=
&
\sum_{i=1}^Q x_t^i
\label{eq:proportional-fair-potential-function}
\end{IEEEeqnarray}
\textblue{
Now, we show that we can derive an equivalent MOCP.
It is clear that 
Assumptions \ref{as:differentiability}--\ref{as:convex-state-and-action-sets} hold.
By taking the gradient of the constraints of \eqref{eq:proportional-fair-scheduling-problem}
and building a matrix with the gradients of the constraints 
(i.e., the gradient of each constraint is a column of this matrix),
it is straightforward to show that the matrix is full rank.
Hence, the \textit{linear independence constraint qualification} holds
(see, e.g., \cite[Sec. 3.3.5]{bertsekas1999nonlinear},
\cite{wang2013constraint}),
meaning that Assumption \ref{as:qualified-constraints} is satisfied.
Finally, since $R_t^i \ge 0$ and $x_0^i = 0$,
we conclude that there exists some scalar $M \ge 0$
such that the level set
$
\{
	\mix_t | 
		\sum_{i\in\Q} x_t^i 
		\ge M
\}
$
is nonempty and bounded,
so that Lemma \ref{lemma:existence}.3 establishes that 
Assumption \ref{as:existence-MOCP-solution} is satisfied.
Thus}, from Theorem \ref{theorem:game-and-mocp-equivalence},
we can find an NE of DPG \eqref{eq:proportional-fair-scheduling-problem}
by solving the following MOCP:
\begin{IEEEeqnarray}{rCl}
\mc{P}_6:
\begin{aligned}
		\underset{ \{ \miu_t \} \in \prod_{t=0}^{\infty}\U }{\rm maximize} 	
				&\quad 	
				\sum_{t=0}^{\infty} 
					\beta^t 
					\sum_{i=1}^Q 
						x_t^i
\\
		{\rm s.t.} 		
			&\quad
				x_{t+1}^i
				= 
				\left(
					1 
					- 
					\frac{1}{t}
				\right)
				x_t^i
				+
				\frac{R_t^i}{t}
\\
			&\quad
				x_0^i = 0
			,\quad 
				0 \le u_t^{i} \le P^i_{\max}
	\end{aligned}
\quad\;
\label{eq:proportional-fair-mocp-problem}
\end{IEEEeqnarray}

\subsubsection{Equal rate scheduling}
\label{ssec:equal-rate}

In this problem,
the aim of each user is to maximize its rate, 
while at the same time keeping the users' cumulative rates as close as possible.
Let $x_t^i$ denote the cumulative rate of user $i$.
The state of the system is the vector of all users' cumulative rate 
$\mix_t = \left( x_t^i \right)_{i \in \Q}$. 
Again $S=Q$ and $\X(i) = \{ i \} $.
This problem is modeled by the following game:
\begin{IEEEeqnarray}{rCl}
\begin{aligned}
	\mc{G}_7
	:
\\
\forall i\in\Q
\end{aligned}
	\begin{aligned}
		\underset{ \{ u_t^i \} \in \prod_{t=0}^{\infty}\U^i }{\rm maximize} 	
			&\quad 	
				\sum_{t=0}^{\infty} 
					\beta^t 
					\Bigg(
						(1 - \alpha)
						R_t^i
\\
&\quad\quad
						-
						\alpha
						\sum_{j \in \Q : j \neq i} 
							\left(
								x_t^i - x_t^j
							\right)^2
					\Bigg)	
\quad				
\\
		{\rm s.t.} 		
			&\quad
				x_{t+1}^i
				= 
				x_t^i + R_t^i
\\
			&\quad
				x_0^i = 0
			,\quad 
				0 \le u_t^{i} \le P^i_{\max}
	\end{aligned}
\label{eq:equal-rate-scheduling-problem}
\end{IEEEeqnarray}
where parameter $\alpha$ weights the contribution of both terms.

It is easy to verify that conditions 
\eqref{eq:conservative-field-condition-1}--\eqref{eq:conservative-field-condition-3}
are satisfied.
Hence, from Lemma \ref{lemma:dynamic-potential-games-3}, we know that 
problem \eqref{eq:equal-rate-scheduling-problem} is a DPG.
\textblue{In order to obtain an equivalent MOCP,
let us define $\X^i$ and $\U^i$ as open convex sets that contain the intervals $[0, \infty)$ and $[0, P_{\max}^i]$, respectively.
It follows that 
Assumptions \ref{as:differentiability}--\ref{as:convex-state-and-action-sets} hold.
Similar to the proportional fair scheduling problem \eqref{eq:proportional-fair-scheduling-problem}, 
Assumption \ref{as:qualified-constraints} holds through the 
linear independence constraint qualification.
Finally, let us check Assumption \ref{as:existence-MOCP-solution} as follows. 
Derive the potential $\Pi$ by integrating \eqref{eq:potential-function-dynamic-game}:
\begin{IEEEeqnarray}{rCl}
\Pi(\mix_t, \miu_t, t)
&
=
&
(1 - \alpha)
\log
	\left(
		1
		+
		\sum_{i=1}^Q
			| h_t^i |^2 
			u_t^i
	\right)
\notag\\
&&
-
\:
	\alpha
	\sum_{i=1}^{Q-1}
		\sum_{j = i+1}^Q 
			\left(
				x_t^i - x_t^j
			\right)^2
\label{eq:equal-rate-potential-function}
\end{IEEEeqnarray}
We distinguish two extreme cases:
\textit{i)} all players have exactly the same rate
(i.e., $x_t^i = x_t^j$,
$i,j = 1,\ldots, Q$);
and 
\textit{ii)} each player's rate is different from any other player's rate
(i.e., $x_t^i \neq x_t^j$,
$i \neq j$).
When all players have exactly the same rate,
the terms $( x_t^i - x_t^j)^2$ vanish for all $(i,j)$ pairs,
and \eqref{eq:equal-rate-potential-function} only depends on the actions 
(the state becomes irrelevant).
Since the action constraint set is compact, 
existence of solution is guaranteed by Lemma \ref{lemma:existence}.1.
When each player's rate is different from any other player's rate,
the term $-( x_t^i - x_t^j)^2 $ is coercive,
so that \eqref{eq:equal-rate-potential-function} becomes coercive too
(since the constraint action set is compact,
the term depending on $u_t^i$ is bounded).
Thus, 
existence of optimal solution is guaranteed by 
Lemma \ref{lemma:existence}.2.
Finally,
the case where some player's rate are equal and some are different is a combination of the two cases already mentioned.
so that the equal terms vanish 
and the different terms make \eqref{eq:equal-rate-potential-function} coercive.
Hence,}
Theorem \ref{theorem:game-and-mocp-equivalence} states that 
we can find an NE of DPG \eqref{eq:equal-rate-scheduling-problem}
by solving the following MOCP:
\begin{IEEEeqnarray}{rCl}
\mc{P}_7:
\begin{aligned}
		\underset{ \{ \miu_t \} \in \prod_{t=0}^{\infty}\U }{\rm maximize} 	
				&\quad 	
				\sum_{t=0}^{\infty} 
					\beta^t 
					\Bigg(
						(1 - \alpha)
						\log
						\left(
							1
							+
							\sum_{i=1}^Q
								| h_t^i |^2 
								u_t^i
						\right)
\\
&\qquad
						-
						\alpha
						\sum_{i=1}^{Q-1}
							\sum_{j = i+1}^Q 
								\left(
									x_t^i - x_t^j
								\right)^2
					\Bigg)					
\\
		{\rm s.t.} 		
			&\quad
				x_{t+1}^i
				= 
				x_t^i + R_t^i
			,\quad 
					x_0^i = 0
\\
			&\quad
				0 \le u_t^{i} \le P^i_{\max}
	\end{aligned}
\label{eq:equal-rate-mocp-problem}
\end{IEEEeqnarray}

%
\subsection{Solving the MOCP with dynamic programming and simulation results}
\label{ssec:dynamic-programming}
%

Although Lemma \ref{lemma:existence} establishes 
existence of optimal solution to these MOCP,
these problems are nonconcave and cannot be approximated by finite horizon problems.
Thus, we cannot rely on efficient convex optimization software.
In order to numerically solve these problems,
we can use dynamic programming methods \cite{Bertsekas2007}.

Standard dynamic programming methods assume that the MOCP is stationary. 
One standard method to cope with nonstationary MOCP is to augment the state space 
so that it includes the time as an extra dimension
for some time length $T$.
Let the augmented state-vector at time $t$ be denoted by
$
\widetilde{\mix}_t 
= 
\left( \mix_t, t \right)
	\in 
		\widetilde{\X}
		\defeq
			\X
			\times
			\{0,\ldots,T\}
$.
The state-transition equation in the augmented state space becomes
$\widetilde{f}: \widetilde{\X} \times \U \rightarrow \widetilde{\X}$.
Since we are tackling an infinite horizon problem, 
when augmenting the state space with the time dimension, 
it is convenient to impose a periodic time variation:
\begin{IEEEeqnarray}{rCl}
\widetilde{f}(\widetilde{\mix}_t, \miu_t)
&
\defeq
&
	\begin{bmatrix}
		f (\mix_t, \miu_t, t) \\
		t+1 \;\: ({\rm if }\: t < T) \;\:{\rm or }\;\:  0 \;({\rm if }\: t = T)
	\end{bmatrix}
\label{eq:augmented-state-transition-function}
\end{IEEEeqnarray}
Otherwise, it could be difficult to apply computational dynamic programming methods.

One further difficulty for solving MOCP with continuous state and action spaces is that dynamic programming methods are mainly derived for discrete state-action spaces.
Two common approaches to overcome this limitation are 
\textit{i)} to use a parametric approximation of the value function 
(e.g., consider a neural network with inputs the continuous state action variables
that is trained by minimizing the error in the Bellman equation);
or \textit{ii)} to discretize the continuous spaces, so the value function is approximated in a set of points.
For simplicity, 
we follow the discretization approach here.
We remark that 
it may be problematic to finely discretize the state-action spaces in high-dimensional problems though,
since the computational load increases exponentially with the number of states.
These and other approximation techniques,
usually known as approximate dynamic programming, 
are still an active area of research  
(see, e.g., \cite[Ch. 6]{Bertsekas2007}, 
\cite{BertsekasNP1996}).

Introduce the optimal value function for the augmented set:
\begin{IEEEeqnarray}{rCl}
V^{\star}(\widetilde{\mix}_0)
&
\defeq
&
	\max_{ \{ \miu_t \} \in \prod_{t=0}^{\infty} \U }
	\sum_{t=0}^{\infty}
		\beta^t
		\Pi(\widetilde{\mix}_t, \miu_t, t)
\notag 
\\
&
\textblue{=}
&
\textblue{
	\sum_{t=0}^{\infty}
		\beta^t
		\Pi(\widetilde{\mix}_t, \phi^{\star}( \mix_t, t), t)
}
\notag 
\\
&
=
&
	\sum_{t=0}^{\infty}
		\beta^t
		\Pi(\widetilde{\mix}_t, \miu_t^{\star}, t)
\label{eq:optimal-value-function}
\end{IEEEeqnarray}
\textblue{where
$\phi^{\star} : \widetilde{\X} \rightarrow \U$
is the optimal policy that provides the sequence of actions
$
\{ 
	\miu_t^{\star}
	\defeq
		\phi^{\star}( \widetilde{\mix}_t)
\}_{t=0}^{\infty}$
that is the solution to the MOCP,
as explained by Lemma \ref{lemma:existence}.}
Then,
the Bellman optimality equation is given by
\begin{IEEEeqnarray}{rCl}
V^{\star}(\widetilde{\mix}_t)
&=&
	\Pi(\widetilde{\mix}_t, \miu_t^{\star})
	+
	\beta 
	V^{\star}( \widetilde{f}(\widetilde{\mix}_t, \miu_t^{\star}) )
\label{eq:nonstationary-bellman-equation}
\end{IEEEeqnarray}

Among the available dynamic programming methods, 
we choose \textit{value iteration} (VI)
for its reduced complexity per iteration with respect to \textit{policy iteration} (PI),
which is especially relevant when the state-grid has fine resolution
(i.e., large number of states).
\textblue{VI is obtained by turning the Bellman optimality equation 
\eqref{eq:nonstationary-bellman-equation} into an update rule,
so that it generates a sequence of value functions $V^{k}$ that converge to the optimal value
(i.e., $\lim_{k\rightarrow \infty} V^k = V^{\star}$,
where $V_0$ is arbitrary).
In particular, 
at every iteration $k$, we obtain the policy $\phi$ that maximizes $V^k$
(policy improvement).
Then, we update the value function $V^{k+1}$ for the latest policy
(policy evaluation).}
VI is summarized in Algorithm \ref{alg:value-iteration},
where the operator $\lceil \widetilde{\mix} \rceil$ denotes the closest point to $\widetilde{\mix}$ in the discrete grid.
\begin{algorithm}
\caption{Value Iteration for the non-stationary MOCP}
\label{alg:value-iteration}
\vspace{2pt}
\textbf{Inputs}: number of states $S$, threshold $\epsilon$ \\
\textbf{Discretize} the augmented space $\widetilde{\X}$ into a grid of $S$ states \\
\textbf{Initialize} $\Delta=\infty$, $k=0$ and $V_0(\widetilde{ \mix}_{s})=0$ for $s = 1 \ldots S$\\
\textbf{while} $\Delta > \epsilon$ \\
\hspace{1em} \textbf{for} every state $s=1$ to $S$ \textbf{do} \\
\hspace{3em}$\widetilde{ \mix}_{s} \leftarrow$ the $s$-th point on the grid \\
\hspace{3em}$ 
	\phi (\widetilde{ \mix}_{s}) 
		= 
			\arg\max_{\miu} 
			\: 
			\Pi (\widetilde{ \mix}_{s},\miu) 
			+ 
			\beta V_k (\lceil \widetilde{f} \left(\widetilde{ \mix}_{s}, \miu\right)\rceil) 
$\\		
\vspace{.2em}
\hspace{3em}$
	V_{k+1} (\widetilde{ \mix}_{s}) 
		= 
			\Pi (\widetilde{ \mix}_{s}, \phi(\widetilde{ \mix}_{s})) 
			+ 
			\beta V_k (\lceil \widetilde{f} \left(\widetilde{ \mix}_{s}, \phi(\widetilde{ \mix}_{s})\rceil\right))
$ \\
\hspace{1.5em}\textbf{end for} \\
\hspace{1.5em}$k=k+1$ \\
\hspace{1.5em}$\Delta = \max_s |V_{k+1}(\widetilde{ \mix}_{s})) - V_{k}(\widetilde{ \mix}_{s}))|$ \\
\hspace{0em}\textbf{end while} \\
\textbf{Return}: $
	\phi(\widetilde{ \mix}_{s}) 
	\;\:
	{\rm and}
	\;\: 
	V_{k+1}(\widetilde{ \mix}_{s}) 
	\;\: 
	{\rm for} 
	\;\:
	s=1, \ldots, S$ \\
\vspace{2pt}
\label{algorithm}
\end{algorithm}

\textblue{Note that the output of the value iteration algorithm is a policy (i.e., a function), rather than a sequence of actions. 
This result allows to compute the optimal actions of every player from the current state at every time-step of the game.
When there is no reason to propose a finite-horizon approximation of the game, a policy is a more practical representation of the solution than an infinite sequence of actions.}


We simulate a simple scenario with $Q=2$ users. 
The channel coefficients are sinusoids with different frequency and different amplitude for each user (see Fig. \ref{fig:channel-coefficients}).  
The maximum transmitter power is 
$P_{\max}^{1}= P_{\max}^{2} = 5$,
with $20$ possible power levels per user, 
which amounts to $400$ possible actions. 
We discretize the state-space (i.e., the users' rates) 
into a grid of $30$ points per user. 
The nonstationarity of the environment 
is surmounted by augmenting the state-space with $T=20$ time steps.
Hence, the augmented state space has a total of $30^{2}\times20=18.000$ states.
For the equal-rate problem, the utility function uses $\alpha=0.9$.

The solution of the proportional fair game leads to an efficient scheduler 
(see Figure \ref{fig:proportional-fair}), 
in which both users try to minimize interference so that they approach their respective maximum rates.

For the equal rate problem, 
we observe that the agents achieve much lower rate, 
but very similar between them
(see Figure \ref{fig:equal-rate}).
The trend is that the user with a channel with less gain 
(User 2, red-dashed line) 
tries to achieve its maximum rate, while the user with higher gain channel 
(User 1, blue-continuous line) 
reduces its transmitter power to match the rate of the other user. 
In other words, the user with poorest channel sets a bottleneck for the other user.

%
%
%

\begin{figure}[ht] 
\centering
\input{scheduling_channel_coefficients.tikz} 
\vspace{-0.3cm}
\caption{Periodic time variation of the channel coefficients 
$|h_{t}^{i}|^{2}$
for $Q=2$ users.
All possible combination of coefficients are included in a window of $T=20$ time steps.}
\label{fig:channel-coefficients}
\end{figure}
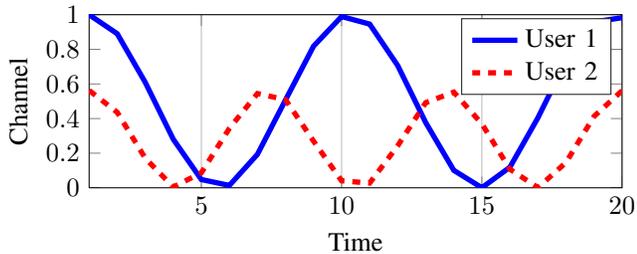
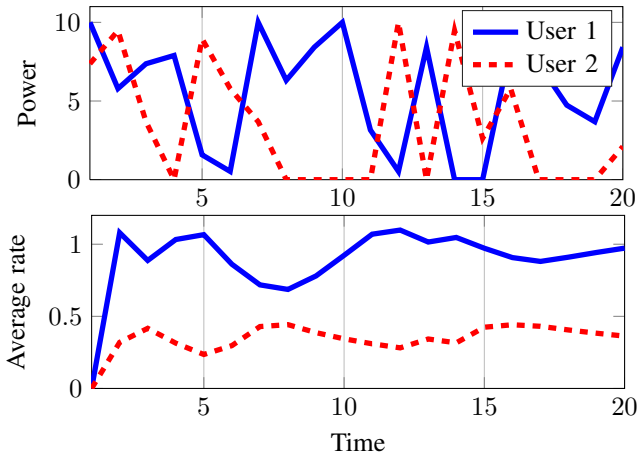
\begin{figure}[h!t]
\centering
\input{scheduling_pf_power.tikz}
\\
\input{scheduling_pf_rate.tikz}
\vspace{-0.3cm}
\caption{Proportional fair scheduling problem for $Q=2$ users.
(Top) Transmitter power $u_t^i$.
(Bottom) Average rate $x_t^i$ given by \eqref{eq:proportional-fair-scheduling-problem}.
Both users achieve near maximum average rates for their channel coefficients $|h_{t}^{i}|^{2}$.}
\label{fig:proportional-fair}
\end{figure}
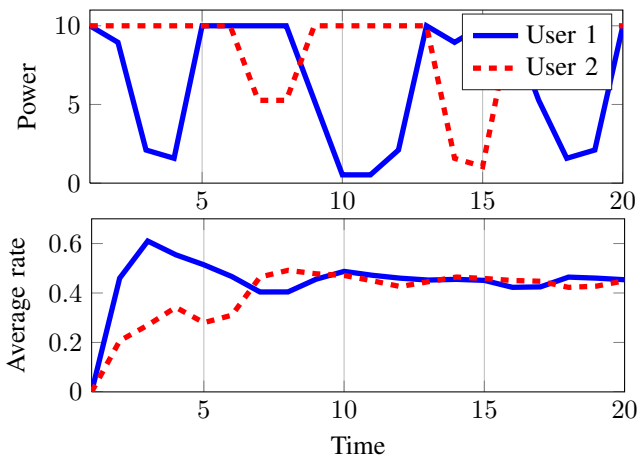
\begin{figure}[h!t]
\centering
\input{scheduling_er_power.tikz}
\\
\input{scheduling_er_rate.tikz}
\vspace{-0.3cm}
\caption{Equal rate problem for $Q=2$ users.
(Top) Transmitter power $u_t^i$.
(Bottom) Average rate $x_t^i/t$ (recall that $x_t^i$ given by \eqref{eq:equal-rate-scheduling-problem} denotes accumulated rate).
User 1 reduces its average rate to match that of User 2,
regardless of having higher channel coefficient.}
\label{fig:equal-rate}
\end{figure}

\textblue{
	Finally, note that Algorithm \ref{alg:value-iteration} is centralized,
	such that the results displayed in Figures \ref{fig:proportional-fair}--\ref{fig:equal-rate}
	have been obtained assuming the existence of a central unit that knows the channel coefficients, transmission power and average rate for all users,
	so that it can update the value and policy functions for all states.
	We remark that the design and analysis of distributed dynamic programming algorithms
	when multiple players share state-vector components 
	and/or have coupled constraints
	is a nontrivial task.
	Nevertheless,
	when the players share no state-vector components and they have uncoupled constraints,
	there are distributed implementations of VI and PI
	that converge to the optimal solution 
	\cite{BertsekasDistributed1982, bertsekas1997parallel, Jalali1992Distributed, bertsekas2010distributed}.
	This is indeed the case
	for problems \eqref{eq:proportional-fair-mocp-problem}
	and \eqref{eq:equal-rate-mocp-problem},
	where each agent $i$ has a unique state-vector component $x^i_t$
	and the constraints are uncoupled.
	Therefore, the agents could solve these problems in a decentralized manner.}

%
\section{Conclusions}
\label{sec:conclusions}
%

DPG provide a useful framework for competitive multiagent applications under time-varying environments.
On one hand, DPG allows nonstationary scenarios, 
thus, more realistic models.
On the other hand, the analysis and solution of DPG is affordable 
through an equivalent MOCP.
We presented a complete description of DPG 
and provided conditions for a dynamic game with constrained state and action sets to be of the potential type.
To the best of our knowledge,
previous works have not dealt with DPG with constraints explicitly.

We also introduced a range of communication and networking examples:
energy demand control in a smart-grid network, 
network flow with relays that have bounded link capacity and limited battery life, 
multiple access communication in which users have to optimize the use of their batteries, 
and two optimal scheduling games with nonstationary channels.
Although these problems have different features 
each---including utilities in separable and nonseparable form,
convex and nonconvex objectives,
closed-form and numerical solutions,
and solution methods based on convex optimization and dynamic programming 
algorithms---the proposed framework allowed us to analyze and solve them in a unified manner.

\textblue{
	The DPG framework is promising in the sense that,
	once the equivalent MOCP has been formulated,
	it is possible to use ideas from optimal control theory to extend the current analysis.
	In particular,
	we have assumed that the agents can observe all the variables 
	that influence their objective functions	and constraints.
	This is known as \textit{perfect information}.
	Although \textit{perfect information} is reasonable in many applications, 
	there are problems in which all the information is not available to all agents.
	An example of games with \textit{imperfect information} 
	is when the agents cannot directly observe 
	the variables that influence their objective and constraints;
	rather, they only have access to another variable that depends on the state. 
	The current framework could possibly be extended to this case
	by using a partially-observable-Markov-decision-process (POMDP) formulation
	\cite{porta2006point,brechtel2013solving}.
	Nevertheless, other forms of imperfect information---like 
	when the agents cannot see other players' actions---would require further study.
	Another possible direction to extend the current analysis
	is to allow \textit{stochastic} state transitions and utilities 
	(i.e., considering $\mix_{t+1}$ and $\pi^i$ random variables).
	This can be done by applying the Euler equation to the stochastic Lagrangian
	in order to derive a set of \textit{stochastic} optimality conditions.
	Finally, we could also consider the case where the agents know
	nothing about the problem; 
	rather, they have to learn the optimal policy from trial-and-error experimentation.
	To this end, 
	we could apply reinforcement learning (RL) and approximate dynamic programming (APD) techniques 
	(such as Q-learning)
	\cite{SuttonBarto1998, Szepesvari2009, busoniu2010reinforcement},
	\cite[Ch. 6]{Bertsekas2007}.
	The main difficulty with standard APD/RL techniques 
	is that they have been developed 
	for unconstrained MOCP,
	and some adaptation of these techniques is necessary.
	}

\bibliographystyle{IEEEtran}
\bibliography{refs_DPG}

\end{document}

%% file: lq_utility.tikz
%
%
\definecolor{mycolor1}{rgb}{0.92900,0.69400,0.12500}%
\definecolor{mycolor2}{rgb}{0.49400,0.18400,0.55600}%
\definecolor{mycolor3}{rgb}{0.00000,0.49804,0.00000}%
\definecolor{mycolor4}{rgb}{0.29804,0.74118,0.92941}%
\definecolor{mycolor5}{rgb}{0.63500,0.07800,0.18400}%
\begin{tikzpicture}

\begin{axis}[%
width=0.8\columnwidth,
height=0.5\columnwidth,
at={(1.625in,0.793in)},
scale only axis,
xmin=0,
xmax=30,
xlabel={Time},
xmajorgrids,
ymin=-20,
ymax=1,
ylabel={Utilities},
axis background/.style={fill=white},
legend style={at={(0.7,0.1)},anchor=south west,legend cell align=left,align=left,draw=white!15!black}
]
\addplot [color=blue,solid,line width=2.0pt]
  table[row sep=crcr]{%
0	-0.720521820601445\\
1	-15.2797172788371\\
2	-7.74276306023061\\
3	-4.7127192062984\\
4	-2.7581005723456\\
5	-1.6535187744097\\
6	-1.00085422635572\\
7	-0.609068649614383\\
8	-0.370893479630659\\
9	-0.225657099705154\\
10	-0.137127365908888\\
11	-0.0832462675740724\\
12	-0.0505009968557089\\
13	-0.0306224139277115\\
14	-0.0185635242817272\\
15	-0.0112515582369391\\
16	-0.00681908819230494\\
17	-0.0041325575411378\\
18	-0.00250438092307004\\
19	-0.0015176658628661\\
20	-0.000919706237505133\\
21	-0.000557340721238072\\
22	-0.000337747262154182\\
23	-0.000204673997028394\\
24	-0.000124031910754301\\
25	-7.51630147810351e-05\\
26	-4.55485927201056e-05\\
27	-2.76023306644709e-05\\
28	-1.67269423797068e-05\\
29	-1.0136484932989e-05\\
30	-6.14268443942893e-06\\
31	-3.72245140365837e-06\\
32	-2.25579625492997e-06\\
33	-1.36700690330064e-06\\
34	-8.28402775448691e-07\\
35	-5.02010017437409e-07\\
36	-3.04216819737932e-07\\
37	-1.84354634841358e-07\\
38	-1.11718449441604e-07\\
39	-6.77011020515474e-08\\
40	-4.10266991903362e-08\\
41	-2.48620775067543e-08\\
42	-1.50663570347482e-08\\
43	-9.13017483103757e-09\\
44	-5.53286320330487e-09\\
45	-3.3529013182112e-09\\
46	-2.03184984637665e-09\\
47	-1.23129594533416e-09\\
48	-7.46162275573382e-10\\
49	-4.52172480221729e-10\\
};
\addlegendentry{User 1};

\addplot [color=red,solid,line width=2.0pt]
  table[row sep=crcr]{%
0	-11.5822908908432\\
1	-19.1378868867058\\
2	-10.6170106338674\\
3	-6.38697673014112\\
4	-3.77636483131266\\
5	-2.26419714890319\\
6	-1.36887516803651\\
7	-0.831074702265587\\
8	-0.505050006510975\\
9	-0.30680973561313\\
10	-0.186253059869729\\
11	-0.112997636939067\\
12	-0.0685238275931944\\
13	-0.0415420431962156\\
14	-0.0251800845433925\\
15	-0.0152609515411597\\
16	-0.00924870078155504\\
17	-0.00560487594561786\\
18	-0.0033965963190211\\
19	-0.00205834461412082\\
20	-0.00124735584885039\\
21	-0.000755895503917584\\
22	-0.00045807096386656\\
23	-0.000277589866740819\\
24	-0.00016821875054385\\
25	-0.000101940131840416\\
26	-6.17754599362505e-05\\
27	-3.74357721772525e-05\\
28	-2.26859835968919e-05\\
29	-1.37476491245418e-05\\
30	-8.33104096490731e-06\\
31	-5.04859001412442e-06\\
32	-3.0594329518733e-06\\
33	-1.85400874055494e-06\\
34	-1.12352467592697e-06\\
35	-6.80853154033345e-07\\
36	-4.12595315023478e-07\\
37	-2.50031733009144e-07\\
38	-1.51518607324335e-07\\
39	-9.18198985822992e-08\\
40	-5.56426298054104e-08\\
41	-3.37192950495485e-08\\
42	-2.04338088013343e-08\\
43	-1.23828372306466e-08\\
44	-7.50396851471054e-09\\
45	-4.54738622667225e-09\\
46	-2.75570472530455e-09\\
47	-1.66995019875865e-09\\
48	-1.01198566041044e-09\\
49	-6.13260789236366e-10\\
};
\addlegendentry{User 2};

\addplot [color=mycolor1,solid,line width=2.0pt]
  table[row sep=crcr]{%
0	-0.194930829610491\\
1	-14.8029696702997\\
2	-7.59361766540287\\
3	-4.65877119571697\\
4	-2.74041998167153\\
5	-1.64798433301456\\
6	-0.999318211150517\\
7	-0.608772568752567\\
8	-0.370934457812317\\
9	-0.225757926390118\\
10	-0.13721430092751\\
11	-0.083307585236795\\
12	-0.0505409812377828\\
13	-0.0306475469359677\\
14	-0.0185790348291924\\
15	-0.0112610414254219\\
16	-0.00682485894528561\\
17	-0.00413606101554219\\
18	-0.0025065055594763\\
19	-0.00151895368122873\\
20	-0.000920486672156431\\
21	-0.000557813642266031\\
22	-0.000338033835158582\\
23	-0.000204847651219227\\
24	-0.000124137141015394\\
25	-7.52267825074422e-05\\
26	-4.55872351965296e-05\\
27	-2.76257476807048e-05\\
28	-1.67411329501778e-05\\
29	-1.01450843539839e-05\\
30	-6.14789565701617e-06\\
31	-3.72560938513786e-06\\
32	-2.25770998272793e-06\\
33	-1.36816661722496e-06\\
34	-8.29105559088657e-07\\
35	-5.02435902540084e-07\\
36	-3.04474905040061e-07\\
37	-1.8451103388912e-07\\
38	-1.11813226874381e-07\\
39	-6.77585369396108e-08\\
40	-4.10615045894245e-08\\
41	-2.48831694919081e-08\\
42	-1.50791387252159e-08\\
43	-9.13792050362392e-09\\
44	-5.53755706159841e-09\\
45	-3.35574578464479e-09\\
46	-2.03357358594976e-09\\
47	-1.23234052722152e-09\\
48	-7.46795289594984e-10\\
49	-4.52556085141921e-10\\
};
\addlegendentry{User 3};

\addplot [color=mycolor2,solid,line width=2.0pt]
  table[row sep=crcr]{%
0	-2.41383068491104\\
1	-15.6403046666322\\
2	-8.03460120068366\\
3	-4.8561102735323\\
4	-2.8349032596999\\
5	-1.69569675925562\\
6	-1.02516240881074\\
7	-0.623425516905186\\
8	-0.379493558491084\\
9	-0.23084156654539\\
10	-0.14026205967186\\
11	-0.0851440570129973\\
12	-0.0516506050407245\\
13	-0.0313189733813689\\
14	-0.0189856199527089\\
15	-0.0115073464508706\\
16	-0.00697409668746869\\
17	-0.00422649335769157\\
18	-0.00256130632819921\\
19	-0.00155216275267881\\
20	-0.000940611363408326\\
21	-0.000570009212994581\\
22	-0.000345424350625419\\
23	-0.000209326298593202\\
24	-0.000126851195699378\\
25	-7.68714950471016e-05\\
26	-4.65839277101085e-05\\
27	-2.82297411559844e-05\\
28	-1.71071515957394e-05\\
29	-1.03668907772377e-05\\
30	-6.28230980898525e-06\\
31	-3.8070640303203e-06\\
32	-2.30707129604522e-06\\
33	-1.39807945051547e-06\\
34	-8.47232661660497e-07\\
35	-5.13420881574079e-07\\
36	-3.11131774978233e-07\\
37	-1.88545080490564e-07\\
38	-1.14257849065914e-07\\
39	-6.9239971897844e-08\\
40	-4.19592504836505e-08\\
41	-2.54272012673951e-08\\
42	-1.54088206259026e-08\\
43	-9.33770691415894e-09\\
44	-5.65862712868944e-09\\
45	-3.42911394370163e-09\\
46	-2.07803450756936e-09\\
47	-1.25928373496562e-09\\
48	-7.63122806369239e-10\\
49	-4.62450519633386e-10\\
};
\addlegendentry{User 4};

\addplot [color=mycolor3,solid,line width=2.0pt]
  table[row sep=crcr]{%
0	-4.58418319080263\\
1	-16.3789474329251\\
2	-8.79002162717194\\
3	-5.42500664715752\\
4	-3.24563125034617\\
5	-1.97055071099818\\
6	-1.20215750005608\\
7	-0.73456373708382\\
8	-0.448263522590441\\
9	-0.273020203223352\\
10	-0.165997603574349\\
11	-0.100799310924032\\
12	-0.06115743189155\\
13	-0.0370864910148491\\
14	-0.0224827376286741\\
15	-0.0136272053317364\\
16	-0.00825890550229228\\
17	-0.00500513396881157\\
18	-0.0030331732146145\\
19	-0.00183811564929494\\
20	-0.00111389849058777\\
21	-0.000675020722930983\\
22	-0.000409060981473233\\
23	-0.000247889889692062\\
24	-0.000150220612750346\\
25	-9.10332937011638e-05\\
26	-5.51659384665772e-05\\
27	-3.34304167852911e-05\\
28	-2.02587475143662e-05\\
29	-1.22767498534499e-05\\
30	-7.43967958198075e-06\\
31	-4.50842720600468e-06\\
32	-2.7320956255156e-06\\
33	-1.65564313286941e-06\\
34	-1.00331561075712e-06\\
35	-6.08006759667434e-07\\
36	-3.68450581409439e-07\\
37	-2.23280134381315e-07\\
38	-1.35307205190037e-07\\
39	-8.19958292717367e-08\\
40	-4.96892682758171e-08\\
41	-3.01115727931244e-08\\
42	-1.82475381012252e-08\\
43	-1.10579626326545e-08\\
44	-6.70109780875284e-09\\
45	-4.06084857891003e-09\\
46	-2.46086412278365e-09\\
47	-1.49127752811358e-09\\
48	-9.03710467093919e-10\\
49	-5.4764629181263e-10\\
};
\addlegendentry{User 5};

\addplot [color=mycolor4,solid,line width=2.0pt]
  table[row sep=crcr]{%
0	-1.04714075932487\\
1	-14.9064599831749\\
2	-7.78947349068938\\
3	-4.74813069528462\\
4	-2.7873332974654\\
5	-1.67086432678961\\
6	-1.01101274650678\\
7	-0.614952242486838\\
8	-0.374336110554645\\
9	-0.227691028705774\\
10	-0.13833947106017\\
11	-0.0839730807498786\\
12	-0.0509386339378026\\
13	-0.030886625182189\\
14	-0.0187232919524672\\
15	-0.0113482613027808\\
16	-0.00687765201064828\\
17	-0.00416803492785072\\
18	-0.00252587636902449\\
19	-0.00153069092609949\\
20	-0.000927599073535659\\
21	-0.000562123675179452\\
22	-0.000340645700153246\\
23	-0.000206430439499274\\
24	-0.000125096310332726\\
25	-7.58080385803782e-05\\
26	-4.59394757617875e-05\\
27	-2.78392049301389e-05\\
28	-1.68704876463835e-05\\
29	-1.02234730283115e-05\\
30	-6.19539901681041e-06\\
31	-3.75439630935054e-06\\
32	-2.27515478933235e-06\\
33	-1.37873812732334e-06\\
34	-8.35511868119845e-07\\
35	-5.06318109932556e-07\\
36	-3.0682751307306e-07\\
37	-1.8593670850368e-07\\
38	-1.12677182141312e-07\\
39	-6.82820916796915e-08\\
40	-4.13787774578018e-08\\
41	-2.5075436059803e-08\\
42	-1.51956517864023e-08\\
43	-9.2085271284531e-09\\
44	-5.58034450035409e-09\\
45	-3.38167486594151e-09\\
46	-2.04928654462291e-09\\
47	-1.24186254103452e-09\\
48	-7.52565606245176e-10\\
49	-4.56052882657395e-10\\
};
\addlegendentry{User 6};

\addplot [color=mycolor5,solid,line width=2.0pt]
  table[row sep=crcr]{%
0	-0.121485189443746\\
1	-14.6965832360907\\
2	-7.53874440856525\\
3	-4.61729718671212\\
4	-2.71330236087364\\
5	-1.63036157653559\\
6	-0.988196721224098\\
7	-0.601851690630692\\
8	-0.366673253301353\\
9	-0.223150987488381\\
10	-0.135625808510131\\
11	-0.0823419657473445\\
12	-0.0499548138365011\\
13	-0.0302920027989605\\
14	-0.0183634717948261\\
15	-0.011130378732155\\
16	-0.00674566820827912\\
17	-0.00408806896688071\\
18	-0.00247742180551582\\
19	-0.00150132884105719\\
20	-0.000909806036841715\\
21	-0.000551341204534865\\
22	-0.000334111557455744\\
23	-0.000202470762927449\\
24	-0.00012269675372608\\
25	-7.43539118751643e-05\\
26	-4.50582779612569e-05\\
27	-2.73052009885348e-05\\
28	-1.65468824810242e-05\\
29	-1.00273690635904e-05\\
30	-6.07656048779082e-06\\
31	-3.68238045149102e-06\\
32	-2.23151335702584e-06\\
33	-1.352291527438e-06\\
34	-8.19485294265328e-07\\
35	-4.96606046034809e-07\\
36	-3.00942026524361e-07\\
37	-1.82370118310424e-07\\
38	-1.10515837367216e-07\\
39	-6.69723221302711e-08\\
40	-4.05850603735242e-08\\
41	-2.45944454839486e-08\\
42	-1.49041726956355e-08\\
43	-9.03189152555771e-09\\
44	-5.47330376502018e-09\\
45	-3.31680844698052e-09\\
46	-2.00997765632252e-09\\
47	-1.21804145264756e-09\\
48	-7.38130086023785e-10\\
49	-4.47304993363788e-10\\
};
\addlegendentry{User 7};

\addplot [color=blue,solid,line width=2.0pt]
  table[row sep=crcr]{%
0	-3.65174139053351\\
1	-15.5947440129442\\
2	-8.16810004237649\\
3	-4.88574999112413\\
4	-2.84116881167666\\
5	-1.69206402924865\\
6	-1.02002797829947\\
7	-0.619103002501837\\
8	-0.376410095628399\\
9	-0.228799273313173\\
10	-0.138961580187553\\
11	-0.084333798542687\\
12	-0.0511519726595824\\
13	-0.0310142501620094\\
14	-0.0188001246257517\\
15	-0.0113946719221717\\
16	-0.00690573473594068\\
17	-0.00418504207693086\\
18	-0.00253618017724534\\
19	-0.00153693460035979\\
20	-0.000931382736634092\\
21	-0.000564416620498761\\
22	-0.000342035254669169\\
23	-0.000207272523436816\\
24	-0.000125606618659737\\
25	-7.61172872199707e-05\\
26	-4.6126880849985e-05\\
27	-2.79527723864924e-05\\
28	-1.69393094004718e-05\\
29	-1.02651788953843e-05\\
30	-6.22067268733456e-06\\
31	-3.76971209704741e-06\\
32	-2.28443612057331e-06\\
33	-1.38436259158081e-06\\
34	-8.38920279645783e-07\\
35	-5.08383598884215e-07\\
36	-3.08079194248365e-07\\
37	-1.86695224181895e-07\\
38	-1.13136840753646e-07\\
39	-6.8560643653829e-08\\
40	-4.15475792602066e-08\\
41	-2.51777295315294e-08\\
42	-1.52576413754282e-08\\
43	-9.24609266497226e-09\\
44	-5.60310912190149e-09\\
45	-3.39547016988868e-09\\
46	-2.05764646444872e-09\\
47	-1.24692863162327e-09\\
48	-7.55635644521678e-10\\
49	-4.57913318205115e-10\\
};
\addlegendentry{User 8};

\end{axis}
\end{tikzpicture}%

%% file: lq_coefficients.tikz
%
%
\definecolor{mycolor1}{rgb}{0.00000,0.44700,0.74100}%
\definecolor{mycolor2}{rgb}{0.85000,0.32500,0.09800}%
\definecolor{mycolor3}{rgb}{0.92900,0.69400,0.12500}%
\definecolor{mycolor4}{rgb}{0.49400,0.18400,0.55600}%
\definecolor{mycolor5}{rgb}{0.46600,0.67400,0.18800}%
\definecolor{mycolor6}{rgb}{0.30100,0.74500,0.93300}%
\definecolor{mycolor7}{rgb}{0.63500,0.07800,0.18400}%
\begin{tikzpicture}

\begin{axis}[%
width=0.8\columnwidth,
height=0.4\columnwidth,
at={(1.625in,0.793in)},
scale only axis,
xmin=0,
xmax=30,
xlabel={Time},
xmajorgrids,
ymin=-3,
ymax=3,
ylabel={Decision Coefficients},
axis background/.style={fill=white}
]
\addplot [color=mycolor1,solid,line width=2.0pt,forget plot]
  table[row sep=crcr]{%
0	0.0652562140248817\\
1	0.0829327092619945\\
2	0.0419787298483572\\
3	0.0240553550245546\\
4	0.0116743532619231\\
5	0.00534078000641024\\
6	0.00205152933199577\\
7	0.000503645305444136\\
8	-0.000173556223901737\\
9	-0.000423971499637918\\
10	-0.000479866829183591\\
11	-0.000453986290056899\\
12	-0.000399330253625456\\
13	-0.0003392825156037\\
14	-0.000283277577753316\\
15	-0.000234420811565926\\
16	-0.000193119378817948\\
17	-0.000158742008114924\\
18	-0.000130346995739078\\
19	-0.000106980859334536\\
20	-8.77866364804472e-05\\
21	-7.20316171674566e-05\\
22	-5.91035190570454e-05\\
23	-4.84961295035505e-05\\
24	-3.97929547033787e-05\\
25	-3.26520243972444e-05\\
26	-2.67927737445058e-05\\
27	-2.198506501701e-05\\
28	-1.80401219704458e-05\\
29	-1.48030841235133e-05\\
30	-1.21469029581529e-05\\
31	-9.96734012597214e-06\\
32	-8.17886804328874e-06\\
33	-6.7113090683555e-06\\
34	-5.50707941982175e-06\\
35	-4.51892853188887e-06\\
36	-3.70808450581622e-06\\
37	-3.04273255677126e-06\\
38	-2.49676658690479e-06\\
39	-2.0487648184383e-06\\
40	-1.68114925513925e-06\\
41	-1.37949597435003e-06\\
42	-1.13196917972581e-06\\
43	-9.28856805344788e-07\\
44	-7.62189448423233e-07\\
45	-6.25427678284533e-07\\
46	-5.13205452453718e-07\\
47	-4.21119572348447e-07\\
48	-3.45556917528636e-07\\
49	-2.83552679785023e-07\\
};
\addplot [color=mycolor2,solid,line width=2.0pt,forget plot]
  table[row sep=crcr]{%
0	0.00553070392732058\\
1	-0.0347821512182301\\
2	0.0110431208642914\\
3	0.0193421442222254\\
4	0.0240685352725116\\
5	0.0234761833511864\\
6	0.0212421012029047\\
7	0.0183689412022249\\
8	0.0155216067471516\\
9	0.0129423426222961\\
10	0.0107127813854001\\
11	0.00883109743368583\\
12	0.00726375386681696\\
13	0.0059674826983185\\
14	0.0048995107855936\\
15	0.00402141430308353\\
16	0.00330019246278089\\
17	0.00270813023416027\\
18	0.00222221989593911\\
19	0.00182347504735475\\
20	0.00149627505606793\\
21	0.00122778750566234\\
22	0.00100747790871481\\
23	0.000826700899047459\\
24	0.000678362340086549\\
25	0.000556641220444217\\
26	0.000456761201194055\\
27	0.000374803106293562\\
28	0.000307551066696465\\
29	0.000252366289067092\\
30	0.000207083488914827\\
31	0.000169925916308664\\
32	0.00013943563415238\\
33	0.000114416309749585\\
34	9.3886272955108e-05\\
35	7.70399979919238e-05\\
36	6.32164970610612e-05\\
37	5.18733853478752e-05\\
38	4.256559969638e-05\\
39	3.49279358869121e-05\\
40	2.86607193142849e-05\\
41	2.35180468226386e-05\\
42	1.9298138343262e-05\\
43	1.5835419765919e-05\\
44	1.29940264029661e-05\\
45	1.06624721450849e-05\\
46	8.74927514525108e-06\\
47	7.1793683983861e-06\\
48	5.89115438067285e-06\\
49	4.8340881831176e-06\\
};
\addplot [color=mycolor3,solid,line width=2.0pt,forget plot]
  table[row sep=crcr]{%
0	-0.94425601424445\\
1	-1.04202079809722\\
2	-0.583370916237417\\
3	-0.393823576683722\\
4	-0.257097864659653\\
5	-0.178846715326878\\
6	-0.129455148283338\\
7	-0.0977600357206991\\
8	-0.0760935836805142\\
9	-0.0605048182208323\\
10	-0.048758955004296\\
11	-0.039612481784564\\
12	-0.0323315335890573\\
13	-0.0264570145694051\\
14	-0.0216798960081246\\
15	-0.0177781617935346\\
16	-0.0145839071181663\\
17	-0.0119656601889313\\
18	-0.0098182444530158\\
19	-0.00805647968313943\\
20	-0.00661091882073364\\
21	-0.00542474612277206\\
22	-0.00445140016050931\\
23	-0.00365269241366727\\
24	-0.00299729061017433\\
25	-0.00245948427208429\\
26	-0.00201817516686896\\
27	-0.00165604986957151\\
28	-0.00135890095189077\\
29	-0.00111506991173545\\
30	-0.000914989971524319\\
31	-0.000750810846482953\\
32	-0.000616090799632792\\
33	-0.000505543928091955\\
34	-0.000414832781522521\\
35	-0.000340398184946315\\
36	-0.000279319593604526\\
37	-0.000229200503129656\\
38	-0.000188074420135216\\
39	-0.000154327704418512\\
40	-0.000126636255647156\\
41	-0.000103913560462264\\
42	-8.52680616059956e-05\\
43	-6.99681764134539e-05\\
44	-5.74135921292975e-05\\
45	-4.71117117838054e-05\\
46	-3.86583264503012e-05\\
47	-3.17217555329815e-05\\
48	-2.6029832806945e-05\\
49	-2.13592275892308e-05\\
};
\addplot [color=mycolor4,solid,line width=2.0pt,forget plot]
  table[row sep=crcr]{%
0	0.405072859322644\\
1	0.423223292691437\\
2	0.24757887830292\\
3	0.17205564113402\\
4	0.117296607142208\\
5	0.0847090213149585\\
6	0.063322965778255\\
7	0.0489460091003576\\
8	0.0386997696262239\\
9	0.0310714531402018\\
10	0.0251827383188865\\
11	0.0205244787819987\\
12	0.0167810990656636\\
13	0.0137445106615685\\
14	0.0112679387911977\\
15	0.00924208854400688\\
16	0.00758229913883049\\
17	0.00622131188578497\\
18	0.00510488137961437\\
19	0.004188886526849\\
20	0.00343727823261656\\
21	0.00282053415884065\\
22	0.00231444944670862\\
23	0.00189916840156678\\
24	0.00155839929118091\\
25	0.00127877347403684\\
26	0.00104932066522523\\
27	0.000861038669557485\\
28	0.000706540361116183\\
29	0.00057976397886179\\
30	0.000475735370946462\\
31	0.000390372875365347\\
32	0.00032032719602498\\
33	0.000262850001574556\\
34	0.000215686097253216\\
35	0.000176984942254605\\
36	0.000145228042594169\\
37	0.00011916937150725\\
38	9.77864801142888e-05\\
39	8.0240380313517e-05\\
40	6.58426259435154e-05\\
41	5.40283006407653e-05\\
42	4.43338525511129e-05\\
43	3.63789062165425e-05\\
44	2.98513380040961e-05\\
45	2.44950294914891e-05\\
46	2.00998182965889e-05\\
47	1.64932520574337e-05\\
48	1.35338220185223e-05\\
49	1.11054107335212e-05\\
};
\addplot [color=mycolor5,solid,line width=2.0pt,forget plot]
  table[row sep=crcr]{%
0	0.22462826219916\\
1	0.274066149188949\\
2	0.127817122747219\\
3	0.079062807167332\\
4	0.0446525656477694\\
5	0.0274282689608401\\
6	0.0176334203799741\\
7	0.0121598477444536\\
8	0.00887562080890242\\
9	0.00677797344082394\\
10	0.00533458854883503\\
11	0.0042781980972156\\
12	0.00346842847876651\\
13	0.00282881059816098\\
14	0.00231444671644651\\
15	0.00189664957541341\\
16	0.00155548236472124\\
17	0.00127613956437347\\
18	0.00104712053002892\\
19	0.00085924873797022\\
20	0.000705093760984788\\
21	0.000578593779296852\\
22	0.00047478573218677\\
23	0.00038959968410767\\
24	0.000319695994697726\\
25	0.000262333733172558\\
26	0.000215263296777508\\
27	0.000176638407782081\\
28	0.000144943876538497\\
29	0.000118936280632101\\
30	9.75952519979884e-05\\
31	8.00834815741844e-05\\
32	6.57138870602049e-05\\
33	5.39226646139487e-05\\
34	4.42471722101651e-05\\
35	3.63077795783771e-05\\
36	2.97929739504479e-05\\
37	2.44471378977766e-05\\
38	2.00605200286233e-05\\
39	1.64610051830183e-05\\
40	1.35073612860339e-05\\
41	1.10836979206562e-05\\
42	9.09491920766256e-06\\
43	7.46299258528679e-06\\
44	6.12388709115326e-06\\
45	5.02506101646961e-06\\
46	4.12340035731706e-06\\
47	3.38352717534087e-06\\
48	2.77641149398847e-06\\
49	2.27823226607155e-06\\
};
\addplot [color=mycolor6,solid,line width=2.0pt,forget plot]
  table[row sep=crcr]{%
0	-0.132899700806061\\
1	-0.119821170746872\\
2	-0.0782165056115626\\
3	-0.0569759032884668\\
4	-0.0416264469528748\\
5	-0.0316320005008242\\
6	-0.0246187040643396\\
7	-0.0195483577295454\\
8	-0.0157248369405206\\
9	-0.0127557391943976\\
10	-0.0103994878867233\\
11	-0.00850347661425116\\
12	-0.0069646773894525\\
13	-0.00570952824617212\\
14	-0.00468284206079428\\
15	-0.00384173362495039\\
16	-0.00315209288979849\\
17	-0.00258640481509804\\
18	-0.00212229384871691\\
19	-0.00174148287115288\\
20	-0.00142900734240092\\
21	-0.00117260022950846\\
22	-0.000962199861445524\\
23	-0.000789551209499749\\
24	-0.0006478807034755\\
25	-0.000531630124408332\\
26	-0.000436238485683538\\
27	-0.000357963116086541\\
28	-0.000293732854100119\\
29	-0.000241027577387055\\
30	-0.000197779341060822\\
31	-0.000162291249512777\\
32	-0.000133170881922542\\
33	-0.000109275661513381\\
34	-8.96680263411368e-05\\
35	-7.35786433954995e-05\\
36	-6.03762230415353e-05\\
37	-4.95427496170384e-05\\
38	-4.06531564151933e-05\\
39	-3.33586476153868e-05\\
40	-2.73730127946745e-05\\
41	-2.24613970594475e-05\\
42	-1.8431086181465e-05\\
43	-1.51239451816778e-05\\
44	-1.24102136796363e-05\\
45	-1.01834145604746e-05\\
46	-8.35617619388477e-06\\
47	-6.85680428393843e-06\\
48	-5.62646884141822e-06\\
49	-4.61689590552084e-06\\
};
\addplot [color=mycolor7,solid,line width=2.0pt,forget plot]
  table[row sep=crcr]{%
0	-1.49557573577703\\
1	-0.985981214337011\\
2	-0.870735454383935\\
3	-0.68808091014878\\
4	-0.556046248379233\\
5	-0.448028310799407\\
6	-0.36325700203095\\
7	-0.295569217537022\\
8	-0.241242764487517\\
9	-0.197300964773634\\
10	-0.161579817468967\\
11	-0.132435466734151\\
12	-0.108602056921291\\
13	-0.0890836248080967\\
14	-0.0730851848566227\\
15	-0.0599653608972299\\
16	-0.0492031519673081\\
17	-0.0403735186206994\\
18	-0.0331288198845123\\
19	-0.027184297232586\\
20	-0.0223065057521345\\
21	-0.0183039800954354\\
22	-0.0150196473248032\\
23	-0.0123246329417544\\
24	-0.0101131919678881\\
25	-0.00829855522866028\\
26	-0.00680952321655993\\
27	-0.0055876719863065\\
28	-0.00458506070559596\\
29	-0.0037623506386896\\
30	-0.00308726166614944\\
31	-0.00253330575434469\\
32	-0.00207874767673071\\
33	-0.00170575221123026\\
34	-0.00139968435533876\\
35	-0.00114853510380273\\
36	-0.000942450259632273\\
37	-0.000773343791355174\\
38	-0.000634580566397767\\
39	-0.000520716012372397\\
40	-0.000427282491604345\\
41	-0.000350614007043627\\
42	-0.000287702361669766\\
43	-0.000236079127608794\\
44	-0.000193718793857214\\
45	-0.000158959292477978\\
46	-0.000130436785001542\\
47	-0.000107032150284024\\
48	-8.78270741975858e-05\\
49	-7.20680182696561e-05\\
};
\addplot [color=mycolor1,solid,line width=2.0pt,forget plot]
  table[row sep=crcr]{%
0	2.95653766981278\\
1	1.9448466872732\\
2	1.71068579555842\\
3	1.36069199091804\\
4	1.10627794198811\\
5	0.896265708395235\\
6	0.729578215264421\\
7	0.595262160319481\\
8	0.486710826043206\\
9	0.398494443969151\\
10	0.326561637374584\\
11	0.26776144205817\\
12	0.219621828308763\\
13	0.180171874281446\\
14	0.147824425489761\\
15	0.121291891877537\\
16	0.0995248534756996\\
17	0.0816655260625452\\
18	0.0670115642437689\\
19	0.054987324521682\\
20	0.0451207464454038\\
21	0.0370245960412304\\
22	0.03038117174881\\
23	0.0249297973652429\\
24	0.0204565775014028\\
25	0.016785998857205\\
26	0.0137740415409105\\
27	0.0113025275932018\\
28	0.00927448397352429\\
29	0.00761033770109291\\
30	0.00624479373855904\\
31	0.00512427308368335\\
32	0.00420481054399842\\
33	0.00345032971337411\\
34	0.00283122747023007\\
35	0.00232321246076262\\
36	0.001906351995038\\
37	0.00156428996048197\\
38	0.00128360506698041\\
39	0.00105328424367674\\
40	0.000864290525565527\\
41	0.000709208475358464\\
42	0.000581953228273432\\
43	0.000477531743718218\\
44	0.00039184689624481\\
45	0.000321536719006779\\
46	0.000263842492209219\\
47	0.00021650050081441\\
48	0.000177653214463007\\
49	0.000145776404628725\\
};
\addplot [color=mycolor2,solid,line width=2.0pt,forget plot]
  table[row sep=crcr]{%
0	1.52165381980036\\
1	0.994662286028017\\
2	0.866032831938692\\
3	0.696106492501547\\
4	0.571648140529934\\
5	0.467403342703672\\
6	0.383020641394656\\
7	0.313946321616726\\
8	0.257456971911532\\
9	0.211181064957436\\
10	0.173251377780875\\
11	0.142147278258133\\
12	0.116633748159803\\
13	0.0957024829161627\\
14	0.0785289020235488\\
15	0.0644376614525703\\
16	0.0528752047693148\\
17	0.0433875756060384\\
18	0.035602394039973\\
19	0.0292141496321499\\
20	0.0239721723774504\\
21	0.0196707794501787\\
22	0.0161411976108082\\
23	0.0132449382564177\\
24	0.0108683625487673\\
25	0.00891822226765106\\
26	0.00731800097974099\\
27	0.00600491178959207\\
28	0.00492743381230606\\
29	0.00404329068737\\
30	0.00331779181248864\\
31	0.00272247121287513\\
32	0.0022339706407754\\
33	0.00183312308276171\\
34	0.0015042007154054\\
35	0.0012342978021074\\
36	0.00101282431833808\\
37	0.000831090437043929\\
38	0.000681965570961416\\
39	0.000559598593901369\\
40	0.000459188263498911\\
41	0.000376794837644243\\
42	0.000309185493099327\\
43	0.000253707481081092\\
44	0.000208184042890529\\
45	0.000170829001689537\\
46	0.000140176679312502\\
47	0.000115024388299078\\
48	9.43852427412832e-05\\
49	7.74494364114133e-05\\
};
\addplot [color=mycolor3,solid,line width=2.0pt,forget plot]
  table[row sep=crcr]{%
0	-1.38009808963736\\
1	-0.89569416707839\\
2	-0.800997534411909\\
3	-0.639073974111863\\
4	-0.521252726971612\\
5	-0.422896025262773\\
6	-0.344540151007089\\
7	-0.281228925161234\\
8	-0.22999464271277\\
9	-0.188327597754243\\
10	-0.154339586594047\\
11	-0.126551870082128\\
12	-0.103800368315489\\
13	-0.0851551508586475\\
14	-0.0698666343819271\\
15	-0.0573264291429624\\
16	-0.0470385804512165\\
17	-0.0385976685997993\\
18	-0.0316717325109505\\
19	-0.0259886963370375\\
20	-0.0213254439709311\\
21	-0.0174989533810867\\
22	-0.0143590672548594\\
23	-0.0117825811077683\\
24	-0.00966840098139778\\
25	-0.00793357371304471\\
26	-0.00651003100865579\\
27	-0.0053419183183143\\
28	-0.00438340321030671\\
29	-0.00359687706352689\\
30	-0.00295147945307147\\
31	-0.00242188731112569\\
32	-0.00198732134985286\\
33	-0.00163073076221321\\
34	-0.00133812421249832\\
35	-0.00109802086806834\\
36	-0.000900999933415371\\
37	-0.0007393310122768\\
38	-0.00060667079474627\\
39	-0.000497814168582587\\
40	-0.000408489989271275\\
41	-0.000335193495614315\\
42	-0.000275048795448413\\
43	-0.000225696025929456\\
44	-0.000185198760959128\\
45	-0.000151968032753596\\
46	-0.00012469998643296\\
47	-0.000102324721420834\\
48	-8.39643123736978e-05\\
49	-6.88983625315131e-05\\
};
\addplot [color=mycolor4,solid,line width=2.0pt,forget plot]
  table[row sep=crcr]{%
0	-2.61425024962473\\
1	-1.71721983004516\\
2	-1.49710007602977\\
3	-1.19919088044484\\
4	-0.981375510541939\\
5	-0.800018579942816\\
6	-0.654178396606031\\
7	-0.53542021260599\\
8	-0.438670003428355\\
9	-0.359614929669286\\
10	-0.294923911691362\\
11	-0.241927435404473\\
12	-0.198482307655797\\
13	-0.162852253626746\\
14	-0.133624348756337\\
15	-0.109644862539811\\
16	-0.0899698014863862\\
17	-0.0738258120236917\\
18	-0.0605788746269841\\
19	-0.0497089841018963\\
20	-0.0407895485004877\\
21	-0.0334705648960243\\
22	-0.0274648500524995\\
23	-0.0225367574938553\\
24	-0.0184929257686161\\
25	-0.0151746895071699\\
26	-0.0124518532527166\\
27	-0.0102175828648557\\
28	-0.00838421370471412\\
29	-0.00687981100565381\\
30	-0.0056453474316042\\
31	-0.00463238707111992\\
32	-0.00380118499512471\\
33	-0.00311912781353616\\
34	-0.0025594540454308\\
35	-0.00210020409576925\\
36	-0.00172335864019168\\
37	-0.00141413161155452\\
38	-0.00116039004772409\\
39	-0.000952178037628184\\
40	-0.000781326086960116\\
41	-0.00064113057646604\\
42	-0.000526090735916157\\
43	-0.000431692813564371\\
44	-0.000354232972680301\\
45	-0.000290671966248135\\
46	-0.000238515887787847\\
47	-0.000195718319387798\\
48	-0.00016060003758768\\
49	-0.000131783126657962\\
};
\addplot [color=mycolor5,solid,line width=2.0pt,forget plot]
  table[row sep=crcr]{%
0	0.858993947906489\\
1	0.562200208813633\\
2	0.502123226159488\\
3	0.398178044621712\\
4	0.322817399760831\\
5	0.260592061045394\\
6	0.211541382127784\\
7	0.172246104498432\\
8	0.140645234479495\\
9	0.11505385041084\\
10	0.0942354964832023\\
11	0.0772433615673388\\
12	0.0633446709559227\\
13	0.0519609938357264\\
14	0.0426297213274912\\
15	0.0349772117461592\\
16	0.0286997621443761\\
17	0.0235495265790152\\
18	0.0193237573564649\\
19	0.01585636694576\\
20	0.0130111917512051\\
21	0.0106765522051137\\
22	0.00876082890228055\\
23	0.00718885025074995\\
24	0.00589893595625734\\
25	0.00484047420971417\\
26	0.00397193490036122\\
27	0.00325923982057181\\
28	0.00267442548728594\\
29	0.00219454595372116\\
30	0.00180077250645712\\
31	0.00147765490698566\\
32	0.00121251518838433\\
33	0.000994950222471279\\
34	0.000816423541160765\\
35	0.00066993039751319\\
36	0.000549722925293118\\
37	0.000451084613638261\\
38	0.000370145248226806\\
39	0.000303729057997357\\
40	0.000249230109292421\\
41	0.000204510058361878\\
42	0.000167814250411993\\
43	0.000137702873232158\\
44	0.000112994464116362\\
45	9.27195534941608e-05\\
46	7.60826264135115e-05\\
47	6.2430909380344e-05\\
48	5.12287578621958e-05\\
49	4.20366395132183e-05\\
};
\addplot [color=mycolor6,solid,line width=2.0pt,forget plot]
  table[row sep=crcr]{%
0	0.0208858532794561\\
1	0.0144474924230532\\
2	0.00921591845665313\\
3	0.0073188982211389\\
4	0.00600389755965001\\
5	0.00504805637428132\\
6	0.00423741451510349\\
7	0.00354099020036859\\
8	0.0029428059311836\\
9	0.00243515947213827\\
10	0.00200882773043182\\
11	0.00165371166520101\\
12	0.00135957811124891\\
13	0.00111685548606873\\
14	0.000917023064220785\\
15	0.000752734813560839\\
16	0.000617781655524719\\
17	0.000506979148610636\\
18	0.000416030149312702\\
19	0.000341388476773153\\
20	0.000280135094302861\\
21	0.000229870667632111\\
22	0.000188624622943785\\
23	0.000154779234845783\\
24	0.000127006753270118\\
25	0.000104217555524005\\
26	8.55174918467681e-05\\
27	7.01728391251403e-05\\
28	5.75815249790339e-05\\
29	4.72495083734666e-05\\
30	3.87713959535985e-05\\
31	3.18145357328186e-05\\
32	2.61059647484371e-05\\
33	2.14216987532465e-05\\
34	1.75779437340844e-05\\
35	1.44238844105936e-05\\
36	1.18357667388129e-05\\
37	9.71204222224063e-06\\
38	7.96938350064299e-06\\
39	6.53941487698425e-06\\
40	5.36602949707537e-06\\
41	4.40318791614346e-06\\
42	3.61311167519661e-06\\
43	2.96480100922093e-06\\
44	2.43281852721314e-06\\
45	1.9962911399276e-06\\
46	1.63809107451909e-06\\
47	1.34416384201108e-06\\
48	1.10297678943129e-06\\
49	9.05066599770706e-07\\
};
\addplot [color=mycolor7,solid,line width=2.0pt,forget plot]
  table[row sep=crcr]{%
0	0.096951128654181\\
1	0.332520498034373\\
2	0.227649128398152\\
3	0.204765862379189\\
4	0.169921047111462\\
5	0.142459862326571\\
6	0.118157949261246\\
7	0.0976918706452468\\
8	0.0805145582246317\\
9	0.0662420748590281\\
10	0.0544378657855062\\
11	0.0447077865436729\\
12	0.0367027279390189\\
13	0.0301244962058477\\
14	0.0247223361632886\\
15	0.0202876383389228\\
16	0.0166478827416635\\
17	0.0136608969887693\\
18	0.0112097487968058\\
19	0.00919837085682399\\
20	0.00754788473590859\\
21	0.00619354575182985\\
22	0.0050822190168715\\
23	0.00417030109915991\\
24	0.00342201160297596\\
25	0.00280799010460932\\
26	0.00230414439678341\\
27	0.00189070524142343\\
28	0.00155145068502076\\
29	0.00127306954830282\\
30	0.00104463912824626\\
31	0.000857196620736158\\
32	0.000703387446687348\\
33	0.000577176681690777\\
34	0.000473612265628432\\
35	0.000388630700781802\\
36	0.00031889761434699\\
37	0.000261676929399925\\
38	0.000214723510946343\\
39	0.000176195074821181\\
40	0.000144579903033581\\
41	0.000118637529354955\\
42	9.73500678596556e-05\\
43	7.98822747220348e-05\\
44	6.55487762367197e-05\\
45	5.37871772064531e-05\\
46	4.41359945666876e-05\\
47	3.62165504412501e-05\\
48	2.97181141773403e-05\\
49	2.438571038647e-05\\
};
\addplot [color=mycolor1,solid,line width=2.0pt,forget plot]
  table[row sep=crcr]{%
0	0.0485141583087164\\
1	0.303478054351326\\
2	0.208930929769453\\
3	0.191464987515286\\
4	0.159864110703636\\
5	0.1345363226504\\
6	0.111776738443242\\
7	0.0924950261072698\\
8	0.0762606221358632\\
9	0.062752560873537\\
10	0.0515733170855197\\
11	0.0423559347545169\\
12	0.0347719462527705\\
13	0.0285395910257413\\
14	0.0234214949398278\\
15	0.0192200405089012\\
16	0.0157717601318701\\
17	0.0129419366416079\\
18	0.0106197729521588\\
19	0.00871424641748179\\
20	0.00715062371763894\\
21	0.00586756444108321\\
22	0.00481472860033356\\
23	0.0039508068804247\\
24	0.00324190170029128\\
25	0.00266019776785705\\
26	0.00218287083019458\\
27	0.00179119212164672\\
28	0.00146979347929732\\
29	0.00120606432180553\\
30	0.000989656835957286\\
31	0.000812079954156059\\
32	0.000666366189107689\\
33	0.000546798251315047\\
34	0.00044868472147043\\
35	0.000368175975156702\\
36	0.000302113136984091\\
37	0.000247904137448983\\
38	0.000203422009322818\\
39	0.000166921433033889\\
40	0.000136970256557327\\
41	0.000112393302887684\\
42	9.22262602955043e-05\\
43	7.56778461843419e-05\\
44	6.20987599924775e-05\\
45	5.09562069618952e-05\\
46	4.18129931782268e-05\\
47	3.43103716457758e-05\\
48	2.81539663389592e-05\\
49	2.31022219403067e-05\\
};
\addplot [color=mycolor2,solid,line width=2.0pt,forget plot]
  table[row sep=crcr]{%
0	-0.0728997001262312\\
1	-0.118910590495511\\
2	-0.0841688115942048\\
3	-0.0742359192295582\\
4	-0.0620832795602287\\
5	-0.0522637932458658\\
6	-0.0435488174475146\\
7	-0.0361187183009457\\
8	-0.0298331760002208\\
9	-0.0245790731648326\\
10	-0.020216723997096\\
11	-0.0166119181752262\\
12	-0.0136416782343681\\
13	-0.01119863156193\\
14	-0.00919129654481\\
15	-0.00754295629704166\\
16	-0.00618986461116873\\
17	-0.00507934277043925\\
18	-0.00416799534721872\\
19	-0.00342013847668285\\
20	-0.0028064587486734\\
21	-0.00230288897514206\\
22	-0.00188967493971792\\
23	-0.00155060487880145\\
24	-0.00127237519842448\\
25	-0.00104406916629189\\
26	-0.000856728811372254\\
27	-0.000703003513754962\\
28	-0.000576861605694176\\
29	-0.000473353707947297\\
30	-0.000388418528733326\\
31	-0.000318723509092107\\
32	-0.00026153406257056\\
33	-0.000214606278316212\\
34	-0.000176098877220519\\
35	-0.000144500966288115\\
36	-0.0001185727563884\\
37	-9.72969172625677e-05\\
38	-7.98386610753778e-05\\
39	-6.55129883048416e-05\\
40	-5.37578108017593e-05\\
41	-4.41118974574294e-05\\
42	-3.61967771433028e-05\\
43	-2.97018888572541e-05\\
44	-2.43723964207024e-05\\
45	-1.9999189618623e-05\\
46	-1.64106794628392e-05\\
47	-1.34660656540376e-05\\
48	-1.10498120817857e-05\\
49	-9.06711360093069e-06\\
};
\addplot [color=mycolor3,solid,line width=2.0pt,forget plot]
  table[row sep=crcr]{%
0	-0.0547654730292545\\
1	-0.350331553943812\\
2	-0.239248900997969\\
3	-0.218477822898384\\
4	-0.181780526461458\\
5	-0.15267455462404\\
6	-0.126675962600617\\
7	-0.104740130702785\\
8	-0.0863147283718719\\
9	-0.0710058741310289\\
10	-0.0583469901027399\\
11	-0.0479147385903895\\
12	-0.0393335317581965\\
13	-0.0322827596041208\\
14	-0.026493044152148\\
15	-0.0217404613756136\\
16	-0.0178399364861969\\
17	-0.0146390140894334\\
18	-0.0120123384875131\\
19	-0.00985693957297642\\
20	-0.00808827995772301\\
21	-0.00663697423844832\\
22	-0.00544608096318684\\
23	-0.0044688738844836\\
24	-0.00366701047735129\\
25	-0.00300902808379445\\
26	-0.00246910954764772\\
27	-0.00202607022813552\\
28	-0.00166252675340111\\
29	-0.00136421492855763\\
30	-0.00111943003973892\\
31	-0.000918567591351649\\
32	-0.000753746453677602\\
33	-0.000618499632236749\\
34	-0.000507520524405324\\
35	-0.000416454706509797\\
36	-0.000341729081440431\\
37	-0.00028041168292903\\
38	-0.000230096635603754\\
39	-0.000188809757012829\\
40	-0.000154931097753777\\
41	-0.000127131380448771\\
42	-0.000104319843653915\\
43	-8.5601444281892e-05\\
44	-7.02417393132575e-05\\
45	-5.76380688800038e-05\\
46	-4.72959100485583e-05\\
47	-3.88094735092084e-05\\
48	-3.18457818554544e-05\\
49	-2.61316047419299e-05\\
};
\addplot [color=mycolor4,solid,line width=2.0pt,forget plot]
  table[row sep=crcr]{%
0	0.0217474802522881\\
1	-0.166314973876075\\
2	-0.110230714445917\\
3	-0.103178380287579\\
4	-0.0856851784793827\\
5	-0.0719720488282045\\
6	-0.0596209283098874\\
7	-0.0492383840958185\\
8	-0.0405376388305219\\
9	-0.033326126029112\\
10	-0.0273729901186696\\
11	-0.0224727487860533\\
12	-0.0184450224040973\\
13	-0.0151371889648049\\
14	-0.0124217434684209\\
15	-0.0101930956845969\\
16	-0.00836418009717848\\
17	-0.00686337996760561\\
18	-0.00563185921168007\\
19	-0.00462131248320004\\
20	-0.00379209272005101\\
21	-0.00311166400156685\\
22	-0.00255332779450102\\
23	-0.00209517619161772\\
24	-0.00171923244430205\\
25	-0.00141074556152471\\
26	-0.00115761145599455\\
27	-0.00094989796479673\\
28	-0.000779455110405174\\
29	-0.000639595303964379\\
30	-0.000524830937092495\\
31	-0.00043065906204065\\
32	-0.000353384709294331\\
33	-0.00028997590882098\\
34	-0.000237944725771501\\
35	-0.000195249642524343\\
36	-0.000160215456707209\\
37	-0.000131467552192697\\
38	-0.000107877964055687\\
39	-8.85211212556411e-05\\
40	-7.26375305367665e-05\\
41	-5.96039766286447e-05\\
42	-4.89090695083521e-05\\
43	-4.01331792855241e-05\\
44	-3.29319714268237e-05\\
45	-2.70228963008483e-05\\
46	-2.21741029415312e-05\\
47	-1.81953420457749e-05\\
48	-1.49305012714895e-05\\
49	-1.22514799478412e-05\\
};
\addplot [color=mycolor5,solid,line width=2.0pt,forget plot]
  table[row sep=crcr]{%
0	0.089568906461844\\
1	0.072381321663862\\
2	0.0490325160091872\\
3	0.0299922027735382\\
4	0.0177420784824722\\
5	0.0104414637458059\\
6	0.00629371785742477\\
7	0.00396818066819582\\
8	0.00265145922926085\\
9	0.00187949985210364\\
10	0.00140118671528678\\
11	0.00108429484544116\\
12	0.000860110005771602\\
13	0.000692716268262835\\
14	0.000562822028819616\\
15	0.000459510532265819\\
16	0.00037613317714242\\
17	0.00030829241892573\\
18	0.000252852366719763\\
19	0.000207445275714776\\
20	0.000170214934286945\\
21	0.000139673483618573\\
22	0.000114613760892994\\
23	9.40502153870952e-05\\
24	7.71757841692564e-05\\
25	6.33286565513381e-05\\
26	5.19658229650386e-05\\
27	4.26416610190177e-05\\
28	3.49904547561566e-05\\
29	2.87120715729045e-05\\
30	2.35602091106467e-05\\
31	1.93327465557126e-05\\
32	1.58638231807366e-05\\
33	1.30173355278183e-05\\
34	1.06815997915673e-05\\
35	8.76497105310305e-06\\
36	7.19224809910688e-06\\
37	5.90172309803502e-06\\
38	4.84276054566384e-06\\
39	3.97381057225467e-06\\
40	3.26077870208759e-06\\
41	2.67568812007391e-06\\
42	2.19558196652573e-06\\
43	1.80162259405952e-06\\
44	1.47835244643926e-06\\
45	1.21308755962571e-06\\
46	9.95419888483128e-07\\
47	8.1680893234103e-07\\
48	6.7024663630197e-07\\
49	5.49982420235817e-07\\
};
\addplot [color=mycolor6,solid,line width=2.0pt,forget plot]
  table[row sep=crcr]{%
0	-0.76937821728172\\
1	-0.433068715371825\\
2	-0.336828926491806\\
3	-0.24683165446178\\
4	-0.191270075760949\\
5	-0.150485489055005\\
6	-0.120525786027541\\
7	-0.0974681596205839\\
8	-0.0793287792913885\\
9	-0.0648014992172719\\
10	-0.0530464172559307\\
11	-0.0434740326855159\\
12	-0.0356512347818677\\
13	-0.0292455641810425\\
14	-0.0239947565243893\\
15	-0.0196882408011935\\
16	-0.0161552283507562\\
17	-0.013256407240944\\
18	-0.010877796231524\\
19	-0.00892599303202671\\
20	-0.00732439965369262\\
21	-0.00601017621210961\\
22	-0.00493176152072493\\
23	-0.00404684609566174\\
24	-0.00332071143705337\\
25	-0.00272486803910201\\
26	-0.00223593791714454\\
27	-0.00183473760889707\\
28	-0.00150552565405743\\
29	-0.00123538505301137\\
30	-0.00101371650234505\\
31	-0.000831822543067633\\
32	-0.000682566316849811\\
33	-0.000560091547712853\\
34	-0.000459592765756067\\
35	-0.000377126759106201\\
36	-0.000309457857015006\\
37	-0.000253930973986874\\
38	-0.00020836743383788\\
39	-0.000170979486283897\\
40	-0.000140300162033899\\
41	-0.000115125714170203\\
42	-9.44683874284389e-05\\
43	-7.75176622156108e-05\\
44	-6.36084527212602e-05\\
45	-5.21950113297879e-05\\
46	-4.28295154365576e-05\\
47	-3.51444964910874e-05\\
48	-2.8838421845848e-05\\
49	-2.36638636939995e-05\\
};
\addplot [color=mycolor7,solid,line width=2.0pt,forget plot]
  table[row sep=crcr]{%
0	0.601439760323562\\
1	0.0618488400610952\\
2	0.0783761711657787\\
3	0.0686026832510509\\
4	0.0781488153758804\\
5	0.0779627388517248\\
6	0.0729899022199323\\
7	0.0649203960662874\\
8	0.0559778251538172\\
9	0.0473170257605559\\
10	0.0395128800951362\\
11	0.0327529244620479\\
12	0.0270306160753703\\
13	0.0222510756307139\\
14	0.0182899587229781\\
15	0.0150217707260472\\
16	0.0123321010996614\\
17	0.0101216389715933\\
18	0.00830638228903882\\
19	0.00681626927144711\\
20	0.00559331205495743\\
21	0.00458971533417109\\
22	0.00376617215465745\\
23	0.00309039376618118\\
24	0.00253587205200944\\
25	0.00208085083326289\\
26	0.00170747637942549\\
27	0.00140109825018816\\
28	0.00114969482268685\\
29	0.000943401644507682\\
30	0.00077412434732423\\
31	0.000635221007999436\\
32	0.000521241511182097\\
33	0.00042771368663016\\
34	0.00035096782620384\\
35	0.000287992691774221\\
36	0.000236317362016262\\
37	0.000193914280817843\\
38	0.000159119702523139\\
39	0.000130568412174538\\
40	0.000107140159211416\\
41	8.79157027792972e-05\\
42	7.21407439787328e-05\\
43	5.91963298631675e-05\\
44	4.8574567934966e-05\\
45	3.98586982591039e-05\\
46	3.27067412937564e-05\\
47	2.68380798364203e-05\\
48	2.20224486088446e-05\\
49	1.80708994713676e-05\\
};
\end{axis}
\end{tikzpicture}%

%% file: flow_aggregated_2.tikz
%
%
\definecolor{mycolor1}{rgb}{0.00000,0.44700,0.74100}%
\definecolor{mycolor2}{rgb}{0.85000,0.32500,0.09800}%
\definecolor{mycolor3}{rgb}{0.46600,0.67400,0.18800}%
\definecolor{mycolor4}{rgb}{0.49400,0.18400,0.55600}%
\definecolor{mycolor5}{rgb}{0.30100,0.74500,0.93300}%
\definecolor{mycolor6}{rgb}{0.92900,0.69400,0.12500}%
\begin{tikzpicture}

\begin{axis}[%
width=0.85\columnwidth,
height=0.35\columnwidth,
at={(0.772in,0.484in)},
scale only axis,
xmin=0,
xmax=180,
xticklabels={,,},
xmajorgrids,
ymin=0,
ymax=0.35,
ylabel={Aggregated flow rate},
axis background/.style={fill=white},
scaled ticks=false, tick label style={/pgf/number format/fixed},
legend style={legend cell align=left,draw=white!15!black,
	at={(0.98,0.98)}, anchor=north east}
]
\addplot [color=mycolor1,mark=star,mark repeat={20},solid,line width=1.0pt]
  table[row sep=crcr]{%
0	0.334512794152713\\
1	0.334466937466419\\
2	0.334415974759417\\
3	0.334359333291013\\
4	0.334296389702897\\
5	0.33422645474288\\
6	0.334148767972921\\
7	0.334062481293342\\
8	0.333966648735213\\
9	0.333860216943548\\
10	0.333742010854915\\
11	0.33361071892087\\
12	0.333464880921269\\
13	0.333302875202648\\
14	0.333122904574231\\
15	0.332922984245681\\
16	0.3327009248974\\
17	0.332454311845755\\
18	0.332180474469502\\
19	0.331876449780482\\
20	0.331538945681553\\
21	0.331164307899722\\
22	0.330748489587385\\
23	0.330287017091474\\
24	0.329774948883118\\
25	0.329206816609349\\
26	0.328576554890033\\
27	0.327877446816399\\
28	0.327102075881023\\
29	0.326242258167858\\
30	0.325288966591296\\
31	0.324232260563203\\
32	0.323061227295267\\
33	0.321763941355525\\
34	0.320327427391019\\
35	0.318737592339035\\
36	0.316979110349673\\
37	0.315035289612986\\
38	0.312887952261801\\
39	0.310517319519917\\
40	0.307901924749002\\
41	0.305018645964075\\
42	0.301842782645006\\
43	0.298348035340683\\
44	0.294506465276582\\
45	0.290288573332353\\
46	0.285663688298895\\
47	0.280600792990919\\
48	0.275069313420091\\
49	0.269039236117615\\
50	0.262480932277371\\
51	0.255365443245677\\
52	0.247665348699995\\
53	0.239356240325227\\
54	0.230418276102192\\
55	0.220836786408913\\
56	0.210602851225376\\
57	0.199715813858361\\
58	0.188187433166749\\
59	0.176043110737424\\
60	0.163318507281759\\
61	0.15005784839746\\
62	0.136314606290614\\
63	0.122149875550353\\
64	0.107630033837102\\
65	0.0928272966376579\\
66	0.0778189627814598\\
67	0.0626831489876955\\
68	0.0475079565271277\\
69	0.032399193249741\\
70	0.0174736792928515\\
71	0.00289371368303956\\
72	1.44906628408507e-05\\
73	6.8755672401981e-06\\
74	4.66763765382752e-06\\
75	3.62349784872832e-06\\
76	3.01822196133801e-06\\
77	2.62533960081187e-06\\
78	2.35131209353185e-06\\
79	2.15048687233966e-06\\
80	1.99793140272153e-06\\
81	1.87887525342223e-06\\
82	1.78401064334262e-06\\
83	1.70718010313652e-06\\
84	1.64414708730924e-06\\
85	1.59190068770476e-06\\
86	1.5482420254868e-06\\
87	1.51152761214687e-06\\
88	1.48050431546345e-06\\
89	1.45419992542187e-06\\
90	1.43184862779054e-06\\
91	1.41283904930317e-06\\
92	1.39667728083247e-06\\
93	1.38296007042787e-06\\
94	1.37135506452214e-06\\
95	1.36158602455782e-06\\
96	1.35342161472442e-06\\
97	1.3466667919672e-06\\
98	1.34115611879393e-06\\
99	1.33674851498789e-06\\
100	1.33332309881217e-06\\
101	1.33077586247134e-06\\
102	1.32901699242025e-06\\
103	1.32796869059835e-06\\
104	1.32756338854912e-06\\
105	1.3277422774271e-06\\
106	1.32845409453918e-06\\
107	1.32965410864051e-06\\
108	1.33130324456445e-06\\
109	1.33336736414982e-06\\
110	1.33581682653399e-06\\
111	1.33862595512227e-06\\
112	1.34177235346043e-06\\
113	1.34523663270814e-06\\
114	1.34900218383165e-06\\
115	1.35305441198171e-06\\
116	1.35738124520521e-06\\
117	1.36197499904623e-06\\
118	1.36682924328885e-06\\
119	1.37193778357286e-06\\
120	1.37729401919313e-06\\
121	1.38290016087893e-06\\
122	1.38877954942695e-06\\
123	1.39490162382661e-06\\
124	1.40133750397441e-06\\
125	1.40797582959735e-06\\
126	1.41511738829405e-06\\
127	1.42263743720121e-06\\
128	1.43035586605347e-06\\
129	1.4382847042256e-06\\
130	1.44643819456628e-06\\
131	1.45482819813463e-06\\
132	1.46346679046492e-06\\
133	1.4723670277634e-06\\
134	1.48154319282797e-06\\
135	1.49101094531991e-06\\
136	1.5007874718338e-06\\
137	1.51089165416278e-06\\
138	1.52134426048019e-06\\
139	1.53216817056388e-06\\
140	1.54338866296253e-06\\
141	1.55503380893886e-06\\
142	1.56713501239908e-06\\
143	1.57972767309282e-06\\
144	1.59285180849693e-06\\
145	1.60655227767511e-06\\
146	1.6208781374762e-06\\
147	1.63588084712312e-06\\
148	1.6516116891477e-06\\
149	1.66811975416313e-06\\
150	1.68545255146696e-06\\
151	1.70366096474034e-06\\
152	1.72280855816992e-06\\
153	1.74298292685016e-06\\
154	1.7643055069814e-06\\
155	1.78693720857989e-06\\
156	1.81107990839746e-06\\
157	1.83697640177763e-06\\
158	1.8649124062777e-06\\
159	1.89522370870744e-06\\
160	1.92831065268439e-06\\
161	1.96466190345545e-06\\
162	2.00489029731958e-06\\
163	2.04978586505998e-06\\
164	2.10039549424159e-06\\
165	2.15814702259993e-06\\
166	2.22505235118874e-06\\
167	2.3040606083335e-06\\
168	2.39971867738604e-06\\
169	2.51952458937679e-06\\
170	2.6770574762743e-06\\
171	2.9005851781361e-06\\
172	3.26426256354303e-06\\
173	4.08307425522903e-06\\
};
\addlegendentry{$L_1$};

\addplot [color=mycolor2,mark=o,mark repeat={20},mark phase={10},solid,line width=1.0pt]
  table[row sep=crcr]{%
0	0.14999960993337\\
1	0.149999609933366\\
2	0.14999960993336\\
3	0.149999609933354\\
4	0.149999609933344\\
5	0.149999609933334\\
6	0.149999609933322\\
7	0.149999609933308\\
8	0.149999609933291\\
9	0.149999609933274\\
10	0.149999609933254\\
11	0.149999609933231\\
12	0.149999609933205\\
13	0.149999609933177\\
14	0.149999609933145\\
15	0.14999960993311\\
16	0.149999609933071\\
17	0.14999960993303\\
18	0.149999609932982\\
19	0.14999960993293\\
20	0.149999609932873\\
21	0.149999609932811\\
22	0.149999609932743\\
23	0.149999609932667\\
24	0.149999609932584\\
25	0.149999609932495\\
26	0.149999609932396\\
27	0.149999609932288\\
28	0.14999960993217\\
29	0.14999960993204\\
30	0.149999609931899\\
31	0.149999609931743\\
32	0.149999609931574\\
33	0.149999609931389\\
34	0.149999609931185\\
35	0.149999609930962\\
36	0.149999609930717\\
37	0.14999960993045\\
38	0.149999609930157\\
39	0.149999609929834\\
40	0.14999960992948\\
41	0.149999609929091\\
42	0.149999609928663\\
43	0.149999609928191\\
44	0.149999609927671\\
45	0.149999609927096\\
46	0.149999609926457\\
47	0.14999960992575\\
48	0.149999609924962\\
49	0.149999609924081\\
50	0.149999609923093\\
51	0.149999609921981\\
52	0.14999960992072\\
53	0.149999609919284\\
54	0.149999609917636\\
55	0.149999609915732\\
56	0.149999609913516\\
57	0.149999609910909\\
58	0.149999609907813\\
59	0.149999609904101\\
60	0.149999609899602\\
61	0.149999609894091\\
62	0.14999960988727\\
63	0.149999609878733\\
64	0.149999609867884\\
65	0.149999609853769\\
66	0.149999609834683\\
67	0.149999609807263\\
68	0.149999609763991\\
69	0.149999609684541\\
70	0.149999609490342\\
71	0.149999608428716\\
72	0.149999574407703\\
73	0.149999527185354\\
74	0.149999474481721\\
75	0.149999415816509\\
76	0.149999350539024\\
77	0.149999277905107\\
78	0.149999197078688\\
79	0.149999107123633\\
80	0.149999006992358\\
81	0.149998895512254\\
82	0.149998771369985\\
83	0.149998633093462\\
84	0.149998479031152\\
85	0.149998307328301\\
86	0.149998115899514\\
87	0.149997902397076\\
88	0.149997664174176\\
89	0.149997398242091\\
90	0.149997101220112\\
91	0.14999676927671\\
92	0.149996398060104\\
93	0.149995982615886\\
94	0.149995517288752\\
95	0.1499949956046\\
96	0.149994410128167\\
97	0.149993752289972\\
98	0.149993012174463\\
99	0.149992178258665\\
100	0.149991237087178\\
101	0.149990172864545\\
102	0.149988966939314\\
103	0.14998759714391\\
104	0.14998603693991\\
105	0.1499842542988\\
106	0.149982210219874\\
107	0.149979856736861\\
108	0.149977134167883\\
109	0.149973967273867\\
110	0.14997026005942\\
111	0.149965888202749\\
112	0.149960687367042\\
113	0.149954436293941\\
114	0.149946831434288\\
115	0.149937439279167\\
116	0.149925619540789\\
117	0.149910424444339\\
118	0.149890395414972\\
119	0.149863249025578\\
120	0.149824830379405\\
121	0.149766837659698\\
122	0.149675175671668\\
123	0.149518115253816\\
124	0.149184528179521\\
125	0.147431920073295\\
126	0.13803564674017\\
127	0.125305495686381\\
128	0.112617267158695\\
129	0.100544598433394\\
130	0.0892259307013116\\
131	0.0787191298729954\\
132	0.069050748260726\\
133	0.0602267273354262\\
134	0.0522365828449294\\
135	0.0450562443150897\\
136	0.038650617744755\\
137	0.0329760805052359\\
138	0.0279828889092627\\
139	0.0236174211578502\\
140	0.0198241829633293\\
141	0.0165475293450142\\
142	0.013733084056945\\
143	0.0113288613199064\\
144	0.00928611164090246\\
145	0.00755992104926684\\
146	0.00610958681240994\\
147	0.00489877503773481\\
148	0.0038954492405787\\
149	0.00307156092538347\\
150	0.00240252376586797\\
151	0.00186654843751236\\
152	0.00144397498112258\\
153	0.00111676622891651\\
154	0.000868281416167626\\
155	0.000683335101782822\\
156	0.000548424690871787\\
157	0.000451957872170593\\
158	0.000384351752776926\\
159	0.000337963045934444\\
160	0.000306882421879313\\
161	0.000286657801740017\\
162	0.000274008035576828\\
163	0.000266569537862516\\
164	0.000262698876563087\\
165	0.00026134117595627\\
166	0.000261970792756855\\
167	0.000264621181180797\\
168	0.000270056906192714\\
169	0.000280242059854618\\
170	0.000299577118954183\\
171	0.000338611458065845\\
172	0.000428598798490659\\
173	0.000721856922106661\\
};
\addlegendentry{$L_2$};

\addplot [color=mycolor3,mark=diamond,mark repeat={20},mark phase={10},solid,line width=1.0pt]
  table[row sep=crcr]{%
0	0.347551199633813\\
1	0.347501326962925\\
2	0.347445921648498\\
3	0.347384370996853\\
4	0.347315994731651\\
5	0.347240037576807\\
6	0.347155661065652\\
7	0.347061934475793\\
8	0.346957824776713\\
9	0.346842185545517\\
10	0.34671374465958\\
11	0.346571090737627\\
12	0.346412658161616\\
13	0.346236710529902\\
14	0.346041322413593\\
15	0.345824359289003\\
16	0.34558345539952\\
17	0.345315989456622\\
18	0.345019057940516\\
19	0.344689445839216\\
20	0.344323594605535\\
21	0.343917567141278\\
22	0.343467009620863\\
23	0.342967109957655\\
24	0.342412552733545\\
25	0.341797470448352\\
26	0.341115390973008\\
27	0.340359181127882\\
28	0.33952098641648\\
29	0.338592166967247\\
30	0.337563229922207\\
31	0.33642375862959\\
32	0.335162339156613\\
33	0.333766484939471\\
34	0.332222560631861\\
35	0.330515706551756\\
36	0.328629765612632\\
37	0.326547215038186\\
38	0.324249105843526\\
39	0.321715013687695\\
40	0.318923005469137\\
41	0.315849626993263\\
42	0.312469917868063\\
43	0.308757460911717\\
44	0.304684474320414\\
45	0.300221955868376\\
46	0.295339889354103\\
47	0.29000752412933\\
48	0.284193738957828\\
49	0.277867501258451\\
50	0.270998431884176\\
51	0.263557483771829\\
52	0.255517739747483\\
53	0.246855330287638\\
54	0.237550465863563\\
55	0.227588570604727\\
56	0.216961494211598\\
57	0.205668767752697\\
58	0.193718856494421\\
59	0.181130350086116\\
60	0.167933018426799\\
61	0.154168651856587\\
62	0.13989159869356\\
63	0.125168913694078\\
64	0.110080039495691\\
65	0.0947159611698623\\
66	0.0791778028735498\\
67	0.0635748776858648\\
68	0.0480222688976973\\
69	0.032638193726746\\
70	0.017542465199244\\
71	0.0028954516101829\\
72	1.44911415982534e-05\\
73	6.87568200840767e-06\\
74	4.66769155735978e-06\\
75	3.62353081778915e-06\\
76	3.018245194141e-06\\
77	2.62535747816537e-06\\
78	2.35132668881873e-06\\
79	2.15049929548112e-06\\
80	1.9979423017677e-06\\
81	1.87888503229276e-06\\
82	1.7840195681229e-06\\
83	1.70718835721008e-06\\
84	1.64415480310634e-06\\
85	1.59190796482357e-06\\
86	1.5482489412955e-06\\
87	1.51153422790418e-06\\
88	1.48051068059855e-06\\
89	1.45420608009631e-06\\
90	1.43185460512759e-06\\
91	1.41284487672628e-06\\
92	1.3966829812657e-06\\
93	1.38296566328041e-06\\
94	1.37136056651655e-06\\
95	1.36159145027539e-06\\
96	1.35342697678352e-06\\
97	1.34667210099004e-06\\
98	1.34116138360182e-06\\
99	1.33675374334017e-06\\
100	1.33332829819247e-06\\
101	1.33078104028542e-06\\
102	1.32902215596269e-06\\
103	1.32797384735389e-06\\
104	1.32756854713016e-06\\
105	1.32774744731678e-06\\
106	1.32845927361959e-06\\
107	1.32965928418097e-06\\
108	1.33130842087795e-06\\
109	1.33337254986904e-06\\
110	1.33582202852472e-06\\
111	1.33863117980331e-06\\
112	1.34177760801743e-06\\
113	1.34524191626144e-06\\
114	1.34900748796945e-06\\
115	1.35305974142793e-06\\
116	1.35738660594275e-06\\
117	1.36198039544838e-06\\
118	1.36683467856551e-06\\
119	1.37194325729485e-06\\
120	1.37729953307552e-06\\
121	1.38290571767169e-06\\
122	1.3887851522771e-06\\
123	1.39490727538117e-06\\
124	1.40134320688927e-06\\
125	1.40798158497568e-06\\
126	1.41512319853035e-06\\
127	1.42264330519789e-06\\
128	1.43036179315761e-06\\
129	1.43829069223457e-06\\
130	1.44644424548242e-06\\
131	1.4548343140749e-06\\
132	1.4634729736529e-06\\
133	1.47237328053969e-06\\
134	1.48154951766517e-06\\
135	1.4910173448397e-06\\
136	1.50079394882628e-06\\
137	1.51089821160853e-06\\
138	1.52135090157565e-06\\
139	1.53217489875098e-06\\
140	1.54339548196404e-06\\
141	1.55504072280225e-06\\
142	1.56714202555332e-06\\
143	1.57973479042296e-06\\
144	1.59285903543778e-06\\
145	1.60655962031571e-06\\
146	1.62088560265141e-06\\
147	1.63588844245467e-06\\
148	1.65161942299855e-06\\
149	1.66812763549815e-06\\
150	1.68546058966725e-06\\
151	1.70366916947321e-06\\
152	1.72281693943262e-06\\
153	1.74299149525938e-06\\
154	1.76431427429064e-06\\
155	1.78694618832784e-06\\
156	1.81108911657234e-06\\
157	1.83698585743029e-06\\
158	1.86492213209519e-06\\
159	1.89523373162823e-06\\
160	1.92832100468161e-06\\
161	1.9646726226451e-06\\
162	2.00490142958971e-06\\
163	2.04979746648912e-06\\
164	2.10040763471558e-06\\
165	2.15815979130844e-06\\
166	2.22506586522473e-06\\
167	2.30407502675589e-06\\
168	2.39973422567611e-06\\
169	2.51954160564233e-06\\
170	2.67707650898186e-06\\
171	2.90060722949269e-06\\
172	3.26428987762571e-06\\
173	4.08311463814045e-06\\
};
\addlegendentry{$L_3$};

\addplot [color=mycolor4,mark=square,mark repeat={20},solid,line width=1.0pt]
  table[row sep=crcr]{%
0	0.149999609940435\\
1	0.149999609940427\\
2	0.149999609940418\\
3	0.149999609940406\\
4	0.149999609940392\\
5	0.149999609940375\\
6	0.149999609940358\\
7	0.149999609940337\\
8	0.149999609940313\\
9	0.149999609940287\\
10	0.149999609940257\\
11	0.149999609940224\\
12	0.149999609940187\\
13	0.149999609940146\\
14	0.1499996099401\\
15	0.149999609940049\\
16	0.149999609939994\\
17	0.149999609939933\\
18	0.149999609939864\\
19	0.149999609939789\\
20	0.149999609939707\\
21	0.149999609939616\\
22	0.149999609939517\\
23	0.149999609939406\\
24	0.149999609939286\\
25	0.149999609939153\\
26	0.149999609939009\\
27	0.14999960993885\\
28	0.149999609938676\\
29	0.149999609938486\\
30	0.149999609938278\\
31	0.14999960993805\\
32	0.149999609937801\\
33	0.149999609937528\\
34	0.14999960993723\\
35	0.149999609936906\\
36	0.149999609936551\\
37	0.149999609936164\\
38	0.149999609935742\\
39	0.149999609935283\\
40	0.149999609934782\\
41	0.149999609934235\\
42	0.149999609933641\\
43	0.149999609932993\\
44	0.149999609932288\\
45	0.149999609931519\\
46	0.149999609930678\\
47	0.149999609929759\\
48	0.149999609928755\\
49	0.14999960992765\\
50	0.149999609926434\\
51	0.149999609925087\\
52	0.14999960992359\\
53	0.149999609921917\\
54	0.149999609920032\\
55	0.149999609917893\\
56	0.149999609915441\\
57	0.149999609912604\\
58	0.149999609909285\\
59	0.149999609905355\\
60	0.149999609900645\\
61	0.14999960989493\\
62	0.149999609887916\\
63	0.149999609879194\\
64	0.149999609868171\\
65	0.149999609853891\\
66	0.149999609834652\\
67	0.149999609807087\\
68	0.149999609763681\\
69	0.149999609684108\\
70	0.149999609489793\\
71	0.149999608428057\\
72	0.149999574406889\\
73	0.149999527184367\\
74	0.14999947448055\\
75	0.149999415815144\\
76	0.149999350537472\\
77	0.149999277903384\\
78	0.149999197076822\\
79	0.149999107121659\\
80	0.149999006990312\\
81	0.149998895510171\\
82	0.149998771367895\\
83	0.149998633091388\\
84	0.149998479029111\\
85	0.149998307326302\\
86	0.14999811589756\\
87	0.149997902395167\\
88	0.149997664172309\\
89	0.149997398240264\\
90	0.14999710121832\\
91	0.14999676927495\\
92	0.149996398058373\\
93	0.149995982614182\\
94	0.149995517287071\\
95	0.149994995602941\\
96	0.149994410126525\\
97	0.149993752288346\\
98	0.149993012172852\\
99	0.149992178257069\\
100	0.149991237085595\\
101	0.149990172862971\\
102	0.149988966937746\\
103	0.149987597142342\\
104	0.149986036938332\\
105	0.149984254297202\\
106	0.149982210218268\\
107	0.149979856735278\\
108	0.14997713416632\\
109	0.149973967272312\\
110	0.149970260057866\\
111	0.149965888201187\\
112	0.149960687365463\\
113	0.149954436292349\\
114	0.149946831432706\\
115	0.149937439277589\\
116	0.149925619539208\\
117	0.149910424442748\\
118	0.14989039541337\\
119	0.14986324902397\\
120	0.149824830377791\\
121	0.149766837658076\\
122	0.149675175670037\\
123	0.149518115252176\\
124	0.149184528177871\\
125	0.147431920071637\\
126	0.138035646738502\\
127	0.125305495684703\\
128	0.112617267157007\\
129	0.100544598431696\\
130	0.0892259306996027\\
131	0.0787191298712756\\
132	0.0690507482589951\\
133	0.0602267273336839\\
134	0.0522365828431754\\
135	0.0450562443133236\\
136	0.0386506177429765\\
137	0.0329760805034447\\
138	0.0279828889074584\\
139	0.0236174211560324\\
140	0.0198241829614975\\
141	0.0165475293431681\\
142	0.0137330840550839\\
143	0.0113288613180298\\
144	0.00928611163900959\\
145	0.00755992104735692\\
146	0.00610958681048203\\
147	0.00489877503578784\\
148	0.00389544923861151\\
149	0.00307156092339488\\
150	0.00240252376385689\\
151	0.00186654843547784\\
152	0.00144397497906387\\
153	0.00111676622683297\\
154	0.000868281414058631\\
155	0.000683335099647646\\
156	0.000548424688709476\\
157	0.000451957869979903\\
158	0.000384351750556316\\
159	0.000337963043682108\\
160	0.000306882419593227\\
161	0.000286657799417999\\
162	0.00027400803321659\\
163	0.000266569535461733\\
164	0.00026269887411949\\
165	0.000261341173467849\\
166	0.000261970790222303\\
167	0.000264621178600506\\
168	0.000270056903571126\\
169	0.000280242057206076\\
170	0.000299577116319284\\
171	0.00033861145556638\\
172	0.00042859879658412\\
173	0.000721856923941808\\
};
\addlegendentry{$L_4$};

\addplot [color=mycolor5,mark=|,mark repeat={20},mark phase={10},solid,line width=1.0pt]
  table[row sep=crcr]{%
0	0.248775404786686\\
1	0.248750468452493\\
2	0.248722765795483\\
3	0.24869199046808\\
4	0.248657802335261\\
5	0.248619823758176\\
6	0.248577635503195\\
7	0.248530772208473\\
8	0.248478717358821\\
9	0.248420897742073\\
10	0.248356677300082\\
11	0.248285350338669\\
12	0.248206134051233\\
13	0.248118160234949\\
14	0.248020466177452\\
15	0.247911984614767\\
16	0.247791532670053\\
17	0.247657799698495\\
18	0.247509333940244\\
19	0.247344527889282\\
20	0.247161602272109\\
21	0.246958588540728\\
22	0.2467333097803\\
23	0.246483359948548\\
24	0.246206081336837\\
25	0.245898540193517\\
26	0.245557500456471\\
27	0.245179395533367\\
28	0.244760298177452\\
29	0.24429588845265\\
30	0.243781419930134\\
31	0.243211684284103\\
32	0.242580974547348\\
33	0.241883047438368\\
34	0.241111085284566\\
35	0.240257658244361\\
36	0.239314687774989\\
37	0.2382734124869\\
38	0.237124357888987\\
39	0.235857311811467\\
40	0.23446130770147\\
41	0.232924618463808\\
42	0.231234763900604\\
43	0.22937853542238\\
44	0.227342042126431\\
45	0.225110782899108\\
46	0.222669749641823\\
47	0.220003567028939\\
48	0.217096674443073\\
49	0.213933555592826\\
50	0.210499020905114\\
51	0.206778546848492\\
52	0.20275867483571\\
53	0.198427470104706\\
54	0.193775037891591\\
55	0.18879409026143\\
56	0.183480552063652\\
57	0.177834188832864\\
58	0.171859233202056\\
59	0.165564979995645\\
60	0.1589663141638\\
61	0.152084130875814\\
62	0.144945604290797\\
63	0.137584261786625\\
64	0.130039824681933\\
65	0.122357785511864\\
66	0.114588706354082\\
67	0.106787243746457\\
68	0.0990109393306915\\
69	0.0913189017054299\\
70	0.0837710373445187\\
71	0.0764475300190915\\
72	0.0750070327742294\\
73	0.0750032014331885\\
74	0.0750020710860535\\
75	0.0750015196729807\\
76	0.0750011843913328\\
77	0.075000951630431\\
78	0.0750007742017553\\
79	0.0750006288104771\\
80	0.075000502466307\\
81	0.0750003871976019\\
82	0.075000277693732\\
83	0.0750001701398733\\
84	0.0750000615919578\\
85	0.0749999496171337\\
86	0.0749998320732512\\
87	0.0749997069646969\\
88	0.0749995723414936\\
89	0.0749994262231702\\
90	0.0749992665364602\\
91	0.0749990910599117\\
92	0.0749988973706768\\
93	0.0749986827899242\\
94	0.0749984443238243\\
95	0.074998178597207\\
96	0.0749978817767714\\
97	0.0749975494802549\\
98	0.0749971766671618\\
99	0.0749967575054643\\
100	0.0749962852070133\\
101	0.074995751822063\\
102	0.0749951479799559\\
103	0.0749944625579889\\
104	0.0749936822531745\\
105	0.0749927910218781\\
106	0.0749917693381339\\
107	0.0749905931964016\\
108	0.0749892327360812\\
109	0.0749876503206566\\
110	0.0749857979384297\\
111	0.0749836134161493\\
112	0.0749810145732342\\
113	0.074977890769772\\
114	0.0749740902226993\\
115	0.0749693961713988\\
116	0.074963488466027\\
117	0.0749558932116719\\
118	0.0749458811176802\\
119	0.074932310469122\\
120	0.0749131038151866\\
121	0.0748841102668583\\
122	0.0748382822446994\\
123	0.0747597551102808\\
124	0.0745929647473202\\
125	0.0737166640608854\\
126	0.069018530964004\\
127	0.0626534591714886\\
128	0.0563093487600618\\
129	0.0502730183615566\\
130	0.0446136885732798\\
131	0.0393602923553715\\
132	0.034526105870035\\
133	0.0301140998593545\\
134	0.0261190322043006\\
135	0.0225288676753929\\
136	0.0193260592803742\\
137	0.0164887957141183\\
138	0.0139922051432459\\
139	0.0118094766796799\\
140	0.00991286319229613\\
141	0.00827454220492937\\
142	0.00686732561047846\\
143	0.00566522053721323\\
144	0.0046438522587995\\
145	0.0037807638124492\\
146	0.0030556038564502\\
147	0.00245020547020152\\
148	0.00194855043688605\\
149	0.00153661453307666\\
150	0.00120210461919332\\
151	0.000934126058298739\\
152	0.000722848902582548\\
153	0.000559254612077593\\
154	0.000435022865336822\\
155	0.000342561022467124\\
156	0.000275117887102422\\
157	0.000226897425068619\\
158	0.000193108332764348\\
159	0.000169929134655429\\
160	0.000154405365970725\\
161	0.000144311231554263\\
162	0.000138006462819512\\
163	0.000134309662004355\\
164	0.000132399636542116\\
165	0.0001317496625166\\
166	0.000132097924282558\\
167	0.000133462623585953\\
168	0.000136228316467642\\
169	0.000141380798170889\\
170	0.000151127097020543\\
171	0.000170756035002898\\
172	0.000215931552293048\\
173	0.000362970041770101\\
};
\addlegendentry{$L_5$};

\addplot [color=mycolor6,mark=Mercedes star,mark repeat={20},solid,line width=1.0pt]
  table[row sep=crcr]{%
0	0.248775404787561\\
1	0.248750468450858\\
2	0.248722765793433\\
3	0.248691990469179\\
4	0.248657802336782\\
5	0.248619823759006\\
6	0.248577635502814\\
7	0.248530772207658\\
8	0.248478717358206\\
9	0.24842089774373\\
10	0.248356677299755\\
11	0.248285350339182\\
12	0.24820613405057\\
13	0.2481181602351\\
14	0.248020466176241\\
15	0.247911984614285\\
16	0.247791532669461\\
17	0.247657799698059\\
18	0.247509333940135\\
19	0.247344527889724\\
20	0.247161602273133\\
21	0.246958588540166\\
22	0.246733309780079\\
23	0.246483359948514\\
24	0.246206081335993\\
25	0.245898540193988\\
26	0.245557500455545\\
27	0.245179395533365\\
28	0.244760298177704\\
29	0.244295888453084\\
30	0.243781419930351\\
31	0.243211684283537\\
32	0.242580974547065\\
33	0.24188304743863\\
34	0.241111085284525\\
35	0.240257658244301\\
36	0.239314687774194\\
37	0.238273412487449\\
38	0.237124357890282\\
39	0.23585731181151\\
40	0.234461307702449\\
41	0.23292461846369\\
42	0.231234763901101\\
43	0.229378535422331\\
44	0.227342042126271\\
45	0.225110782900786\\
46	0.222669749642958\\
47	0.22000356703015\\
48	0.217096674443509\\
49	0.213933555593276\\
50	0.210499020905496\\
51	0.206778546848424\\
52	0.202758674835363\\
53	0.198427470104849\\
54	0.193775037892004\\
55	0.18879409026119\\
56	0.183480552063387\\
57	0.177834188832437\\
58	0.17185923320165\\
59	0.165564979995826\\
60	0.158966314163643\\
61	0.152084130875702\\
62	0.144945604290679\\
63	0.137584261786647\\
64	0.130039824681928\\
65	0.122357785511889\\
66	0.114588706354119\\
67	0.106787243746495\\
68	0.0990109393306872\\
69	0.0913189017054242\\
70	0.0837710373445188\\
71	0.0764475300191484\\
72	0.0750070327742581\\
73	0.0750032014331871\\
74	0.0750020710860537\\
75	0.0750015196729814\\
76	0.075001184391333\\
77	0.0750009516304316\\
78	0.0750007742017555\\
79	0.0750006288104771\\
80	0.0750005024663066\\
81	0.0750003871976012\\
82	0.0750002776937307\\
83	0.0750001701398721\\
84	0.0750000615919568\\
85	0.0749999496171329\\
86	0.0749998320732505\\
87	0.0749997069646978\\
88	0.0749995723414959\\
89	0.0749994262231734\\
90	0.0749992665364644\\
91	0.0749990910599151\\
92	0.0749988973706778\\
93	0.0749986827899208\\
94	0.0749984443238136\\
95	0.074998178597184\\
96	0.0749978817767309\\
97	0.0749975494801922\\
98	0.0749971766670736\\
99	0.0749967575053481\\
100	0.0749962852068795\\
101	0.0749957518219486\\
102	0.0749951479799458\\
103	0.0749944625582001\\
104	0.0749936822537049\\
105	0.0749927910227715\\
106	0.0749917693394076\\
107	0.0749905931981607\\
108	0.0749892327386596\\
109	0.0749876503242057\\
110	0.074985797941465\\
111	0.0749836134162178\\
112	0.0749810145698365\\
113	0.0749778907644937\\
114	0.0749740902174948\\
115	0.0749693961659316\\
116	0.0749634884597865\\
117	0.074955893211472\\
118	0.0749458811303687\\
119	0.0749323104981055\\
120	0.0749131038621377\\
121	0.0748841102969358\\
122	0.0748382822104903\\
123	0.0747597550491709\\
124	0.074592964773758\\
125	0.0737166639923364\\
126	0.0690185308976965\\
127	0.0626534591565199\\
128	0.056309348758738\\
129	0.0502730183608315\\
130	0.0446136885705684\\
131	0.0393602923502182\\
132	0.0345261058619337\\
133	0.0301140998476099\\
134	0.0261190321883924\\
135	0.0225288676552755\\
136	0.0193260592565511\\
137	0.016488795687538\\
138	0.0139922051151141\\
139	0.0118094766512513\\
140	0.00991286316468337\\
141	0.00827454217896149\\
142	0.00686732558663099\\
143	0.00566522051560697\\
144	0.00464385223924553\\
145	0.00378076379452803\\
146	0.00305560383963448\\
147	0.00245020545402877\\
148	0.00194855042114845\\
149	0.00153661451795372\\
150	0.00120210460525324\\
151	0.000934126046348576\\
152	0.000722848893420753\\
153	0.000559254606250635\\
154	0.000435022862996099\\
155	0.00034256102336885\\
156	0.000275117890723626\\
157	0.000226897430768714\\
158	0.000193108339924063\\
159	0.000169929142758308\\
160	0.000154405374627184\\
161	0.000144311240486381\\
162	0.000138006471826668\\
163	0.000134309670923867\\
164	0.00013239964521209\\
165	0.000131749670742558\\
166	0.00013209793180497\\
167	0.000133462630041309\\
168	0.000136228321329159\\
169	0.000141380800640829\\
170	0.000151127095807723\\
171	0.000170756027792974\\
172	0.000215931534168698\\
173	0.000362969996809847\\
};
\addlegendentry{$\text{L}_\text{6}$};

\end{axis}
\end{tikzpicture}%

%% file: flow_flows_2.tikz
%
%
\definecolor{mycolor1}{HTML}{4180FF}%
\definecolor{mycolor2}{HTML}{FF2600}%
\definecolor{mycolor3}{HTML}{008F00}%
\definecolor{mycolor4}{rgb}{0.49400,0.18400,0.55600}%
\definecolor{mycolor5}{rgb}{0.92900,0.69400,0.12500}%
\definecolor{mycolor6}{rgb}{0.30100,0.74500,0.93300}%
\definecolor{mycolor7}{rgb}{0.63500,0.07800,0.18400}%
\definecolor{mycolor8}{rgb}{0,0,0}%
\begin{tikzpicture}

\begin{axis}[%
width=0.85\columnwidth,
height=0.48\columnwidth,
at={(0.772in,0.484in)},
scale only axis,
xmin=0,
xmax=180,
xticklabels={,,},
xmajorgrids,
ymin=0,
ymax=0.14,
ylabel={Individual flow rate},
axis background/.style={fill=white},
scaled ticks=false, tick label style={/pgf/number format/fixed},
legend style={legend cell align=left,draw=white!15!black,
	at={(0.98,0.98)}, anchor=north east}
]
\addplot [color=mycolor1,mark=star,mark repeat={20},solid,line width=1.0pt]
  table[row sep=crcr]{%
0	0.127245410618718\\
1	0.127221134424656\\
2	0.127194171252168\\
3	0.127164214857437\\
4	0.127130936304678\\
5	0.127093975676451\\
6	0.127052932822235\\
7	0.127007358259124\\
8	0.126956746879365\\
9	0.12690053537617\\
10	0.126838100451079\\
11	0.126768753759438\\
12	0.126691733794748\\
13	0.126606197739793\\
14	0.126511214546484\\
15	0.126405757782749\\
16	0.126288694389033\\
17	0.126158764905749\\
18	0.126014558476508\\
19	0.125854494983964\\
20	0.125676816284335\\
21	0.125479577361769\\
22	0.12526063555477\\
23	0.125017640511725\\
24	0.124748023141494\\
25	0.124448977609243\\
26	0.12411743171856\\
27	0.123750007834697\\
28	0.123342983109473\\
29	0.122892256795773\\
30	0.122393323443804\\
31	0.121841230341291\\
32	0.121230479648215\\
33	0.120554949545917\\
34	0.119807931772585\\
35	0.118982126025309\\
36	0.118069563459877\\
37	0.117061571882537\\
38	0.115948840345403\\
39	0.114721615583344\\
40	0.113369927991376\\
41	0.111883616639038\\
42	0.110252108419185\\
43	0.108464166737926\\
44	0.106507734746142\\
45	0.104369774940862\\
46	0.102036451822134\\
47	0.0994940068716688\\
48	0.0967289625522193\\
49	0.0937274040096917\\
50	0.0904747108310421\\
51	0.0869573990291085\\
52	0.0831680552365331\\
53	0.0791090953070626\\
54	0.0747902949326147\\
55	0.0702245089989606\\
56	0.0654271824207195\\
57	0.0604195273310382\\
58	0.0552345851126162\\
59	0.0499169132251453\\
60	0.0445189860695234\\
61	0.039113862656968\\
62	0.03381279313053\\
63	0.0287446051642284\\
64	0.0240102135508194\\
65	0.0196593568506453\\
66	0.0156884021281124\\
67	0.012063762570244\\
68	0.00876092849280428\\
69	0.00575591466925898\\
70	0.00300700462907146\\
71	0.000488698007723923\\
72	3.48168619441487e-06\\
73	1.65837518311303e-06\\
74	1.12673156529158e-06\\
75	8.74961009907479e-07\\
76	7.28924816372598e-07\\
77	6.3410257350329e-07\\
78	5.67952826808418e-07\\
79	5.19467546877658e-07\\
80	4.82632591201081e-07\\
81	4.53884160201489e-07\\
82	4.30976052655235e-07\\
83	4.1242210230238e-07\\
84	3.97199661253344e-07\\
85	3.84581868745955e-07\\
86	3.74037855780428e-07\\
87	3.65170815026988e-07\\
88	3.57678171759113e-07\\
89	3.51325170786886e-07\\
90	3.45926887336681e-07\\
91	3.41335685409387e-07\\
92	3.37432291204595e-07\\
93	3.34119321267085e-07\\
94	3.31316511979118e-07\\
95	3.28957150165119e-07\\
96	3.26985365878283e-07\\
97	3.25354053503751e-07\\
98	3.24023257097154e-07\\
99	3.22958903135774e-07\\
100	3.22131796563427e-07\\
101	3.21516818611526e-07\\
102	3.21092279815675e-07\\
103	3.20839392725415e-07\\
104	3.20741839527597e-07\\
105	3.20785417340237e-07\\
106	3.20957745763737e-07\\
107	3.21248023124337e-07\\
108	3.21646825087743e-07\\
109	3.22145938688051e-07\\
110	3.22738213986286e-07\\
111	3.23417428533264e-07\\
112	3.24178187262108e-07\\
113	3.25015840780867e-07\\
114	3.25926320870315e-07\\
115	3.26906170531749e-07\\
116	3.27953030440459e-07\\
117	3.29064934287274e-07\\
118	3.30239814977501e-07\\
119	3.31475856086506e-07\\
120	3.32772662940715e-07\\
121	3.34134361830852e-07\\
122	3.35561343897233e-07\\
123	3.37040209819724e-07\\
124	3.38613390736804e-07\\
125	3.40218207308165e-07\\
126	3.41944167942299e-07\\
127	3.4375894128056e-07\\
128	3.45624010694265e-07\\
129	3.47540148533655e-07\\
130	3.49510619610521e-07\\
131	3.51538387593128e-07\\
132	3.53626449294783e-07\\
133	3.55778018221678e-07\\
134	3.57996597250064e-07\\
135	3.60286019280375e-07\\
136	3.62650481638073e-07\\
137	3.65094581638895e-07\\
138	3.67623356586813e-07\\
139	3.70242332374296e-07\\
140	3.72957587105916e-07\\
141	3.75775836294321e-07\\
142	3.7870454001408e-07\\
143	3.81752017449971e-07\\
144	3.84927534483446e-07\\
145	3.88241319417874e-07\\
146	3.917044809579e-07\\
147	3.95328862038658e-07\\
148	3.99126945609942e-07\\
149	4.03111985882223e-07\\
150	4.0729851819929e-07\\
151	4.11703286910551e-07\\
152	4.16346470795481e-07\\
153	4.21252974815095e-07\\
154	4.26453576977143e-07\\
155	4.31985871213068e-07\\
156	4.37895146533852e-07\\
157	4.44235490372964e-07\\
158	4.5107146428875e-07\\
159	4.58480711453285e-07\\
160	4.66557887568007e-07\\
161	4.75420427410189e-07\\
162	4.85216938201809e-07\\
163	4.96139559249846e-07\\
164	5.08442665579722e-07\\
165	5.22472295092611e-07\\
166	5.38714727263009e-07\\
167	5.57881416347324e-07\\
168	5.81068263474338e-07\\
169	6.10082230419942e-07\\
170	6.48196763652742e-07\\
171	7.02229233580436e-07\\
172	7.90071148170588e-07\\
173	9.87748462559185e-07\\
};
\addlegendentry{$u_t^{11}$};

\addplot [color=mycolor2,mark=x,mark repeat={20},mark phase={5},solid,line width=1.0pt]
  table[row sep=crcr]{%
0	0.0465302662958487\\
1	0.0465296049853614\\
2	0.0465288641993443\\
3	0.0465280438256918\\
4	0.0465271326427936\\
5	0.0465261129128098\\
6	0.0465249655348348\\
7	0.046523674608608\\
8	0.0465222287050559\\
9	0.0465206178882862\\
10	0.0465188293692314\\
11	0.046516845770353\\
12	0.0465146457524908\\
13	0.0465122038910548\\
14	0.0465094884788861\\
15	0.0465064586354604\\
16	0.0465030644897027\\
17	0.04649925479628\\
18	0.0464949885904759\\
19	0.0464902384088515\\
20	0.0464849830439688\\
21	0.0464791988763708\\
22	0.0464728515600017\\
23	0.0464658852975956\\
24	0.0464582113592261\\
25	0.0464497017030421\\
26	0.0464401923246745\\
27	0.0464294941224483\\
28	0.0464174025323993\\
29	0.0464036981929325\\
30	0.0463881399373737\\
31	0.0463704719459063\\
32	0.0463504848772359\\
33	0.0463280570333334\\
34	0.0463030787550225\\
35	0.0462754202467954\\
36	0.0462449715325667\\
37	0.0462116431366279\\
38	0.0461752712212603\\
39	0.0461353965835149\\
40	0.0460910219835009\\
41	0.0460405809630349\\
42	0.0459821661551662\\
43	0.0459138053120966\\
44	0.0458336641590362\\
45	0.0457402789163994\\
46	0.0456324768754501\\
47	0.0455086412040553\\
48	0.0453666889012891\\
49	0.0452050187379498\\
50	0.0450230619137235\\
51	0.0448197794301963\\
52	0.0445891268283851\\
53	0.0443167545011917\\
54	0.043982993291659\\
55	0.0435677019992524\\
56	0.043051362536254\\
57	0.0424125306644629\\
58	0.041622400394988\\
59	0.0406457123109058\\
60	0.0394448807056338\\
61	0.0379677460515907\\
62	0.0361302372726402\\
63	0.0338370597363024\\
64	0.0310270271260812\\
65	0.0276959031183876\\
66	0.0238978973142577\\
67	0.019721279003358\\
68	0.0152481547408229\\
69	0.0105617671303619\\
70	0.00576434111639287\\
71	0.000971488007701984\\
72	6.96037339554107e-06\\
73	3.31601976195562e-06\\
74	2.25312060697082e-06\\
75	1.74971416540717e-06\\
76	1.45770488296421e-06\\
77	1.26809537111667e-06\\
78	1.13581745195339e-06\\
79	1.03886122265587e-06\\
80	9.65201357064897e-07\\
81	9.07711829699478e-07\\
82	8.61901141226638e-07\\
83	8.24797509794397e-07\\
84	7.9435599192373e-07\\
85	7.69123100720299e-07\\
86	7.480372597606e-07\\
87	7.30304968932543e-07\\
88	7.15321161981071e-07\\
89	7.02616390245922e-07\\
90	6.91820850869707e-07\\
91	6.8263930771827e-07\\
92	6.74833241174184e-07\\
93	6.68207906425271e-07\\
94	6.62602793900872e-07\\
95	6.57884491497234e-07\\
96	6.53941270922101e-07\\
97	6.50678930460442e-07\\
98	6.48017566070225e-07\\
99	6.45889037210499e-07\\
100	6.44234959162874e-07\\
101	6.43005098871391e-07\\
102	6.42156081104482e-07\\
103	6.41650333985594e-07\\
104	6.41455224316216e-07\\
105	6.41542348220392e-07\\
106	6.41886946275888e-07\\
107	6.42467415964205e-07\\
108	6.43264908795719e-07\\
109	6.44262998275564e-07\\
110	6.45447383120997e-07\\
111	6.46805615916462e-07\\
112	6.48326902501879e-07\\
113	6.50001939436344e-07\\
114	6.518225860476e-07\\
115	6.5378191811842e-07\\
116	6.55875187095399e-07\\
117	6.58098402651766e-07\\
118	6.60447365193543e-07\\
119	6.62918533287791e-07\\
120	6.65511312773579e-07\\
121	6.68233494572335e-07\\
122	6.71084898823458e-07\\
123	6.74040516964689e-07\\
124	6.77184369034536e-07\\
125	6.80389393241554e-07\\
126	6.83836368862884e-07\\
127	6.87460289873202e-07\\
128	6.9118394458868e-07\\
129	6.95008728343782e-07\\
130	6.98941070187602e-07\\
131	7.02986797680913e-07\\
132	7.07151794585352e-07\\
133	7.1144236913905e-07\\
134	7.15865402908274e-07\\
135	7.2042843685604e-07\\
136	7.25139747236086e-07\\
137	7.30008427910144e-07\\
138	7.35044488598424e-07\\
139	7.40258982773131e-07\\
140	7.45664185548769e-07\\
141	7.51273839323926e-07\\
142	7.57103457700874e-07\\
143	7.63170617331357e-07\\
144	7.6949508908504e-07\\
145	7.76098620467912e-07\\
146	7.83004260692135e-07\\
147	7.90235360114425e-07\\
148	7.97814705298912e-07\\
149	8.05764475831179e-07\\
150	8.14107618407922e-07\\
151	8.22870765752067e-07\\
152	8.32088173692912e-07\\
153	8.41805690192864e-07\\
154	8.52083830558881e-07\\
155	8.62999614557342e-07\\
156	8.74647578098605e-07\\
157	8.87140900678523e-07\\
158	9.00613735769764e-07\\
159	9.15225729069028e-07\\
160	9.31169604699865e-07\\
161	9.4868280002565e-07\\
162	9.68064601259951e-07\\
163	9.89701296504404e-07\\
164	1.01410395516448e-06\\
165	1.04196749934745e-06\\
166	1.07426790224955e-06\\
167	1.11243196736851e-06\\
168	1.15865575220028e-06\\
169	1.21655775824335e-06\\
170	1.29268994920693e-06\\
171	1.40069313474305e-06\\
172	1.57635809196245e-06\\
173	1.9717399476902e-06\\
};
\addlegendentry{$u_t^{12}$};

\addplot [color=mycolor3,mark=Mercedes star,mark repeat={20},mark phase={10},solid,line width=1.0pt]
  table[row sep=crcr]{%
0	0.0465302662959088\\
1	0.0465296049854215\\
2	0.0465288641994038\\
3	0.0465280438257514\\
4	0.0465271326428532\\
5	0.0465261129128694\\
6	0.0465249655348945\\
7	0.0465236746086682\\
8	0.0465222287051156\\
9	0.0465206178883462\\
10	0.0465188293692914\\
11	0.046516845770413\\
12	0.046514645752551\\
13	0.0465122038911156\\
14	0.0465094884789458\\
15	0.0465064586355202\\
16	0.0465030644897626\\
17	0.0464992547963396\\
18	0.0464949885905352\\
19	0.0464902384089111\\
20	0.0464849830440284\\
21	0.0464791988764305\\
22	0.0464728515600615\\
23	0.0464658852976555\\
24	0.0464582113592862\\
25	0.0464497017031017\\
26	0.0464401923247341\\
27	0.0464294941225085\\
28	0.0464174025324593\\
29	0.0464036981929928\\
30	0.0463881399374334\\
31	0.0463704719459668\\
32	0.0463504848772958\\
33	0.0463280570333957\\
34	0.0463030787550885\\
35	0.0462754202468549\\
36	0.0462449715326262\\
37	0.0462116431366875\\
38	0.0461752712213202\\
39	0.0461353965835748\\
40	0.0460910219835607\\
41	0.0460405809630946\\
42	0.0459821661552261\\
43	0.0459138053121567\\
44	0.0458336641590965\\
45	0.0457402789164588\\
46	0.0456324768755101\\
47	0.0455086412041151\\
48	0.0453666889013153\\
49	0.0452050187380095\\
50	0.0450230619137833\\
51	0.0448197794302558\\
52	0.0445891268284452\\
53	0.0443167545012196\\
54	0.0439829932917104\\
55	0.0435677019993124\\
56	0.0430513625363138\\
57	0.0424125306645227\\
58	0.0416224003950477\\
59	0.0406457123109651\\
60	0.0394448807056935\\
61	0.0379677460516506\\
62	0.0361302372727016\\
63	0.033837059736362\\
64	0.0310270271261413\\
65	0.0276959031184478\\
66	0.0238978973143174\\
67	0.0197212790034178\\
68	0.0152481547408826\\
69	0.010561767130422\\
70	0.00576434111645323\\
71	0.000971488007763577\\
72	6.96037346675879e-06\\
73	3.31601984796939e-06\\
74	2.25312070946847e-06\\
75	1.74971428523902e-06\\
76	1.45770501985823e-06\\
77	1.26809552315932e-06\\
78	1.13581761691042e-06\\
79	1.03886139734618e-06\\
80	9.65201538455749e-07\\
81	9.07712014667131e-07\\
82	8.61901327422253e-07\\
83	8.24797695321047e-07\\
84	7.94356175463509e-07\\
85	7.69123281435969e-07\\
86	7.4803743723361e-07\\
87	7.30305142976816e-07\\
88	7.15321332582141e-07\\
89	7.02616557537864e-07\\
90	6.91821015030764e-07\\
91	6.82639468989922e-07\\
92	6.74833399805867e-07\\
93	6.68208062677415e-07\\
94	6.62602948017626e-07\\
95	6.57884643720081e-07\\
96	6.53941421478726e-07\\
97	6.50679079561155e-07\\
98	6.48017713911948e-07\\
99	6.45889183968613e-07\\
100	6.44235105005571e-07\\
101	6.43005243947561e-07\\
102	6.42156225553155e-07\\
103	6.41650477933201e-07\\
104	6.41455367878107e-07\\
105	6.41542491502785e-07\\
106	6.41887089376444e-07\\
107	6.42467558972524e-07\\
108	6.43265051794184e-07\\
109	6.44263141340837e-07\\
110	6.45447526324029e-07\\
111	6.46805759323363e-07\\
112	6.48327046174446e-07\\
113	6.50002083432686e-07\\
114	6.51822730422127e-07\\
115	6.53782062922977e-07\\
116	6.55875332380426e-07\\
117	6.58098548467632e-07\\
118	6.60447511589225e-07\\
119	6.6291868030168e-07\\
120	6.65511460437539e-07\\
121	6.68233642935681e-07\\
122	6.71085047964601e-07\\
123	6.7404066690835e-07\\
124	6.77184519801111e-07\\
125	6.80389544895418e-07\\
126	6.83836521473257e-07\\
127	6.87460443488323e-07\\
128	6.91184099241796e-07\\
129	6.95008884066698e-07\\
130	6.98941227012807e-07\\
131	7.02986955642169e-07\\
132	7.07151953718029e-07\\
133	7.11442529480368e-07\\
134	7.1586556449752e-07\\
135	7.20428599734773e-07\\
136	7.25139911448353e-07\\
137	7.30008593502722e-07\\
138	7.35044655621103e-07\\
139	7.4025915127903e-07\\
140	7.45664355594707e-07\\
141	7.51274010970807e-07\\
142	7.57103631014127e-07\\
143	7.63170792381403e-07\\
144	7.6949526594783e-07\\
145	7.76098799225511e-07\\
146	7.8300444143335e-07\\
147	7.90235542935426e-07\\
148	7.97814890303858e-07\\
149	8.05764663132947e-07\\
150	8.14107808129126e-07\\
151	8.22870958026605e-07\\
152	8.32088368668218e-07\\
153	8.41805888032944e-07\\
154	8.52084031448055e-07\\
155	8.62999818704622e-07\\
156	8.74647785742658e-07\\
157	8.8714111209338e-07\\
158	9.00613951271653e-07\\
159	9.15225949024918e-07\\
160	9.31169829538624e-07\\
161	9.48683030253055e-07\\
162	9.68064837479361e-07\\
163	9.8970153944583e-07\\
164	1.01410420572643e-06\\
165	1.04196775865897e-06\\
166	1.07426817176572e-06\\
167	1.11243224902474e-06\\
168	1.15865604867733e-06\\
169	1.21655807346096e-06\\
170	1.29269028934967e-06\\
171	1.40069351076092e-06\\
172	1.57635852746456e-06\\
173	1.97174052088514e-06\\
};
\addlegendentry{$u_t^{13}$};

\addplot [color=mycolor4,mark=square,mark repeat={20},solid,line width=1.0pt]
  table[row sep=crcr]{%
0	0.0284694509986931\\
1	0.0284701123091207\\
2	0.0284708530952185\\
3	0.0284716734688295\\
4	0.0284725846517063\\
5	0.0284736043817203\\
6	0.0284747517596282\\
7	0.0284760426858671\\
8	0.02847748858937\\
9	0.0284790994060793\\
10	0.0284808879251694\\
11	0.0284828715240454\\
12	0.0284850715418544\\
13	0.0284875134033046\\
14	0.0284902288153279\\
15	0.0284932586587505\\
16	0.028496652804549\\
17	0.0285004624978692\\
18	0.028504728703647\\
19	0.0285094788852064\\
20	0.0285147342500131\\
21	0.0285205184174763\\
22	0.0285268657337725\\
23	0.0285338319960963\\
24	0.0285415059343568\\
25	0.0285500155903674\\
26	0.0285595249687525\\
27	0.0285702231708225\\
28	0.0285823147606607\\
29	0.0285960190999647\\
30	0.0286115773553341\\
31	0.0286292453465944\\
32	0.0286492324151897\\
33	0.028671660258379\\
34	0.0286966385368444\\
35	0.0287242970448619\\
36	0.0287547457587896\\
37	0.0287880741544321\\
38	0.0288244460693895\\
39	0.028864320706744\\
40	0.0289086953063798\\
41	0.0289591363264332\\
42	0.0290175511337525\\
43	0.0290859119762362\\
44	0.0291660531286897\\
45	0.0292594383706707\\
46	0.0293672404110091\\
47	0.0294910760815297\\
48	0.0296330283833548\\
49	0.0297946985458131\\
50	0.0299766553689394\\
51	0.0301799378513443\\
52	0.0304105904518479\\
53	0.0306829627774847\\
54	0.0310167239853099\\
55	0.0314320152757572\\
56	0.0319483547364961\\
57	0.0325871866057072\\
58	0.0333773168720371\\
59	0.034354004952439\\
60	0.0355548365532649\\
61	0.0370319712018364\\
62	0.0388694799741167\\
63	0.0411626575020374\\
64	0.0439726901017373\\
65	0.0473038140956468\\
66	0.051101819881247\\
67	0.0552784381655543\\
68	0.0597515623863168\\
69	0.0644379499204607\\
70	0.0692353757499527\\
71	0.0740282279059007\\
72	0.0749927309495436\\
73	0.0749963411097653\\
74	0.0749973658060189\\
75	0.0749978266767885\\
76	0.0749980713503344\\
77	0.0749982082854593\\
78	0.0749982819439873\\
79	0.0749983136563505\\
80	0.0749983146871732\\
81	0.0749982913107179\\
82	0.0749982470642265\\
83	0.0749981838495725\\
84	0.0749981025110196\\
85	0.0749980031531969\\
86	0.0749978853205692\\
87	0.0749977480980308\\
88	0.0749975901635581\\
89	0.0749974098095555\\
90	0.0749972049419761\\
91	0.0749969730620613\\
92	0.0749967112329602\\
93	0.0749964160317968\\
94	0.0749960834864713\\
95	0.0749957089953601\\
96	0.0749952872269405\\
97	0.0749948119950929\\
98	0.0749942761042145\\
99	0.0749936711562201\\
100	0.0749929873088292\\
101	0.0749922129709066\\
102	0.0749913344150174\\
103	0.0749903352787553\\
104	0.0749891959155714\\
105	0.074987892541661\\
106	0.0749863961009861\\
107	0.0749846707290871\\
108	0.0749826716485595\\
109	0.0749803422576622\\
110	0.0749776099896603\\
111	0.0749743802494792\\
112	0.0749705275967543\\
113	0.0749658830017049\\
114	0.0749602116547968\\
115	0.0749531748573262\\
116	0.0749442846333541\\
117	0.074932826305638\\
118	0.0749177484529342\\
119	0.0748972568174682\\
120	0.0748679318685067\\
121	0.074824055346746\\
122	0.0747579799090784\\
123	0.0746410113812627\\
124	0.0744253461256167\\
125	0.073301901553034\\
126	0.068566274028743\\
127	0.0623160725144034\\
128	0.0560367023918095\\
129	0.0500436923506226\\
130	0.0444199390666752\\
131	0.0391985251461468\\
132	0.0343940462584685\\
133	0.0300099057533146\\
134	0.0260409919501159\\
135	0.0224752965290072\\
136	0.019295268039869\\
137	0.0164790846182346\\
138	0.0140018664180321\\
139	0.0118368026681511\\
140	0.00995616134314349\\
141	0.0083321568459838\\
142	0.00693766404235002\\
143	0.00574678034621774\\
144	0.00473524593527441\\
145	0.00388073219402023\\
146	0.00316300286545678\\
147	0.00256394923461283\\
148	0.00206750675187952\\
149	0.00165947514432556\\
150	0.00132728005812816\\
151	0.00105972400998363\\
152	0.000846773362364587\\
153	0.000679414019239601\\
154	0.000549583195795126\\
155	0.000450157097975824\\
156	0.000374958227368225\\
157	0.000318748448787741\\
158	0.000277189732291143\\
159	0.000246772117176352\\
160	0.000224719758391328\\
161	0.000208889414150875\\
162	0.000197673959726052\\
163	0.000189919683245801\\
164	0.000184862761961961\\
165	0.000182089291009061\\
166	0.000181526231712698\\
167	0.000183480476086037\\
168	0.000188767756950519\\
169	0.000199038377975863\\
170	0.000217609417715398\\
171	0.000251892053781247\\
172	0.0003226265278838\\
173	0.000525242472940518\\
};
\addlegendentry{$u_t^{14}$};

\addplot [color=mycolor5,mark=diamond,mark repeat={20},mark phase={15},solid,line width=1.0pt]
  table[row sep=crcr]{%
0	0.0465302662958546\\
1	0.0465296049854445\\
2	0.0465288641992996\\
3	0.0465280438256936\\
4	0.0465271326428103\\
5	0.0465261129127635\\
6	0.0465249655348646\\
7	0.0465236746085949\\
8	0.0465222287050866\\
9	0.046520617888373\\
10	0.0465188293692307\\
11	0.0465168457703159\\
12	0.0465146457524862\\
13	0.0465122038909789\\
14	0.0465094884789667\\
15	0.0465064586354851\\
16	0.0465030644896002\\
17	0.0464992547962518\\
18	0.0464949885904036\\
19	0.0464902384087833\\
20	0.0464849830439115\\
21	0.0464791988763994\\
22	0.0464728515600162\\
23	0.0464658852975974\\
24	0.0464582113592415\\
25	0.0464497017031508\\
26	0.0464401923245841\\
27	0.0464294941223938\\
28	0.0464174025324432\\
29	0.0464036981929846\\
30	0.0463881399374518\\
31	0.0463704719460106\\
32	0.0463504848771523\\
33	0.0463280570339564\\
34	0.0463030787550732\\
35	0.0462754202467549\\
36	0.0462449715325374\\
37	0.0462116431365642\\
38	0.0461752712212834\\
39	0.0461353965835511\\
40	0.0460910219834805\\
41	0.0460405809629522\\
42	0.0459821661551564\\
43	0.045913805312145\\
44	0.0458336641590994\\
45	0.0457402789164653\\
46	0.0456324768753642\\
47	0.0455086412040886\\
48	0.045366688901382\\
49	0.0452050187379684\\
50	0.0450230619137858\\
51	0.0448197794301635\\
52	0.0445891268283155\\
53	0.0443167545011533\\
54	0.0439829932916523\\
55	0.0435677019992565\\
56	0.0430513625362594\\
57	0.0424125306644009\\
58	0.0416224003949788\\
59	0.0406457123108752\\
60	0.0394448807055941\\
61	0.0379677460515929\\
62	0.036130237272619\\
63	0.0338370597363326\\
64	0.0310270271260608\\
65	0.0276959031183911\\
66	0.0238978973142549\\
67	0.0197212790033617\\
68	0.0152481547408188\\
69	0.010561767130362\\
70	0.00576434111639287\\
71	0.000971488007701978\\
72	6.96037339554107e-06\\
73	3.31601976195562e-06\\
74	2.25312060697082e-06\\
75	1.74971416540717e-06\\
76	1.45770488296421e-06\\
77	1.26809537111667e-06\\
78	1.13581745195339e-06\\
79	1.03886122265587e-06\\
80	9.65201357064897e-07\\
81	9.07711829699478e-07\\
82	8.61901141226638e-07\\
83	8.24797509794397e-07\\
84	7.9435599192373e-07\\
85	7.69123100720299e-07\\
86	7.480372597606e-07\\
87	7.30304968932543e-07\\
88	7.15321161981071e-07\\
89	7.02616390245922e-07\\
90	6.91820850869707e-07\\
91	6.8263930771827e-07\\
92	6.74833241174184e-07\\
93	6.68207906425271e-07\\
94	6.62602793900872e-07\\
95	6.57884491497234e-07\\
96	6.53941270922101e-07\\
97	6.50678930460442e-07\\
98	6.48017566070225e-07\\
99	6.45889037210499e-07\\
100	6.44234959162874e-07\\
101	6.43005098871391e-07\\
102	6.42156081104482e-07\\
103	6.41650333985594e-07\\
104	6.41455224316216e-07\\
105	6.41542348220392e-07\\
106	6.41886946275888e-07\\
107	6.42467415964205e-07\\
108	6.43264908795719e-07\\
109	6.44262998275564e-07\\
110	6.45447383120997e-07\\
111	6.46805615916462e-07\\
112	6.48326902501879e-07\\
113	6.50001939436344e-07\\
114	6.518225860476e-07\\
115	6.5378191811842e-07\\
116	6.55875187095399e-07\\
117	6.58098402651766e-07\\
118	6.60447365193543e-07\\
119	6.62918533287791e-07\\
120	6.65511312773579e-07\\
121	6.68233494572335e-07\\
122	6.71084898823458e-07\\
123	6.74040516964689e-07\\
124	6.77184369034536e-07\\
125	6.80389393241554e-07\\
126	6.83836368862884e-07\\
127	6.87460289873202e-07\\
128	6.9118394458868e-07\\
129	6.95008728343782e-07\\
130	6.98941070187602e-07\\
131	7.02986797680913e-07\\
132	7.07151794585352e-07\\
133	7.1144236913905e-07\\
134	7.15865402908274e-07\\
135	7.2042843685604e-07\\
136	7.25139747236086e-07\\
137	7.30008427910144e-07\\
138	7.35044488598424e-07\\
139	7.40258982773131e-07\\
140	7.45664185548769e-07\\
141	7.51273839323926e-07\\
142	7.57103457700874e-07\\
143	7.63170617331357e-07\\
144	7.6949508908504e-07\\
145	7.76098620467912e-07\\
146	7.83004260692135e-07\\
147	7.90235360114425e-07\\
148	7.97814705298912e-07\\
149	8.05764475831179e-07\\
150	8.14107618407922e-07\\
151	8.22870765752067e-07\\
152	8.32088173692912e-07\\
153	8.41805690192864e-07\\
154	8.52083830558881e-07\\
155	8.62999614557342e-07\\
156	8.74647578098605e-07\\
157	8.87140900678523e-07\\
158	9.00613735769764e-07\\
159	9.15225729069028e-07\\
160	9.31169604699865e-07\\
161	9.4868280002565e-07\\
162	9.68064601259951e-07\\
163	9.89701296504404e-07\\
164	1.01410395516448e-06\\
165	1.04196749934745e-06\\
166	1.07426790224955e-06\\
167	1.11243196736851e-06\\
168	1.15865575220028e-06\\
169	1.21655775824335e-06\\
170	1.29268994920693e-06\\
171	1.40069313474305e-06\\
172	1.57635809196245e-06\\
173	1.9717399476902e-06\\
};
\addlegendentry{$u_t^{21}$};

\addplot [color=mycolor6,mark=o,mark repeat={20},mark phase={10},solid,line width=1.0pt]
  table[row sep=crcr]{%
0	0.127245410618766\\
1	0.127221134425279\\
2	0.127194171251833\\
3	0.127164214857454\\
4	0.127130936304804\\
5	0.12709397567611\\
6	0.12705293282246\\
7	0.127007358259026\\
8	0.126956746879594\\
9	0.126900535376819\\
10	0.12683810045107\\
11	0.12676875375916\\
12	0.126691733794713\\
13	0.126606197739229\\
14	0.126511214547078\\
15	0.126405757782927\\
16	0.126288694388276\\
17	0.126158764905541\\
18	0.126014558475976\\
19	0.125854494983465\\
20	0.125676816283915\\
21	0.125479577361978\\
22	0.125260635554873\\
23	0.125017640511738\\
24	0.124748023141605\\
25	0.124448977610025\\
26	0.124117431717915\\
27	0.12375000783431\\
28	0.123342983109782\\
29	0.122892256796136\\
30	0.122393323444348\\
31	0.12184123034201\\
32	0.121230479647641\\
33	0.120554949543754\\
34	0.119807931772843\\
35	0.118982126025042\\
36	0.118069563459686\\
37	0.117061571882127\\
38	0.115948840345547\\
39	0.114721615583568\\
40	0.113369927991254\\
41	0.111883616638554\\
42	0.110252108419128\\
43	0.108464166738196\\
44	0.106507734746483\\
45	0.104369774941207\\
46	0.102036451821704\\
47	0.0994940068718324\\
48	0.0967289625529446\\
49	0.0937274040097708\\
50	0.0904747108312939\\
51	0.0869573990289846\\
52	0.0831680552362933\\
53	0.0791090953069405\\
54	0.0747902949326188\\
55	0.0702245089989727\\
56	0.0654271824207335\\
57	0.0604195273309117\\
58	0.0552345851125995\\
59	0.0499169132251008\\
60	0.0445189860694739\\
61	0.0391138626569751\\
62	0.0338127931305117\\
63	0.0287446051642503\\
64	0.0240102135508074\\
65	0.019659356850647\\
66	0.0156884021281112\\
67	0.0120637625702454\\
68	0.00876092849280309\\
69	0.005755914669259\\
70	0.00300700462907147\\
71	0.000488698007723922\\
72	3.48168619441487e-06\\
73	1.65837518311303e-06\\
74	1.12673156529158e-06\\
75	8.74961009907479e-07\\
76	7.28924816372598e-07\\
77	6.3410257350329e-07\\
78	5.67952826808418e-07\\
79	5.19467546877658e-07\\
80	4.82632591201081e-07\\
81	4.53884160201489e-07\\
82	4.30976052655235e-07\\
83	4.1242210230238e-07\\
84	3.97199661253344e-07\\
85	3.84581868745955e-07\\
86	3.74037855780428e-07\\
87	3.65170815026988e-07\\
88	3.57678171759113e-07\\
89	3.51325170786886e-07\\
90	3.45926887336681e-07\\
91	3.41335685409387e-07\\
92	3.37432291204595e-07\\
93	3.34119321267085e-07\\
94	3.31316511979118e-07\\
95	3.28957150165119e-07\\
96	3.26985365878283e-07\\
97	3.25354053503751e-07\\
98	3.24023257097154e-07\\
99	3.22958903135774e-07\\
100	3.22131796563427e-07\\
101	3.21516818611526e-07\\
102	3.21092279815675e-07\\
103	3.20839392725415e-07\\
104	3.20741839527597e-07\\
105	3.20785417340237e-07\\
106	3.20957745763737e-07\\
107	3.21248023124337e-07\\
108	3.21646825087743e-07\\
109	3.22145938688051e-07\\
110	3.22738213986286e-07\\
111	3.23417428533264e-07\\
112	3.24178187262108e-07\\
113	3.25015840780867e-07\\
114	3.25926320870315e-07\\
115	3.26906170531749e-07\\
116	3.27953030440459e-07\\
117	3.29064934287274e-07\\
118	3.30239814977501e-07\\
119	3.31475856086506e-07\\
120	3.32772662940715e-07\\
121	3.34134361830852e-07\\
122	3.35561343897233e-07\\
123	3.37040209819724e-07\\
124	3.38613390736804e-07\\
125	3.40218207308165e-07\\
126	3.41944167942299e-07\\
127	3.4375894128056e-07\\
128	3.45624010694265e-07\\
129	3.47540148533655e-07\\
130	3.49510619610521e-07\\
131	3.51538387593128e-07\\
132	3.53626449294783e-07\\
133	3.55778018221678e-07\\
134	3.57996597250064e-07\\
135	3.60286019280375e-07\\
136	3.62650481638073e-07\\
137	3.65094581638895e-07\\
138	3.67623356586813e-07\\
139	3.70242332374296e-07\\
140	3.72957587105916e-07\\
141	3.75775836294321e-07\\
142	3.7870454001408e-07\\
143	3.81752017449971e-07\\
144	3.84927534483446e-07\\
145	3.88241319417874e-07\\
146	3.917044809579e-07\\
147	3.95328862038658e-07\\
148	3.99126945609942e-07\\
149	4.03111985882223e-07\\
150	4.0729851819929e-07\\
151	4.11703286910551e-07\\
152	4.16346470795481e-07\\
153	4.21252974815095e-07\\
154	4.26453576977143e-07\\
155	4.31985871213068e-07\\
156	4.37895146533852e-07\\
157	4.44235490372964e-07\\
158	4.5107146428875e-07\\
159	4.58480711453285e-07\\
160	4.66557887568007e-07\\
161	4.75420427410189e-07\\
162	4.85216938201809e-07\\
163	4.96139559249846e-07\\
164	5.08442665579722e-07\\
165	5.22472295092611e-07\\
166	5.38714727263009e-07\\
167	5.57881416347324e-07\\
168	5.81068263474338e-07\\
169	6.10082230419942e-07\\
170	6.48196763652742e-07\\
171	7.02229233580436e-07\\
172	7.90071148170588e-07\\
173	9.87748462559185e-07\\
};
\addlegendentry{$u_t^{22}$};

\addplot [color=mycolor7,mark=triangle,mark repeat={20},solid,line width=1.0pt]
  table[row sep=crcr]{%
0	0.0465302662959145\\
1	0.0465296049855046\\
2	0.0465288641993591\\
3	0.0465280438257534\\
4	0.0465271326428701\\
5	0.0465261129128232\\
6	0.0465249655349242\\
7	0.046523674608655\\
8	0.0465222287051465\\
9	0.0465206178884327\\
10	0.0465188293692906\\
11	0.0465168457703758\\
12	0.0465146457525465\\
13	0.0465122038910393\\
14	0.0465094884790266\\
15	0.0465064586355445\\
16	0.04650306448966\\
17	0.0464992547963114\\
18	0.0464949885904634\\
19	0.0464902384088429\\
20	0.0464849830439708\\
21	0.046479198876459\\
22	0.0464728515600758\\
23	0.0464658852976572\\
24	0.0464582113593016\\
25	0.0464497017032105\\
26	0.0464401923246435\\
27	0.0464294941224538\\
28	0.046417402532503\\
29	0.0464036981930446\\
30	0.0463881399375113\\
31	0.046370471946071\\
32	0.0463504848772122\\
33	0.0463280570340137\\
34	0.0463030787551269\\
35	0.0462754202468146\\
36	0.0462449715325972\\
37	0.0462116431366239\\
38	0.0461752712213432\\
39	0.046135396583611\\
40	0.0460910219835406\\
41	0.0460405809630122\\
42	0.0459821661552163\\
43	0.0459138053122051\\
44	0.045833664159159\\
45	0.0457402789165251\\
46	0.0456324768754241\\
47	0.0455086412041486\\
48	0.0453666889014754\\
49	0.045205018738028\\
50	0.0450230619138457\\
51	0.0448197794302232\\
52	0.0445891268283755\\
53	0.0443167545012451\\
54	0.0439829932917205\\
55	0.0435677019993167\\
56	0.0430513625363193\\
57	0.0424125306644603\\
58	0.0416224003950385\\
59	0.0406457123109356\\
60	0.0394448807056541\\
61	0.0379677460516528\\
62	0.0361302372726778\\
63	0.0338370597363922\\
64	0.0310270271261208\\
65	0.0276959031184511\\
66	0.0238978973143148\\
67	0.0197212790034215\\
68	0.015248154740879\\
69	0.0105617671304221\\
70	0.00576434111645323\\
71	0.000971488007763571\\
72	6.96037346675879e-06\\
73	3.31601984796939e-06\\
74	2.25312070946847e-06\\
75	1.74971428523902e-06\\
76	1.45770501985823e-06\\
77	1.26809552315932e-06\\
78	1.13581761691042e-06\\
79	1.03886139734618e-06\\
80	9.65201538455749e-07\\
81	9.07712014667131e-07\\
82	8.61901327422253e-07\\
83	8.24797695321047e-07\\
84	7.94356175463509e-07\\
85	7.69123281435969e-07\\
86	7.4803743723361e-07\\
87	7.30305142976816e-07\\
88	7.15321332582141e-07\\
89	7.02616557537864e-07\\
90	6.91821015030764e-07\\
91	6.82639468989922e-07\\
92	6.74833399805867e-07\\
93	6.68208062677415e-07\\
94	6.62602948017626e-07\\
95	6.57884643720081e-07\\
96	6.53941421478726e-07\\
97	6.50679079561155e-07\\
98	6.48017713911948e-07\\
99	6.45889183968613e-07\\
100	6.44235105005571e-07\\
101	6.43005243947561e-07\\
102	6.42156225553155e-07\\
103	6.41650477933201e-07\\
104	6.41455367878107e-07\\
105	6.41542491502785e-07\\
106	6.41887089376444e-07\\
107	6.42467558972524e-07\\
108	6.43265051794184e-07\\
109	6.44263141340837e-07\\
110	6.45447526324029e-07\\
111	6.46805759323363e-07\\
112	6.48327046174446e-07\\
113	6.50002083432686e-07\\
114	6.51822730422127e-07\\
115	6.53782062922977e-07\\
116	6.55875332380426e-07\\
117	6.58098548467632e-07\\
118	6.60447511589225e-07\\
119	6.6291868030168e-07\\
120	6.65511460437539e-07\\
121	6.68233642935681e-07\\
122	6.71085047964601e-07\\
123	6.7404066690835e-07\\
124	6.77184519801111e-07\\
125	6.80389544895418e-07\\
126	6.83836521473257e-07\\
127	6.87460443488323e-07\\
128	6.91184099241796e-07\\
129	6.95008884066698e-07\\
130	6.98941227012807e-07\\
131	7.02986955642169e-07\\
132	7.07151953718029e-07\\
133	7.11442529480368e-07\\
134	7.1586556449752e-07\\
135	7.20428599734773e-07\\
136	7.25139911448353e-07\\
137	7.30008593502722e-07\\
138	7.35044655621103e-07\\
139	7.4025915127903e-07\\
140	7.45664355594707e-07\\
141	7.51274010970807e-07\\
142	7.57103631014127e-07\\
143	7.63170792381403e-07\\
144	7.6949526594783e-07\\
145	7.76098799225511e-07\\
146	7.8300444143335e-07\\
147	7.90235542935426e-07\\
148	7.97814890303858e-07\\
149	8.05764663132947e-07\\
150	8.14107808129126e-07\\
151	8.22870958026605e-07\\
152	8.32088368668218e-07\\
153	8.41805888032944e-07\\
154	8.52084031448055e-07\\
155	8.62999818704622e-07\\
156	8.74647785742658e-07\\
157	8.8714111209338e-07\\
158	9.00613951271653e-07\\
159	9.15225949024918e-07\\
160	9.31169829538624e-07\\
161	9.48683030253055e-07\\
162	9.68064837479361e-07\\
163	9.8970153944583e-07\\
164	1.01410420572643e-06\\
165	1.04196775865897e-06\\
166	1.07426817176572e-06\\
167	1.11243224902474e-06\\
168	1.15865604867733e-06\\
169	1.21655807346096e-06\\
170	1.29269028934967e-06\\
171	1.40069351076092e-06\\
172	1.57635852746456e-06\\
173	1.97174052088514e-06\\
};
\addlegendentry{$u_t^{23}$};

\addplot [color=mycolor8,mark=|,mark repeat={20},mark phase={10},solid,line width=1.0pt]
  table[row sep=crcr]{%
0	0.0284694509986953\\
1	0.0284701123091518\\
2	0.0284708530952019\\
3	0.0284716734688302\\
4	0.0284725846517126\\
5	0.0284736043817031\\
6	0.0284747517596393\\
7	0.0284760426858621\\
8	0.0284774885893815\\
9	0.0284790994061117\\
10	0.028480887925169\\
11	0.0284828715240314\\
12	0.0284850715418527\\
13	0.0284875134032761\\
14	0.028490228815358\\
15	0.0284932586587596\\
16	0.0284966528045104\\
17	0.0285004624978586\\
18	0.0285047287036198\\
19	0.0285094788851808\\
20	0.0285147342499914\\
21	0.028520518417487\\
22	0.0285268657337779\\
23	0.028533831996097\\
24	0.0285415059343625\\
25	0.0285500155904085\\
26	0.0285595249687183\\
27	0.0285702231708018\\
28	0.0285823147606773\\
29	0.0285960190999843\\
30	0.0286115773553637\\
31	0.0286292453466341\\
32	0.0286492324151577\\
33	0.0286716602586177\\
34	0.0286966385368546\\
35	0.0287242970448464\\
36	0.0287547457587782\\
37	0.0287880741544073\\
38	0.0288244460693984\\
39	0.0288643207067582\\
40	0.0289086953063717\\
41	0.0289591363264005\\
42	0.0290175511337486\\
43	0.0290859119762557\\
44	0.0291660531287151\\
45	0.0292594383706976\\
46	0.0293672404109735\\
47	0.0294910760815439\\
48	0.0296330283834518\\
49	0.0297946985458212\\
50	0.0299766553689671\\
51	0.0301799378513296\\
52	0.0304105904518156\\
53	0.0306829627775242\\
54	0.031016723985315\\
55	0.0314320152757593\\
56	0.0319483547364987\\
57	0.0325871866056707\\
58	0.0333773168720312\\
59	0.0343540049524184\\
60	0.0355548365532323\\
61	0.0370319712018385\\
62	0.0388694799740888\\
63	0.0411626575020806\\
64	0.0439726901016925\\
65	0.0473038140956542\\
66	0.051101819881237\\
67	0.0552784381655832\\
68	0.0597515623863249\\
69	0.0644379499204646\\
70	0.0692353757499526\\
71	0.0740282279058607\\
72	0.0749927309495347\\
73	0.0749963411097656\\
74	0.0749973658060188\\
75	0.0749978266767881\\
76	0.0749980713503346\\
77	0.0749982082854594\\
78	0.0749982819439874\\
79	0.0749983136563504\\
80	0.074998314687173\\
81	0.074998291310718\\
82	0.0749982470642264\\
83	0.0749981838495725\\
84	0.0749981025110195\\
85	0.0749980031531969\\
86	0.0749978853205692\\
87	0.0749977480980308\\
88	0.0749975901635582\\
89	0.0749974098095555\\
90	0.074997204941976\\
91	0.0749969730620615\\
92	0.0749967112329603\\
93	0.0749964160317966\\
94	0.0749960834864716\\
95	0.07499570899536\\
96	0.0749952872269406\\
97	0.0749948119950929\\
98	0.0749942761042144\\
99	0.07499367115622\\
100	0.0749929873088291\\
101	0.0749922129709068\\
102	0.0749913344150175\\
103	0.0749903352787553\\
104	0.0749891959155715\\
105	0.0749878925416609\\
106	0.0749863961009859\\
107	0.0749846707290871\\
108	0.0749826716485595\\
109	0.0749803422576622\\
110	0.0749776099896599\\
111	0.0749743802494793\\
112	0.0749705275967544\\
113	0.0749658830017053\\
114	0.0749602116547968\\
115	0.0749531748573262\\
116	0.074944284633354\\
117	0.074932826305638\\
118	0.074917748452934\\
119	0.0748972568174679\\
120	0.0748679318685069\\
121	0.0748240553467462\\
122	0.074757979909078\\
123	0.0746410113812624\\
124	0.0744253461256163\\
125	0.0733019015530335\\
126	0.0685662740287431\\
127	0.0623160725144033\\
128	0.0560367023918086\\
129	0.0500436923506225\\
130	0.044419939066675\\
131	0.0391985251461467\\
132	0.0343940462584684\\
133	0.0300099057533146\\
134	0.0260409919501159\\
135	0.0224752965290072\\
136	0.019295268039869\\
137	0.0164790846182346\\
138	0.0140018664180321\\
139	0.0118368026681511\\
140	0.00995616134314349\\
141	0.0083321568459838\\
142	0.00693766404235001\\
143	0.00574678034621774\\
144	0.00473524593527441\\
145	0.00388073219402023\\
146	0.00316300286545678\\
147	0.00256394923461283\\
148	0.00206750675187952\\
149	0.00165947514432556\\
150	0.00132728005812816\\
151	0.00105972400998363\\
152	0.000846773362364587\\
153	0.000679414019239601\\
154	0.000549583195795125\\
155	0.000450157097975824\\
156	0.000374958227368225\\
157	0.000318748448787741\\
158	0.000277189732291143\\
159	0.000246772117176352\\
160	0.000224719758391328\\
161	0.000208889414150875\\
162	0.000197673959726052\\
163	0.000189919683245801\\
164	0.000184862761961961\\
165	0.000182089291009061\\
166	0.000181526231712698\\
167	0.000183480476086037\\
168	0.000188767756950519\\
169	0.000199038377975863\\
170	0.000217609417715398\\
171	0.000251892053781247\\
172	0.0003226265278838\\
173	0.000525242472940519\\
};
\addlegendentry{$u_t^{24}$};

\end{axis}
\end{tikzpicture}%

%% file: flow_batteries_2.tikz
%
%
\definecolor{mycolor1}{rgb}{0.00000,0.44700,0.74100}%
\definecolor{mycolor2}{rgb}{0.85000,0.32500,0.09800}%
\definecolor{mycolor3}{rgb}{0.46600,0.67400,0.18800}%
\definecolor{mycolor4}{rgb}{0.49400,0.18400,0.55600}%
\begin{tikzpicture}

\begin{axis}[%
width=0.85\columnwidth,
height=0.26\columnwidth,
at={(0.772in,0.484in)},
scale only axis,
xmin=0,
xmax=180,
xlabel={Time},
xmajorgrids,
ymin=0,
ymax=1,
ylabel={Battery level},
axis background/.style={fill=white},
legend style={legend cell align=left,align=left,draw=white!15!black}
]
\addplot [color=mycolor1,mark=star,mark repeat={20},mark phase={10},solid,line width=1.0pt]
  table[row sep=crcr]{%
0	1.00000000000039\\
1	0.982622432308819\\
2	0.965247358369581\\
3	0.947875054823387\\
4	0.93050582895565\\
5	0.913140022059989\\
6	0.895778013201421\\
7	0.878420223364748\\
8	0.861067120077833\\
9	0.843719222518565\\
10	0.826377107191597\\
11	0.809041414210186\\
12	0.791712854258381\\
13	0.774392216303172\\
14	0.757080376139626\\
15	0.739778305836504\\
16	0.722487084194095\\
17	0.7052079083062\\
18	0.687942106336314\\
19	0.670691151629661\\
20	0.65345667828984\\
21	0.63624049835702\\
22	0.619044620733197\\
23	0.601871272021567\\
24	0.584722919440879\\
25	0.567602295990428\\
26	0.550512428059176\\
27	0.533456665654811\\
28	0.516438715459173\\
29	0.499462676895135\\
30	0.482533081396004\\
31	0.465654935057868\\
32	0.448833764829506\\
33	0.432075668377244\\
34	0.415387367719591\\
35	0.398776266666832\\
36	0.382250512039376\\
37	0.365819058540267\\
38	0.349491737038398\\
39	0.33327932588174\\
40	0.317193624665181\\
41	0.301247529667625\\
42	0.285455109907442\\
43	0.269831682449986\\
44	0.254393885244905\\
45	0.239159745354367\\
46	0.224148739968617\\
47	0.209381847098858\\
48	0.19488158229119\\
49	0.180672017145889\\
50	0.166778774871087\\
51	0.153228997596612\\
52	0.140051279750683\\
53	0.127275561544194\\
54	0.114932976563383\\
55	0.103055647740957\\
56	0.0916764266411167\\
57	0.0808285721454145\\
58	0.0705453663458814\\
59	0.0608596677951189\\
60	0.0518034052415211\\
61	0.0434070185640035\\
62	0.0356988576931491\\
63	0.0287045546528339\\
64	0.0224463881627785\\
65	0.0169426640950893\\
66	0.0122071380981859\\
67	0.00824850815394908\\
68	0.00507000399658875\\
69	0.00266909567322633\\
70	0.00103732749326424\\
71	0.000160192918717802\\
72	1.41743171752114e-05\\
73	1.31301112162158e-05\\
74	1.2632671721709e-05\\
75	1.22946865044827e-05\\
76	1.20322189869513e-05\\
77	1.18135560170176e-05\\
78	1.16233362225556e-05\\
79	1.14529591946794e-05\\
80	1.1297126317726e-05\\
81	1.11523429228994e-05\\
82	1.10161833239093e-05\\
83	1.08868956045212e-05\\
84	1.07631736433115e-05\\
85	1.06440180779938e-05\\
86	1.05286475810472e-05\\
87	1.0416440069493e-05\\
88	1.03068924910971e-05\\
89	1.01995925577231e-05\\
90	1.00941984016198e-05\\
91	9.99042362779916e-06\\
92	9.8880261284864e-06\\
93	9.78679957524852e-06\\
94	9.68656685247929e-06\\
95	9.5871749218913e-06\\
96	9.48849075772506e-06\\
97	9.39039809404502e-06\\
98	9.29279479564861e-06\\
99	9.19559071333187e-06\\
100	9.09870591929724e-06\\
101	9.00206924372461e-06\\
102	8.90561705197632e-06\\
103	8.8092922158843e-06\\
104	8.71304324321321e-06\\
105	8.61682353682883e-06\\
106	8.52059076027276e-06\\
107	8.4243062910688e-06\\
108	8.32793474715995e-06\\
109	8.2314435737716e-06\\
110	8.13480268007524e-06\\
111	8.03798412036451e-06\\
112	7.94096181591954e-06\\
113	7.84371130694315e-06\\
114	7.74620952892142e-06\\
115	7.64843463822963e-06\\
116	7.55036582936461e-06\\
117	7.45198300761103e-06\\
118	7.35326667391712e-06\\
119	7.25419795590002e-06\\
120	7.15475851696258e-06\\
121	7.05493011939116e-06\\
122	6.95469333375084e-06\\
123	6.85402870947877e-06\\
124	6.75292063680032e-06\\
125	6.65134086082319e-06\\
126	6.54928010076822e-06\\
127	6.44670204708771e-06\\
128	6.34358012397233e-06\\
129	6.23989932844404e-06\\
130	6.1356444407563e-06\\
131	6.03079927177649e-06\\
132	5.92534675324908e-06\\
133	5.81926892886107e-06\\
134	5.712546890125e-06\\
135	5.60516069010916e-06\\
136	5.49708924449552e-06\\
137	5.38831022160811e-06\\
138	5.2787999206532e-06\\
139	5.16853313613467e-06\\
140	5.05748300461993e-06\\
141	4.94562082735446e-06\\
142	4.83291585979264e-06\\
143	4.71933506002115e-06\\
144	4.60484279654301e-06\\
145	4.48940053418617e-06\\
146	4.37296654019759e-06\\
147	4.25549566603258e-06\\
148	4.13693924381728e-06\\
149	4.01724507872639e-06\\
150	3.89635743255505e-06\\
151	3.77421681889433e-06\\
152	3.65075941362807e-06\\
153	3.52591594917923e-06\\
154	3.39961008267843e-06\\
155	3.27175634192483e-06\\
156	3.14225779334779e-06\\
157	3.01100352088454e-06\\
158	2.87786588177939e-06\\
159	2.74269736177354e-06\\
160	2.60532671772131e-06\\
161	2.46555396849452e-06\\
162	2.32314364575094e-06\\
163	2.17781549180476e-06\\
164	2.02923140622934e-06\\
165	1.87697674415492e-06\\
166	1.72053276471091e-06\\
167	1.55923450175966e-06\\
168	1.39220316338807e-06\\
169	1.21823076182061e-06\\
170	1.03556676295428e-06\\
171	8.41478091668315e-07\\
172	6.31185854835966e-07\\
173	3.94542930822662e-07\\
174	9.85940897977236e-08\\
};
\addlegendentry{Node 1};

\addplot [color=mycolor2,mark=|,mark repeat={20},mark phase={10},solid,line width=1.0pt]
  table[row sep=crcr]{%
0	0.999999999999772\\
1	0.992500028271301\\
2	0.985000056541005\\
3	0.977500084811903\\
4	0.970000113081858\\
5	0.962500141352426\\
6	0.955000169622349\\
7	0.94750019789308\\
8	0.940000226163285\\
9	0.932500254433725\\
10	0.925000282704518\\
11	0.917500310974942\\
12	0.910000339245351\\
13	0.902500367516213\\
14	0.895000395787206\\
15	0.887500424056635\\
16	0.880000452327624\\
17	0.872500480597959\\
18	0.865000508868358\\
19	0.857500537138669\\
20	0.850000565409002\\
21	0.842500593679433\\
22	0.835000621950244\\
23	0.827500650220945\\
24	0.82000067849112\\
25	0.812500706761758\\
26	0.805000735032687\\
27	0.797500763303235\\
28	0.790000791574304\\
29	0.78250081984505\\
30	0.775000848115442\\
31	0.76750087638623\\
32	0.76000090465692\\
33	0.752500932927699\\
34	0.745000961198751\\
35	0.737500989469606\\
36	0.730001017740373\\
37	0.72250104601137\\
38	0.715001074282281\\
39	0.707501102553295\\
40	0.700001130824135\\
41	0.692501159095307\\
42	0.685001187366121\\
43	0.677501215637022\\
44	0.670001243908307\\
45	0.662501272179418\\
46	0.655001300450712\\
47	0.647501328721832\\
48	0.640001356993474\\
49	0.632501385265259\\
50	0.625001413536714\\
51	0.617501441808477\\
52	0.610001470079989\\
53	0.602501498352213\\
54	0.595001526624175\\
55	0.587501554896514\\
56	0.580001583169262\\
57	0.572501611441848\\
58	0.565001639715012\\
59	0.557501667988213\\
60	0.550001696261768\\
61	0.542501724535973\\
62	0.535001752810525\\
63	0.527501781085731\\
64	0.52000180936194\\
65	0.512501837639091\\
66	0.505001865917786\\
67	0.497501894198281\\
68	0.490001922481258\\
69	0.482501950768328\\
70	0.475001979063326\\
71	0.4675020073768\\
72	0.46000203578538\\
73	0.452502066652977\\
74	0.445002100940117\\
75	0.437502139047643\\
76	0.430002181408615\\
77	0.422502228503104\\
78	0.415002280864901\\
79	0.407502339088791\\
80	0.400002403837077\\
81	0.392502475848181\\
82	0.385002555945921\\
83	0.377502645049267\\
84	0.37000274418455\\
85	0.36250285449781\\
86	0.355002977270138\\
87	0.347503113934393\\
88	0.340003266093985\\
89	0.332503435545528\\
90	0.325003624302933\\
91	0.317503834626615\\
92	0.310004069056508\\
93	0.302504330449856\\
94	0.295004622025859\\
95	0.287504947416936\\
96	0.28000531072895\\
97	0.272505716612122\\
98	0.265006170344681\\
99	0.257506677932491\\
100	0.250007246227963\\
101	0.242507883073561\\
102	0.235008597475928\\
103	0.227509399818805\\
104	0.220010302125877\\
105	0.212511318388749\\
106	0.205012464980346\\
107	0.197513761181536\\
108	0.190015229861861\\
109	0.182516898370504\\
110	0.175018799718407\\
111	0.167520974174676\\
112	0.160023471469155\\
113	0.152526353876767\\
114	0.145029700576388\\
115	0.137533614228642\\
116	0.130038231364695\\
117	0.122543737313825\\
118	0.115050388873409\\
119	0.107558547983363\\
120	0.100068756009749\\
121	0.0925818962717483\\
122	0.0850994239137112\\
123	0.0776235588142932\\
124	0.0701593902721013\\
125	0.062716787941087\\
126	0.0553865297468271\\
127	0.0485298339603029\\
128	0.0422981579628191\\
129	0.0366944186052278\\
130	0.0316899798692763\\
131	0.0272479160684861\\
132	0.0233279932551756\\
133	0.0198885179141333\\
134	0.0168874561945488\\
135	0.014283285412981\\
136	0.0120356837172202\\
137	0.0101060843992423\\
138	0.00845810293655946\\
139	0.00705784279029074\\
140	0.00587408849756055\\
141	0.00487839779681063\\
142	0.00404510698481113\\
143	0.00335126487021302\\
144	0.00277651051851201\\
145	0.00230290897545798\\
146	0.00191475814617604\\
147	0.00159837955918621\\
148	0.00134190561217064\\
149	0.00113507515549366\\
150	0.000969047064594788\\
151	0.000836237648001158\\
152	0.000730182959906992\\
153	0.000645422414833667\\
154	0.000577396832320903\\
155	0.000522353304338245\\
156	0.000477251294558792\\
157	0.000439668007043395\\
158	0.000407704448053412\\
159	0.00037989541342917\\
160	0.000355126679116632\\
161	0.000332561586294545\\
162	0.000311577776576432\\
163	0.000291713574120079\\
164	0.000272622635641554\\
165	0.000254034949024786\\
166	0.000235721823148015\\
167	0.000217461773159569\\
168	0.000199002482326063\\
169	0.000180009841026143\\
170	0.000159984347421211\\
171	0.000138094136620736\\
172	0.000112764861891536\\
173	8.03445732504093e-05\\
174	2.7623151904269e-05\\
};
\addlegendentry{Node 2};

\addplot [color=mycolor3,mark=o,mark repeat={20},solid,line width=1.0pt]
  table[row sep=crcr]{%
0	0.999999999999909\\
1	0.982622432310544\\
2	0.965247358370022\\
3	0.947875054824823\\
4	0.930505828956127\\
5	0.913140022061853\\
6	0.895778013203015\\
7	0.878420223366402\\
8	0.861067120080073\\
9	0.843719222520377\\
10	0.826377107194242\\
11	0.809041414212061\\
12	0.791712854259533\\
13	0.774392216304608\\
14	0.757080376141635\\
15	0.739778305839513\\
16	0.722487084196671\\
17	0.705207908309853\\
18	0.687942106339006\\
19	0.670691151631663\\
20	0.653456678291909\\
21	0.636240498359175\\
22	0.619044620734981\\
23	0.601871272023779\\
24	0.584722919442879\\
25	0.567602295992692\\
26	0.550512428061449\\
27	0.533456665657742\\
28	0.51643871546166\\
29	0.499462676897562\\
30	0.482533081398493\\
31	0.46565493506056\\
32	0.448833764831825\\
33	0.432075668379188\\
34	0.415387367721264\\
35	0.398776266668515\\
36	0.38225051204138\\
37	0.365819058542141\\
38	0.349491737040518\\
39	0.333279325883748\\
40	0.317193624666933\\
41	0.301247529669394\\
42	0.285455109909243\\
43	0.269831682451697\\
44	0.25439388524664\\
45	0.239159745356267\\
46	0.224148739970521\\
47	0.209381847100828\\
48	0.194881582293195\\
49	0.180672017147831\\
50	0.166778774873104\\
51	0.153228997598588\\
52	0.140051279752658\\
53	0.127275561546209\\
54	0.114932976565382\\
55	0.103055647742923\\
56	0.0916764266430899\\
57	0.0808285721473952\\
58	0.0705453663478409\\
59	0.0608596677970716\\
60	0.0518034052434677\\
61	0.0434070185659522\\
62	0.0356988576950884\\
63	0.0287045546547689\\
64	0.0224463881647065\\
65	0.0169426640970116\\
66	0.0122071381001019\\
67	0.00824850815585929\\
68	0.00507000399849297\\
69	0.0026690956751245\\
70	0.0010373274951564\\
71	0.000160192920603936\\
72	1.41743190551866e-05\\
73	1.31301130890693e-05\\
74	1.2632673585961e-05\\
75	1.2294688358485e-05\\
76	1.20322208289704e-05\\
77	1.18135578453473e-05\\
78	1.1623338035681e-05\\
79	1.14529609913091e-05\\
80	1.12971280968868e-05\\
81	1.11523446839211e-05\\
82	1.10161850664342e-05\\
83	1.08868973284265e-05\\
84	1.07631753486641e-05\\
85	1.06440197649924e-05\\
86	1.05286492499742e-05\\
87	1.04164417206728e-05\\
88	1.03068941248725e-05\\
89	1.01995941744383e-05\\
90	1.00942000016059e-05\\
91	9.99042521136912e-06\\
92	9.88802769592919e-06\\
93	9.78680112682814e-06\\
94	9.68656838843369e-06\\
95	9.58717644243402e-06\\
96	9.48849226304549e-06\\
97	9.39039958430978e-06\\
98	9.29279627100329e-06\\
99	9.19559217390239e-06\\
100	9.09870736519195e-06\\
101	9.00207067503506e-06\\
102	8.90561846877914e-06\\
103	8.80929361824226e-06\\
104	8.71304463117639e-06\\
105	8.61682491043582e-06\\
106	8.52059211955152e-06\\
107	8.4243076360375e-06\\
108	8.32793607782782e-06\\
109	8.23144489013962e-06\\
110	8.13480398213673e-06\\
111	8.0379854081057e-06\\
112	7.94096308932003e-06\\
113	7.84371256597638e-06\\
114	7.74621077355503e-06\\
115	7.64843586842578e-06\\
116	7.5503670450803e-06\\
117	7.45198420879822e-06\\
118	7.35326786052273e-06\\
119	7.25419912786606e-06\\
120	7.15475967422724e-06\\
121	7.05493126188941e-06\\
122	6.95469446141275e-06\\
123	6.85402982222657e-06\\
124	6.75292173455377e-06\\
125	6.65134194349998e-06\\
126	6.54928116827962e-06\\
127	6.44670309933807e-06\\
128	6.34358116086118e-06\\
129	6.23990034986758e-06\\
130	6.13564544660754e-06\\
131	6.03080026194521e-06\\
132	5.92534772762168e-06\\
133	5.8192698873204e-06\\
134	5.71254783255019e-06\\
135	5.60516161637544e-06\\
136	5.49709015447392e-06\\
137	5.38831111516528e-06\\
138	5.27880079765112e-06\\
139	5.16853399643032e-06\\
140	5.05748384806499e-06\\
141	4.94562165379493e-06\\
142	4.83291666906841e-06\\
143	4.7193358519656e-06\\
144	4.60484357098246e-06\\
145	4.48940129093933e-06\\
146	4.37296727907499e-06\\
147	4.25549638683587e-06\\
148	4.13693994633846e-06\\
149	4.01724576274708e-06\\
150	3.89635809784556e-06\\
151	3.77421746521272e-06\\
152	3.65076004071901e-06\\
153	3.52591655677264e-06\\
154	3.39961067048783e-06\\
155	3.27175690964531e-06\\
156	3.14225834065354e-06\\
157	3.01100404742589e-06\\
158	2.87786638717925e-06\\
159	2.74269784562321e-06\\
160	2.60532717957539e-06\\
161	2.46555440786473e-06\\
162	2.3231440620984e-06\\
163	2.17781588453029e-06\\
164	2.02923177466072e-06\\
165	1.8769770875301e-06\\
166	1.72053308215495e-06\\
167	1.55923479225207e-06\\
168	1.39220342571487e-06\\
169	1.2182309944997e-06\\
170	1.03556696411161e-06\\
171	8.41478258811367e-07\\
172	6.31185984377231e-07\\
173	3.94543016813716e-07\\
174	9.85941184692838e-08\\
};
\addlegendentry{Node 3};

\addplot [color=mycolor4,mark=diamond,mark repeat={20},solid,line width=1.0pt]
  table[row sep=crcr]{%
0	0.999999999999931\\
1	0.992500028271052\\
2	0.985000056541716\\
3	0.977500084812027\\
4	0.970000113082233\\
5	0.962500141352027\\
6	0.955000169622963\\
7	0.947500197893671\\
8	0.940000226164239\\
9	0.932500254434964\\
10	0.92500028270524\\
11	0.917500310974731\\
12	0.910000339245781\\
13	0.902500367516068\\
14	0.895000395786781\\
15	0.887500424057194\\
16	0.880000452327652\\
17	0.872500480599183\\
18	0.865000508869793\\
19	0.857500537140629\\
20	0.850000565410943\\
21	0.84250059368209\\
22	0.835000621952438\\
23	0.827500650223032\\
24	0.820000678494096\\
25	0.812500706764916\\
26	0.805000735035174\\
27	0.797500763306064\\
28	0.790000791577138\\
29	0.782500819847577\\
30	0.775000848118803\\
31	0.767500876389093\\
32	0.760000904659709\\
33	0.752500932930359\\
34	0.745000961200997\\
35	0.737500989471785\\
36	0.730001017742369\\
37	0.722501046013866\\
38	0.715001074284631\\
39	0.707501102555539\\
40	0.700001130826652\\
41	0.692501159097701\\
42	0.685001187368476\\
43	0.677501215639293\\
44	0.67000124391043\\
45	0.662501272181724\\
46	0.655001300452713\\
47	0.647501328724217\\
48	0.640001356995622\\
49	0.632501385266933\\
50	0.625001413538749\\
51	0.617501441810156\\
52	0.610001470081875\\
53	0.602501498353974\\
54	0.595001526625921\\
55	0.587501554898262\\
56	0.58000158317068\\
57	0.572501611443388\\
58	0.565001639716395\\
59	0.557501667989514\\
60	0.550001696263246\\
61	0.542501724537545\\
62	0.535001752812003\\
63	0.527501781087276\\
64	0.520001809363292\\
65	0.512501837640533\\
66	0.505001865918902\\
67	0.49750189419936\\
68	0.490001922482653\\
69	0.48250195076981\\
70	0.475001979064758\\
71	0.467502007378164\\
72	0.460002035786646\\
73	0.452502066654467\\
74	0.445002100941721\\
75	0.437502139049178\\
76	0.430002181410046\\
77	0.422502228504552\\
78	0.4150022808664\\
79	0.407502339090247\\
80	0.400002403838519\\
81	0.392502475849666\\
82	0.385002555947466\\
83	0.377502645050894\\
84	0.370002744186156\\
85	0.36250285449946\\
86	0.355002977271855\\
87	0.347503113936088\\
88	0.340003266095788\\
89	0.332503435547413\\
90	0.325003624304862\\
91	0.317503834628574\\
92	0.310004069058434\\
93	0.302504330451821\\
94	0.295004622027843\\
95	0.287504947418959\\
96	0.280005310730971\\
97	0.272505716614158\\
98	0.26500617034679\\
99	0.257506677934565\\
100	0.250007246230041\\
101	0.242507883075657\\
102	0.235008597478017\\
103	0.227509399820939\\
104	0.220010302128019\\
105	0.212511318390929\\
106	0.205012464982537\\
107	0.197513761183762\\
108	0.190015229864102\\
109	0.182516898372774\\
110	0.17501879972069\\
111	0.167520974176976\\
112	0.160023471471477\\
113	0.152526353879093\\
114	0.145029700578727\\
115	0.137533614230993\\
116	0.13003823136707\\
117	0.122543737316215\\
118	0.115050388875812\\
119	0.107558547985785\\
120	0.100068756012185\\
121	0.0925818962742038\\
122	0.0850994239161796\\
123	0.0776235588167841\\
124	0.070159390274609\\
125	0.0627167879436101\\
126	0.0553865297493668\\
127	0.0485298339628551\\
128	0.042298157965386\\
129	0.0366944186078099\\
130	0.0316899798718749\\
131	0.0272479160711003\\
132	0.0233279932578068\\
133	0.0198885179167802\\
134	0.0168874561972117\\
135	0.0142832854156598\\
136	0.0120356837199152\\
137	0.0101060844019536\\
138	0.00845810293928739\\
139	0.00705784279303531\\
140	0.00587408850032193\\
141	0.00487839779958905\\
142	0.00404510698760672\\
143	0.00335126487302598\\
144	0.00277651052134247\\
145	0.00230290897830612\\
146	0.00191475814904205\\
147	0.0015983795620703\\
148	0.001341905615073\\
149	0.00113507515841452\\
150	0.000969047067534384\\
151	0.000836237650959727\\
152	0.000730182962884788\\
153	0.00064542241783096\\
154	0.00057739683533798\\
155	0.000522353307375411\\
156	0.000477251297616373\\
157	0.000439668010121741\\
158	0.000407704451152899\\
159	0.000379895416550208\\
160	0.000355126682259666\\
161	0.000332561589460063\\
162	0.000311577779764973\\
163	0.000291713577332242\\
164	0.000272622638878012\\
165	0.0002540349522863\\
166	0.000235721826435459\\
167	0.000217461776473964\\
168	0.000199002485668624\\
169	0.000180009844398352\\
170	0.000159984350824942\\
171	0.000138094140058481\\
172	0.000112764865366882\\
173	8.0344576769306e-05\\
174	2.76231554804851e-05\\
};
\addlegendentry{Node 4};

\end{axis}
\end{tikzpicture}%

%% file: MAC_powers.tikz
%
%
\definecolor{mycolor1}{rgb}{0.49400,0.18400,0.55600}%
\definecolor{mycolor2}{rgb}{0.92900,0.69400,0.12500}%
\definecolor{mycolor3}{rgb}{0.00000,0.44700,0.74100}%
\definecolor{mycolor4}{rgb}{0.85000,0.32500,0.09800}%
\begin{tikzpicture}

\begin{axis}[%
width=0.85\columnwidth,
height=0.26\columnwidth,
at={(0.771875in,0.483542in)},
scale only axis,
xmin=1,
xmax=60,
xlabel={Time},
xmajorgrids,
ymin=0,
ymax=6,
ylabel={Power},
legend style={legend cell align=left,align=left,draw=white!15!black}
]
\addplot [color=mycolor1,solid,line width=2.0pt]
  table[row sep=crcr]{%
1	3.94986754201089\\
2	3.74551589015647\\
3	3.54976990059216\\
4	3.36233871479555\\
5	3.18293471105617\\
6	3.01127415974391\\
7	2.84707780001261\\
8	2.69007134260211\\
9	2.5399859034586\\
10	2.39655837289193\\
11	1.72460568234574\\
12	0\\
13	0\\
14	0\\
15	0\\
16	0\\
17	0\\
18	0\\
19	0\\
20	0\\
21	0\\
22	0\\
23	0\\
24	0\\
25	0\\
26	0\\
27	0\\
28	0\\
29	0\\
30	0\\
31	0\\
32	0\\
33	0\\
34	0\\
35	0\\
36	0\\
37	0\\
38	0\\
39	0\\
40	0\\
41	0\\
42	0\\
43	0\\
44	0\\
45	0\\
46	0\\
47	0\\
48	0\\
49	0\\
50	0\\
51	0\\
52	0\\
53	0\\
54	0\\
55	0\\
56	0\\
57	0\\
58	0\\
59	0\\
60	0\\
61	0\\
62	0\\
63	0\\
64	0\\
65	0\\
66	0\\
67	0\\
68	0\\
69	0\\
70	0\\
71	0\\
72	0\\
73	0\\
74	0\\
75	0\\
76	0\\
77	0\\
78	0\\
79	0\\
80	0\\
81	0\\
82	0\\
83	0\\
84	0\\
85	0\\
86	0\\
87	0\\
88	0\\
89	0\\
90	0\\
91	0\\
92	0\\
93	0\\
94	0\\
95	0\\
96	0\\
97	0\\
98	0\\
99	0\\
};
\addlegendentry{User 1};

\addplot [color=mycolor2,dashed,line width=2.0pt]
  table[row sep=crcr]{%
1	0\\
2	0\\
3	0\\
4	0\\
5	0\\
6	0\\
7	0\\
8	0\\
9	0\\
10	0\\
11	1.07732423400264\\
12	4.3011422396958\\
13	4.06167219197194\\
14	3.83192041814739\\
15	3.61159123814423\\
16	3.40038895331863\\
17	3.19801883703567\\
18	3.00418803086025\\
19	2.81860635013594\\
20	2.64098700308144\\
21	1.05416051305534\\
22	0\\
23	0\\
24	0\\
25	0\\
26	0\\
27	0\\
28	0\\
29	0\\
30	0\\
31	0\\
32	0\\
33	0\\
34	0\\
35	0\\
36	0\\
37	0\\
38	0\\
39	0\\
40	0\\
41	0\\
42	0\\
43	0\\
44	0\\
45	0\\
46	0\\
47	0\\
48	0\\
49	0\\
50	0\\
51	0\\
52	0\\
53	0\\
54	0\\
55	0\\
56	0\\
57	0\\
58	0\\
59	0\\
60	0\\
61	0\\
62	0\\
63	0\\
64	0\\
65	0\\
66	0\\
67	0\\
68	0\\
69	0\\
70	0\\
71	0\\
72	0\\
73	0\\
74	0\\
75	0\\
76	0\\
77	0\\
78	0\\
79	0\\
80	0\\
81	0\\
82	0\\
83	0\\
84	0\\
85	0\\
86	0\\
87	0\\
88	0\\
89	0\\
90	0\\
91	0\\
92	0\\
93	0\\
94	0\\
95	0\\
96	0\\
97	0\\
98	0\\
99	0\\
};
\addlegendentry{User 2};

\addplot [color=mycolor3,solid,line width=1.0pt, mark=triangle, mark repeat={3}]
  table[row sep=crcr]{%
1	0\\
2	0\\
3	0\\
4	0\\
5	0\\
6	0\\
7	0\\
8	0\\
9	0\\
10	0\\
11	0\\
12	0\\
13	0\\
14	0\\
15	0\\
16	0\\
17	0\\
18	0\\
19	0\\
20	0\\
21	2.76330091347757\\
22	4.52251704616271\\
23	4.23710220354548\\
24	3.96266980398394\\
25	3.69893838945355\\
26	3.44562096992344\\
27	3.20242645722687\\
28	2.96906099504976\\
29	2.74522918589696\\
30	1.45313404384711\\
31	0\\
32	0\\
33	0\\
34	0\\
35	0\\
36	0\\
37	0\\
38	0\\
39	0\\
40	0\\
41	0\\
42	0\\
43	0\\
44	0\\
45	0\\
46	0\\
47	0\\
48	0\\
49	0\\
50	0\\
51	0\\
52	0\\
53	0\\
54	0\\
55	0\\
56	0\\
57	0\\
58	0\\
59	0\\
60	0\\
61	0\\
62	0\\
63	0\\
64	0\\
65	0\\
66	0\\
67	0\\
68	0\\
69	0\\
70	0\\
71	0\\
72	0\\
73	0\\
74	0\\
75	0\\
76	0\\
77	0\\
78	0\\
79	0\\
80	0\\
81	0\\
82	0\\
83	0\\
84	0\\
85	0\\
86	0\\
87	0\\
88	0\\
89	0\\
90	0\\
91	0\\
92	0\\
93	0\\
94	0\\
95	0\\
96	0\\
97	0\\
98	0\\
99	0\\
};
\addlegendentry{User 3};

\addplot [color=mycolor4,densely dotted,line width=2pt]
  table[row sep=crcr]{%
1	0\\
2	0\\
3	0\\
4	0\\
5	0\\
6	0\\
7	0\\
8	0\\
9	0\\
10	0\\
11	0\\
12	0\\
13	0\\
14	0\\
15	0\\
16	0\\
17	0\\
18	0\\
19	0\\
20	0\\
21	0\\
22	0\\
23	0\\
24	0\\
25	0\\
26	0\\
27	0\\
28	0\\
29	0\\
30	1.79676753730176\\
31	3.8973465900562\\
32	3.58660715277053\\
33	3.28744679796899\\
34	2.99960197227685\\
35	2.72280004200479\\
36	2.45676095513778\\
37	2.20119880139418\\
38	1.95582326875074\\
39	1.72034099602901\\
40	1.49445682219221\\
41	1.27787493391299\\
42	1.07029991374764\\
43	0.871437691899748\\
44	0.68099640508417\\
45	0.498687166420781\\
46	0.324224750605894\\
47	0.157328198838622\\
48	0\\
49	0\\
50	0\\
51	0\\
52	0\\
53	0\\
54	0\\
55	0\\
56	0\\
57	0\\
58	0\\
59	0\\
60	0\\
61	0\\
62	0\\
63	0\\
64	0\\
65	0\\
66	0\\
67	0\\
68	0\\
69	0\\
70	0\\
71	0\\
72	0\\
73	0\\
74	0\\
75	0\\
76	0\\
77	0\\
78	0\\
79	0\\
80	0\\
81	0\\
82	0\\
83	0\\
84	0\\
85	0\\
86	0\\
87	0\\
88	0\\
89	0\\
90	0\\
91	0\\
92	0\\
93	0\\
94	0\\
95	0\\
96	0\\
97	0\\
98	0\\
99	0\\
};
\addlegendentry{User 4};

\end{axis}
\end{tikzpicture}%

%% file: MAC_utilities.tikz
%
%
\definecolor{mycolor1}{rgb}{0.49400,0.18400,0.55600}%
\definecolor{mycolor2}{rgb}{0.92900,0.69400,0.12500}%
\definecolor{mycolor3}{rgb}{0.00000,0.44700,0.74100}%
\definecolor{mycolor4}{rgb}{0.85000,0.32500,0.09800}%
\begin{tikzpicture}

\begin{axis}[%
width=0.85\columnwidth,
height=0.26\columnwidth,
at={(0.758333in,0.48125in)},
scale only axis,
xmin=1,
xmax=60,
xlabel={Time},
xmajorgrids,
ymin=0,
ymax=3,
ylabel={Rate},
]
\addplot [color=mycolor1,solid,line width=2.0pt]
  table[row sep=crcr]{%
1	2.22732948858191\\
2	2.06750230887801\\
3	1.91809801891622\\
4	1.77847316356871\\
5	1.64802263543019\\
6	1.52617744440065\\
7	1.41240261467886\\
8	1.30619520199812\\
9	1.20708242433225\\
10	1.11461989967565\\
11	0.899130005404072\\
12	-1.1186086361363e-11\\
13	-1.06267820432949e-11\\
14	-1.00954429411301e-11\\
15	-9.59067079407362e-12\\
16	-9.11113725436994e-12\\
17	-8.65558039165144e-12\\
18	-8.22280137206887e-12\\
19	-7.81166130346543e-12\\
20	-7.42107823829215e-12\\
21	-7.05002432637755e-12\\
22	-6.69752311005867e-12\\
23	-6.36264695455573e-12\\
24	-6.04451460682795e-12\\
25	-5.74228887648655e-12\\
26	-5.45517443266222e-12\\
27	-5.18241571102911e-12\\
28	-4.92329492547766e-12\\
29	-4.67713017920377e-12\\
30	-4.44327367024358e-12\\
31	-4.2211099867314e-12\\
32	-4.01005448739483e-12\\
33	-3.80955176302509e-12\\
34	-3.61907417487384e-12\\
35	-3.43812046613014e-12\\
36	-3.26621444282364e-12\\
37	-3.10290372068246e-12\\
38	-2.94775853464833e-12\\
39	-2.80037060791592e-12\\
40	-2.66035207752012e-12\\
41	-2.52733447364411e-12\\
42	-2.40096774996191e-12\\
43	-2.28091936246381e-12\\
44	-2.16687339434062e-12\\
45	-2.05852972462359e-12\\
46	-1.95560323839241e-12\\
47	-1.85782307647279e-12\\
48	-1.76493192264915e-12\\
49	-1.67668532651669e-12\\
50	-1.59285106019086e-12\\
51	-1.51320850718132e-12\\
52	-1.43754808182225e-12\\
53	-1.36567067773114e-12\\
54	-1.29738714384458e-12\\
55	-1.23251778665235e-12\\
56	-1.17089189731973e-12\\
57	-1.11234730245375e-12\\
58	-1.05672993733106e-12\\
59	-1.00389344046451e-12\\
60	-9.53698768441281e-13\\
61	-9.06013830019217e-13\\
62	-8.60713138518256e-13\\
63	-8.17677481592343e-13\\
64	-7.76793607512726e-13\\
65	-7.3795392713709e-13\\
66	-7.01056230780235e-13\\
67	-6.66003419241223e-13\\
68	-6.32703248279162e-13\\
69	-6.01068085865204e-13\\
70	-5.71014681571944e-13\\
71	-5.42463947493347e-13\\
72	-5.15340750118679e-13\\
73	-4.89573712612745e-13\\
74	-4.65095026982108e-13\\
75	-4.41840275633003e-13\\
76	-4.19748261851352e-13\\
77	-3.98760848758785e-13\\
78	-3.78822806320845e-13\\
79	-3.59881666004803e-13\\
80	-3.41887582704563e-13\\
81	-3.24793203569335e-13\\
82	-3.08553543390868e-13\\
83	-2.93125866221325e-13\\
84	-2.78469572910258e-13\\
85	-2.64546094264745e-13\\
86	-2.51318789551508e-13\\
87	-2.38752850073933e-13\\
88	-2.26815207570236e-13\\
89	-2.15474447191724e-13\\
90	-2.04700724832138e-13\\
91	-1.94465688590531e-13\\
92	-1.84742404161005e-13\\
93	-1.75505283952954e-13\\
94	-1.66730019755307e-13\\
95	-1.58393518767541e-13\\
96	-1.50473842829164e-13\\
97	-1.42950150687706e-13\\
98	-1.35802643153321e-13\\
99	-1.29012510995655e-13\\
};

\addplot [color=mycolor2,dashed,line width=2.0pt]
  table[row sep=crcr]{%
1	0.033\\
2	0.03135\\
3	0.0297825\\
4	0.028293375\\
5	0.02687870625\\
6	0.0255347709375\\
7	0.024258032390625\\
8	0.0230451307710937\\
9	0.0218928742325391\\
10	0.0207982305209121\\
11	0.458227329283773\\
12	0.967977418065261\\
13	0.892264473287751\\
14	0.821708448521408\\
15	0.755982464920459\\
16	0.694779475661567\\
17	0.637811097512018\\
18	0.584806509769723\\
19	0.535511416761465\\
20	0.489687070297394\\
21	0.258880994374008\\
22	-3.21805672797212e-12\\
23	-3.05715389157352e-12\\
24	-2.90429619699484e-12\\
25	-2.7590813871451e-12\\
26	-2.62112731778785e-12\\
27	-2.49007095189845e-12\\
28	-2.36556740430353e-12\\
29	-2.24728903408835e-12\\
30	-2.13492458238394e-12\\
31	-2.02817835326474e-12\\
32	-1.9267694356015e-12\\
33	-1.83043096382143e-12\\
34	-1.73890941563035e-12\\
35	-1.65196394484884e-12\\
36	-1.5693657476064e-12\\
37	-1.49089746022608e-12\\
38	-1.41635258721477e-12\\
39	-1.34553495785403e-12\\
40	-1.27825820996133e-12\\
41	-1.21434529946326e-12\\
42	-1.1536280344901e-12\\
43	-1.0959466327656e-12\\
44	-1.04114930112732e-12\\
45	-9.89091836070951e-13\\
46	-9.39637244267403e-13\\
47	-8.92655382054033e-13\\
48	-8.48022612951331e-13\\
49	-8.05621482303764e-13\\
50	-7.65340408188576e-13\\
51	-7.27073387779147e-13\\
52	-6.9071971839019e-13\\
53	-6.5618373247068e-13\\
54	-6.23374545847146e-13\\
55	-5.92205818554789e-13\\
56	-5.6259552762705e-13\\
57	-5.34465751245697e-13\\
58	-5.07742463683412e-13\\
59	-4.82355340499241e-13\\
60	-4.58237573474279e-13\\
61	-4.35325694800565e-13\\
62	-4.13559410060537e-13\\
63	-3.9288143955751e-13\\
64	-3.73237367579635e-13\\
65	-3.54575499200653e-13\\
66	-3.3684672424062e-13\\
67	-3.20004388028589e-13\\
68	-3.0400416862716e-13\\
69	-2.88803960195802e-13\\
70	-2.74363762186012e-13\\
71	-2.60645574076711e-13\\
72	-2.47613295372876e-13\\
73	-2.35232630604232e-13\\
74	-2.2347099907402e-13\\
75	-2.12297449120319e-13\\
76	-2.01682576664303e-13\\
77	-1.91598447831088e-13\\
78	-1.82018525439534e-13\\
79	-1.72917599167557e-13\\
80	-1.64271719209179e-13\\
81	-1.5605813324872e-13\\
82	-1.48255226586284e-13\\
83	-1.4084246525697e-13\\
84	-1.33800341994121e-13\\
85	-1.27110324894415e-13\\
86	-1.20754808649695e-13\\
87	-1.1471706821721e-13\\
88	-1.08981214806349e-13\\
89	-1.03532154066032e-13\\
90	-9.83555463627302e-14\\
91	-9.34377690445937e-14\\
92	-8.8765880592364e-14\\
93	-8.43275865627458e-14\\
94	-8.01112072346085e-14\\
95	-7.61056468728781e-14\\
96	-7.23003645292342e-14\\
97	-6.86853463027725e-14\\
98	-6.52510789876339e-14\\
99	-6.19885250382522e-14\\
};

\addplot [color=mycolor3,solid,line width=1.0pt, mark=triangle, mark repeat={3}]
  table[row sep=crcr]{%
1	0.033\\
2	0.03135\\
3	0.0297825\\
4	0.028293375\\
5	0.02687870625\\
6	0.0255347709375\\
7	0.024258032390625\\
8	0.0230451307710937\\
9	0.0218928742325391\\
10	0.0207982305209121\\
11	0.0197583189948665\\
12	0.0187704030451232\\
13	0.017831882892867\\
14	0.0169402887482237\\
15	0.0160932743108125\\
16	0.0152886105952719\\
17	0.0145241800655083\\
18	0.0137979710622328\\
19	0.0131080725091212\\
20	0.0124526688836651\\
21	0.328690305103639\\
22	0.419419412595434\\
23	0.382384307129184\\
24	0.34800804607119\\
25	0.31611701718771\\
26	0.286548345411019\\
27	0.259149251800124\\
28	0.233776449799065\\
29	0.21029557667084\\
30	0.126362645003109\\
31	-1.83889626785367e-12\\
32	-1.74695145446098e-12\\
33	-1.65960388173793e-12\\
34	-1.57662368765104e-12\\
35	-1.49779250326849e-12\\
36	-1.42290287810506e-12\\
37	-1.35175773419981e-12\\
38	-1.28416984748982e-12\\
39	-1.21996135511533e-12\\
40	-1.15896328735956e-12\\
41	-1.10101512299158e-12\\
42	-1.045964366842e-12\\
43	-9.93666148499903e-13\\
44	-9.43982841074908e-13\\
45	-8.96783699021162e-13\\
46	-8.51944514070104e-13\\
47	-8.09347288366599e-13\\
48	-7.68879923948269e-13\\
49	-7.30435927750856e-13\\
50	-6.93914131363313e-13\\
51	-6.59218424795147e-13\\
52	-6.2625750355539e-13\\
53	-5.9494462837762e-13\\
54	-5.65197396958739e-13\\
55	-5.36937527110802e-13\\
56	-5.10090650755262e-13\\
57	-4.84586118217499e-13\\
58	-4.60356812306624e-13\\
59	-4.37338971691293e-13\\
60	-4.15472023106728e-13\\
61	-3.94698421951392e-13\\
62	-3.74963500853822e-13\\
63	-3.56215325811131e-13\\
64	-3.38404559520574e-13\\
65	-3.21484331544546e-13\\
66	-3.05410114967318e-13\\
67	-2.90139609218952e-13\\
68	-2.75632628758005e-13\\
69	-2.61850997320104e-13\\
70	-2.48758447454099e-13\\
71	-2.36320525081394e-13\\
72	-2.24504498827325e-13\\
73	-2.13279273885958e-13\\
74	-2.0261531019166e-13\\
75	-1.92484544682077e-13\\
76	-1.82860317447973e-13\\
77	-1.73717301575575e-13\\
78	-1.65031436496796e-13\\
79	-1.56779864671956e-13\\
80	-1.48940871438358e-13\\
81	-1.41493827866441e-13\\
82	-1.34419136473118e-13\\
83	-1.27698179649463e-13\\
84	-1.21313270666989e-13\\
85	-1.1524760713364e-13\\
86	-1.09485226776958e-13\\
87	-1.0401096543811e-13\\
88	-9.88104171662045e-14\\
89	-9.38698963078943e-14\\
90	-8.91764014924996e-14\\
91	-8.47175814178746e-14\\
92	-8.04817023469809e-14\\
93	-7.64576172296318e-14\\
94	-7.26347363681502e-14\\
95	-6.90029995497427e-14\\
96	-6.55528495722556e-14\\
97	-6.22752070936428e-14\\
98	-5.91614467389606e-14\\
99	-5.62033744020126e-14\\
};

\addplot [color=mycolor4,dotted,line width=2.0pt]
  table[row sep=crcr]{%
1	0.033\\
2	0.03135\\
3	0.0297825\\
4	0.028293375\\
5	0.02687870625\\
6	0.0255347709375\\
7	0.024258032390625\\
8	0.0230451307710937\\
9	0.0218928742325391\\
10	0.0207982305209121\\
11	0.0197583189948665\\
12	0.0187704030451232\\
13	0.017831882892867\\
14	0.0169402887482237\\
15	0.0160932743108125\\
16	0.0152886105952719\\
17	0.0145241800655083\\
18	0.0137979710622328\\
19	0.0131080725091212\\
20	0.0124526688836651\\
21	0.0118300354394819\\
22	0.0112385336675078\\
23	0.0106766069841324\\
24	0.0101427766349258\\
25	0.00963563780317949\\
26	0.00915385591302051\\
27	0.00869616311736949\\
28	0.00826135496150101\\
29	0.00784828721342596\\
30	0.107028550765839\\
31	0.176054994141954\\
32	0.157386639018928\\
33	0.140147609545\\
34	0.124241434994842\\
35	0.109577733111028\\
36	0.0960718416412544\\
37	0.0836444715122073\\
38	0.0722213804020971\\
39	0.0617330655428163\\
40	0.052114474647923\\
41	0.0433047339243148\\
42	0.0352468921837952\\
43	0.0278876801258922\\
44	0.0211772839154528\\
45	0.0150691322278548\\
46	0.00951969598131665\\
47	0.0044883000198652\\
48	3.23720252853575e-13\\
49	3.07534240210897e-13\\
50	2.92157528200352e-13\\
51	2.77549651790334e-13\\
52	2.63672169200817e-13\\
53	2.50488560740777e-13\\
54	2.37964132703738e-13\\
55	2.26065926068551e-13\\
56	2.14762629765123e-13\\
57	2.04024498276867e-13\\
58	1.93823273363024e-13\\
59	1.84132109694873e-13\\
60	1.74925504210129e-13\\
61	1.66179228999622e-13\\
62	1.57870267549641e-13\\
63	1.49976754172159e-13\\
64	1.42477916463551e-13\\
65	1.35354020640374e-13\\
66	1.28586319608355e-13\\
67	1.22157003627937e-13\\
68	1.1604915344654e-13\\
69	1.10246695774213e-13\\
70	1.04734360985503e-13\\
71	9.94976429362276e-14\\
72	9.45227607894162e-14\\
73	8.97966227499454e-14\\
74	8.53067916124481e-14\\
75	8.10414520318257e-14\\
76	7.69893794302344e-14\\
77	7.31399104587227e-14\\
78	6.94829149357865e-14\\
79	6.60087691889972e-14\\
80	6.27083307295474e-14\\
81	5.957291419307e-14\\
82	5.65942684834165e-14\\
83	5.37645550592457e-14\\
84	5.10763273062834e-14\\
85	4.85225109409692e-14\\
86	4.60963853939207e-14\\
87	4.37915661242247e-14\\
88	4.16019878180135e-14\\
89	3.95218884271128e-14\\
90	3.75457940057571e-14\\
91	3.56685043054693e-14\\
92	3.38850790901958e-14\\
93	3.2190825135686e-14\\
94	3.05812838789017e-14\\
95	2.90522196849566e-14\\
96	2.75996087007088e-14\\
97	2.62196282656734e-14\\
98	2.49086468523897e-14\\
99	2.36632145097702e-14\\
};

\end{axis}
\end{tikzpicture}%

%% file: scheduling_channel_coefficients.tikz
%
%
\begin{tikzpicture}

\begin{axis}[%
width=0.8\columnwidth,
height=0.26\columnwidth,
at={(0.078in,0.044in)},
scale only axis,
xmin=1,
xmax=20,
xlabel={Time},
xmajorgrids,
ymin=0,
ymax=1,
ylabel={Channel},
axis background/.style={fill=white},
legend style={legend cell align=left,align=left,draw=white!15!black}
]
\addplot [color=blue,solid,line width=2.0pt]
  table[row sep=crcr]{%
1	1\\
2	0.890263248439729\\
3	0.609221612330919\\
4	0.280237476472628\\
5	0.0477174424006623\\
6	0.0137254829619863\\
7	0.193182266988475\\
8	0.507315776382049\\
9	0.818238047634393\\
10	0.989470680605613\\
11	0.945851623682214\\
12	0.706527331315781\\
13	0.376548485161326\\
14	0.100758311861233\\
15	0.000214082336288552\\
16	0.119049385131283\\
17	0.4051018198486\\
18	0.732809526641101\\
19	0.958326188690088\\
20	0.982661942331674\\
};
\addlegendentry{User 1};

\addplot [color=red,dashed,line width=2.0pt]
  table[row sep=crcr]{%
1	0.5625\\
2	0.436283008045598\\
3	0.170917216594961\\
4	0.00457984850674627\\
5	0.0865658531122011\\
6	0.343289261460852\\
7	0.544329761616355\\
8	0.509244849111965\\
9	0.269524743462614\\
10	0.0403285603735656\\
11	0.0273697419563235\\
12	0.242279385668008\\
13	0.492166833579508\\
14	0.552747342561567\\
15	0.369647297575733\\
16	0.107206872167448\\
17	0.000977647223946829\\
18	0.146304925714234\\
19	0.412751218287457\\
20	0.561169505807651\\
};
\addlegendentry{User 2};

\end{axis}
\end{tikzpicture}%

%% file: scheduling_pf_power.tikz
%
%
\begin{tikzpicture}

\begin{axis}[%
width=0.8\columnwidth,
height=0.26\columnwidth,
at={(0.078in,0.044in)},
scale only axis,
xmin=1,
xmax=20,
xmajorgrids,
ymin=0,
ymax=11,
ylabel={Power},
axis background/.style={fill=white},
legend style={legend cell align=left,align=left,draw=white!15!black}
]
\addplot [color=blue,solid,line width=2.0pt]
  table[row sep=crcr]{%
1	10\\
2	5.78947368421053\\
3	7.36842105263158\\
4	7.89473684210526\\
5	1.57894736842105\\
6	0.526315789473684\\
7	10\\
8	6.31578947368421\\
9	8.42105263157895\\
10	10\\
11	3.15789473684211\\
12	0.526315789473684\\
13	8.42105263157895\\
14	0\\
15	0\\
16	8.42105263157895\\
17	7.36842105263158\\
18	4.73684210526316\\
19	3.68421052631579\\
20	8.42105263157895\\
};
\addlegendentry{User 1};

\addplot [color=red,dashed,line width=2.0pt]
  table[row sep=crcr]{%
1	7.36842105263158\\
2	9.47368421052632\\
3	3.68421052631579\\
4	0\\
5	8.94736842105263\\
6	5.78947368421053\\
7	3.68421052631579\\
8	0\\
9	0\\
10	0\\
11	0\\
12	10\\
13	0\\
14	9.47368421052632\\
15	2.63157894736842\\
16	5.78947368421053\\
17	0\\
18	0\\
19	0\\
20	2.10526315789474\\
};
\addlegendentry{User 2};

\end{axis}
\end{tikzpicture}%

%% file: scheduling_pf_rate.tikz
%
%
\begin{tikzpicture}

\begin{axis}[%
width=0.8\columnwidth,
height=0.26\columnwidth,
at={(0.078in,0.044in)},
scale only axis,
xmin=1,
xmax=20,
xlabel={Time},
xmajorgrids,
ymin=0,
ymax=1.2,
ylabel={Average rate},
axis background/.style={fill=white}
]
\addplot [color=blue,solid,line width=2.0pt,forget plot]
  table[row sep=crcr]{%
1	0\\
2	1.07967884873652\\
3	0.887432204963994\\
4	1.03260686171209\\
5	1.06620981456124\\
6	0.861284134922346\\
7	0.718139306577907\\
8	0.686460688236822\\
9	0.780160621462712\\
10	0.922992783883425\\
11	1.06952121121139\\
12	1.09802055133064\\
13	1.01511343344036\\
14	1.04688471842558\\
15	0.97210723853804\\
16	0.907300089302171\\
17	0.880690808863767\\
18	0.910210612444868\\
19	0.942846550898784\\
20	0.972742576134146\\
};
\addplot [color=red,dashed,line width=2.0pt,forget plot]
  table[row sep=crcr]{%
1	0\\
2	0.319757795637181\\
3	0.416773242310383\\
4	0.314049736516412\\
5	0.235537302387309\\
6	0.296925989180369\\
7	0.429045342604603\\
8	0.442207981011263\\
9	0.386931983384855\\
10	0.343939540786538\\
11	0.309545586707884\\
12	0.281405078825349\\
13	0.342740267802895\\
14	0.316375631818057\\
15	0.424522211661904\\
16	0.441516169104399\\
17	0.430795479344709\\
18	0.40545456879502\\
19	0.382929314973074\\
20	0.362775140500807\\
};
\end{axis}
\end{tikzpicture}%

%% file: scheduling_er_power.tikz
%
%
\begin{tikzpicture}

\begin{axis}[%
width=0.8\columnwidth,
height=0.26\columnwidth,
at={(0.078in,0.044in)},
scale only axis,
xmin=1,
xmax=20,
xmajorgrids,
ymin=0,
ymax=11,
ylabel={Power},
axis background/.style={fill=white},
legend style={legend cell align=left,align=left,draw=white!15!black}
]
\addplot [color=blue,solid,line width=2.0pt]
  table[row sep=crcr]{%
1	10\\
2	8.94736842105263\\
3	2.10526315789474\\
4	1.57894736842105\\
5	10\\
6	10\\
7	10\\
8	10\\
9	5.26315789473684\\
10	0.526315789473684\\
11	0.526315789473684\\
12	2.10526315789474\\
13	10\\
14	8.94736842105263\\
15	10\\
16	10\\
17	5.26315789473684\\
18	1.57894736842105\\
19	2.10526315789474\\
20	10\\
};
\addlegendentry{User 1};

\addplot [color=red,dashed,line width=2.0pt]
  table[row sep=crcr]{%
1	10\\
2	10\\
3	10\\
4	10\\
5	10\\
6	10\\
7	5.26315789473684\\
8	5.26315789473684\\
9	10\\
10	10\\
11	10\\
12	10\\
13	10\\
14	1.57894736842105\\
15	1.05263157894737\\
16	10\\
17	10\\
18	10\\
19	10\\
20	10\\
};
\addlegendentry{User 2};

\end{axis}
\end{tikzpicture}%

%% file: scheduling_er_rate.tikz
%
%
\begin{tikzpicture}

\begin{axis}[%
width=0.8\columnwidth,
height=0.26\columnwidth,
at={(0.078in,0.044in)},
scale only axis,
xmin=1,
xmax=20,
xlabel={Time},
xmajorgrids,
ymin=0,
ymax=0.7,
ylabel={Average rate},
axis background/.style={fill=white}
]
\addplot [color=blue,solid,line width=2.0pt,forget plot]
  table[row sep=crcr]{%
1	0\\
2	0.460028607334816\\
3	0.610152741980001\\
4	0.554510833751728\\
5	0.514176573105179\\
6	0.466438272774823\\
7	0.404160396166073\\
8	0.404310028453275\\
9	0.455660535934285\\
10	0.487355925827294\\
11	0.471743660390949\\
12	0.459924235310918\\
13	0.452304030701137\\
14	0.455152483605079\\
15	0.451007570163764\\
16	0.422915845123527\\
17	0.424742293130734\\
18	0.46420638551253\\
19	0.46004321832029\\
20	0.453630955547213\\
};
\addplot [color=red,dashed,line width=2.0pt,forget plot]
  table[row sep=crcr]{%
1	0\\
2	0.206506156871774\\
3	0.26983993405211\\
4	0.34211130728479\\
5	0.279940288534928\\
6	0.310155202391974\\
7	0.464551338290405\\
8	0.491690830414029\\
9	0.477676604244613\\
10	0.470981588298288\\
11	0.449548526562838\\
12	0.426071658266448\\
13	0.44561001417443\\
14	0.46445224954768\\
15	0.458671234649156\\
16	0.45050798441669\\
17	0.447441981253095\\
18	0.422757232472778\\
19	0.42775719843514\\
20	0.449468354787179\\
};
\end{axis}
\end{tikzpicture}%